%% file: main.tex
\documentclass[10pt,journal,compsoc]{IEEEtran}

\ifCLASSOPTIONcompsoc
  \usepackage[nocompress]{cite}
\else
  \usepackage{cite}
\fi
\ifCLASSINFOpdf
\else
\fi
\usepackage{float}
\usepackage{threeparttable}
\usepackage[acronyms,nonumberlist,nopostdot,nomain,nogroupskip]{glossaries}
\usepackage{tablefootnote}
\usepackage{booktabs}

\usepackage{adjustbox}
\usepackage{pgfplots, pgfplotstable}
\usepackage{booktabs,makecell,multirow,hhline} 
\usepackage{tikz}
\usepackage{pgfplots}
\pgfplotsset{compat=newest}
\pgfplotsset{plot coordinates/math parser=false}
\newlength\fheight
\newlength\fwidth
\usetikzlibrary{plotmarks,patterns,decorations.pathreplacing,backgrounds,calc,arrows,arrows.meta,spy,matrix}
\usepgfplotslibrary{patchplots,groupplots,colorbrewer}
\usepackage{tikzscale}

\usepackage[pdftex]{pict2e}
\usepackage{comment}

\newcolumntype{M}[1]{>{\centering\arraybackslash}m{#1}}
\usepackage{amsfonts}

\newif\ifexttikz
\exttikzfalse

\ifexttikz
	\usetikzlibrary{external}
\fi

\usepackage{enumitem}
\usepackage{multirow}
\newlength{\starsize}
\newlength{\starspread}
\tikzset{starsize/.code={\setlength{\starsize}{#1}},
         starspread/.code={\setlength{\starspread}{#1}}}
\tikzset{starsize=1mm,
         starspread=3mm}
\pgfdeclarepatternformonly[\starspread,\starsize]% variables
  {custom fivepointed stars}% name
  {\pgfpointorigin}% lower left corner
  {\pgfqpoint{\starspread}{\starspread}}% upper right corner
  {\pgfqpoint{\starspread}{\starspread}}% tilesize
  {% shape description
   \pgftransformshift{\pgfqpoint{\starsize}{\starsize}}
   \pgfpathmoveto{\pgfqpointpolar{18}{\starsize}}
   \pgfpathlineto{\pgfqpointpolar{162}{\starsize}}
   \pgfpathlineto{\pgfqpointpolar{306}{\starsize}}
   \pgfpathlineto{\pgfqpointpolar{90}{\starsize}}
   \pgfpathlineto{\pgfqpointpolar{234}{\starsize}}
   \pgfpathclose%
   \pgfusepath{fill}
  }         

\renewcommand{\figurename}{Fig.~}
\usepackage[font=footnotesize]{subcaption}
\usepackage[font=footnotesize]{caption}
\usepackage[T1]{fontenc}

\usepackage{mathtools}
\usepackage{array}
\newcolumntype{L}{>{\centering\arraybackslash}}
\input{acronyms.tex}

\usepackage{hhline} % used for the top-most horizontal line! in order to fix the top left corner cell empty

\definecolor{Mycolor1}{HTML}{4b006e}
\definecolor{Mycolor2}{HTML}{7e1e9c}
\definecolor{Mycolor3}{HTML}{9a0eea}

\usepackage{ragged2e}
\usepackage[utf8]{inputenc}
\usepackage{pgfplots}
\usepackage{amsmath}
\usetikzlibrary{patterns}
\usepackage{soul}

\def\BibTeX{{\rm B\kern-.05em{\sc i\kern-.025em b}\kern-.08em
    T\kern-.1667em\lower.7ex\hbox{E}\kern-.125emX}}
\begin{document}

\title{SignCRF: Scalable Channel-agnostic Data-driven Radio Authentication System}
%{%Real-Time Radio Fingerprinting Through Radio Frequency Signature Emulator}

\author{Amani Al-shawabka$^\dagger$,~\IEEEmembership{Student,~IEEE,}
        Philip Pietraski$^\ddagger$,~\IEEEmembership{Sr. Member,~IEEE,} 
        Sudhir B Pattar$^\ddagger$,~\IEEEmembership{Member,~IEEE,}
        Pedram~Johari$^\dagger$,~\IEEEmembership{Member,~IEEE,}
        and~Tommaso Melodia$^\dagger$,~\IEEEmembership{Fellow,~IEEE}% <-this % stops a space

\thanks{$^\dagger$Institute for the Wireless Internet of Things, Northeastern University, Boston, MA, USA, E-mail: \{al-shawabka.a, p.johari, t.melodia\}@northeastern.edu. $^\ddagger$InterDigital Inc., United States. E-mail: \{philip.pietraski, sudhir.pattar\}@interdigital.com.}
\thanks{This article is based upon work partially supported by the U.S. National
Science Foundation under grants CNS-2112471, and CNS1923789, and InterDigital Inc.}}

%\maketitle

\IEEEtitleabstractindextext{%

\begin{justify}
\begin{abstract}

Radio Frequency Fingerprinting through Deep Learning
(RFFDL) is a data-driven IoT authentication technique that leverages the unique hardware-level manufacturing imperfections associated with a particular device to recognize (``fingerprint'') the device itself based on variations introduced in the transmitted waveform.
%unmet need/challenge
Key impediments in developing robust and scalable \gls{rffdl} techniques that are practical in dynamic and mobile environments are the non-stationary behavior of the wireless channel and other impairments introduced by the propagation conditions. To date, the existing \gls{rffdl}-based techniques have only been able to demonstrate a desirable performance when the training and testing environment remains the same, which makes the solutions impractical.
%proposed solution
\textit{SignCRF} brings to the RFFDL landscape what it has been missing so far: a scalable, channel-agnostic data-driven radio authentication platform with unmatched precision in fingerprinting wireless devices based on their unique manufacturing impairments that is {\em independent of the dynamic nature of the environment or channel irregularities caused by mobility}. \textit{SignCRF} consists of: (i) a classifier developed in a base-environment with minimum channel dynamics, and finely trained to authenticate devices with high accuracy and at scale; (ii) an environment translator that is carefully designed and trained to remove the dynamic channel impact from RF signals while maintaining the radio's specific ``signature''; and (iii) a Max Rule module that selects the highest precision authentication technique between the baseline classifier and the environment translator per radio. We design, train, and validate the performance of \textit{SignCRF} for multiple technologies in dynamic environments and at scale (100 LoRa and 20 WiFi devices, the largest datasets available in the literature). We demonstrate that \textit{SignCRF} can significantly improve the \gls{rffdl} performance by achieving as high as 100\% correct authentication for WiFi devices and 80\% correctly authenticated LoRa devices, a 5x and 8x improvement when compared to the state-of-the-art respectively. Furthermore, we show that \textit{SignCRF}  is resilient to adversarial actions by reducing the device recognition accuracy from 73\% to 6\%, which translates into zero mis-authentication of adversary radios that try to impersonate legitimate devices, which has not been achieved by any prior \gls{rffdl} techniques.

%results

\end{abstract}
\end{justify}

}

\maketitle
\IEEEdisplaynontitleabstractindextext
\IEEEpeerreviewmaketitle

\begin{picture}(0,0)(10,-430)
\put(0,0){
\put(0,0){\footnotesize This work has been submitted to the IEEE for possible publication.}
\put(0,-10){
\footnotesize Copyright may be transferred without notice, after which this version may no longer be accessible.}}
\end{picture}

\IEEEraisesectionheading{\section{Introduction and Motivation}\label{sec:intro}}

%\section{Introduction and Motivation}\label{sec:intro}
\glsresetall

%\noindent\textbf{Motivation:}

Achieving full-fledged  privacy and security for \gls{iot} systems is still a challenging problem. Applying traditional security techniques such as private- or public-key cryptography to \gls{iot} applications suffer from key limitations including their energy consumption inefficiency, required computation power, and scalability challenges~\cite{yang2017survey}. An alternative approach that has recently gained traction in the research and industry communities is the concept of \gls{rffdl}. The \gls{rffdl} is as an emerging data-driven authentication approach that has shown promising results in authenticating \gls{iot} radios in a more energy-efficient and scalable fashion~\cite{shawabka2020exposing,restuccia2019deepradioid,DARPA, sankhe2019oracle,merchant2018deep}. It relies on the hardware-level imperfections that are uniquely associated with the \gls{rf} analog circuitry of every single device, referred to as \gls{rf} ``\textit{fingerprint}'' or ``\textit{signature}''. %during the transmission process.
These distinctive features are extracted by training deep learning models that operate at the physical layer of the wireless network protocol stack, largely deployed at the base station side, with minimal impact on the user device's computation and power resources.

\begin{figure*}[t!]
    \centering
    \includegraphics[width=1\linewidth]{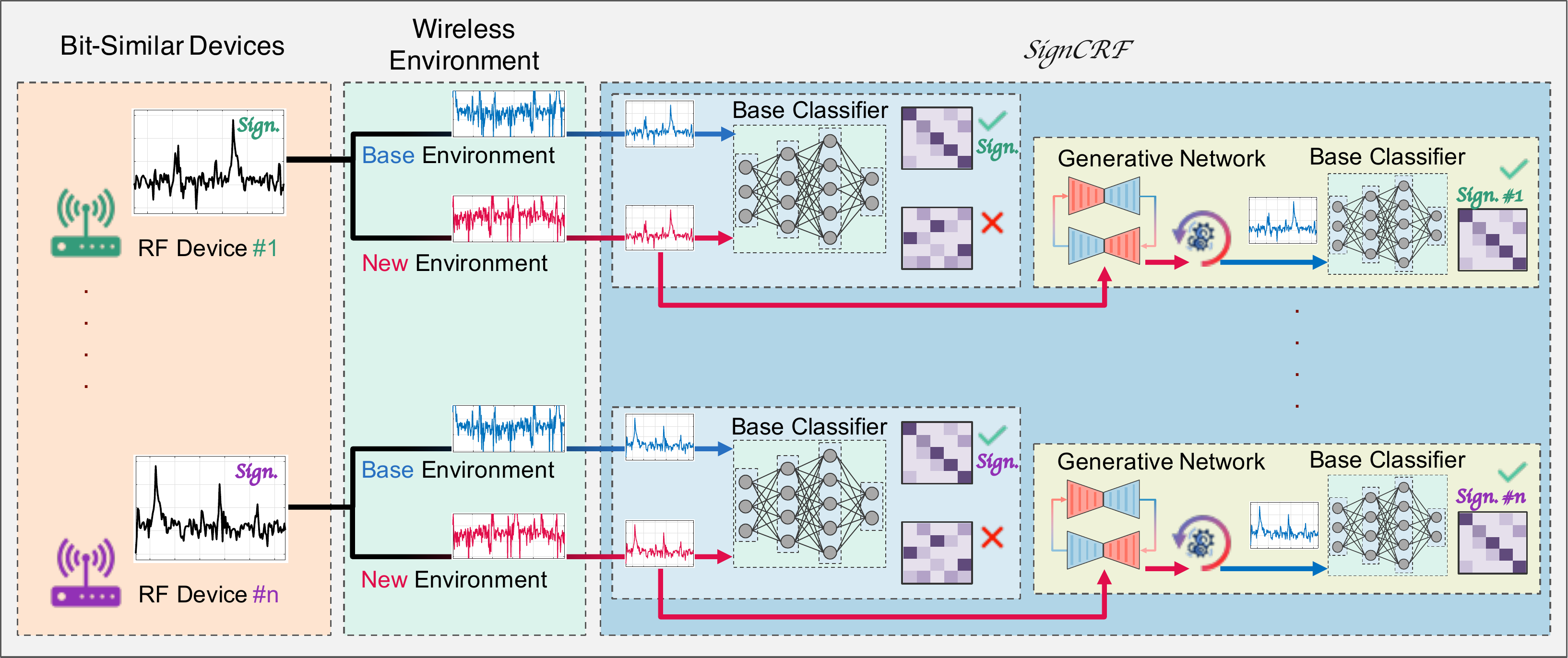}
    %\vspace{-20pt}
    \caption{(i)\textit{Bit-Similar} radio devices (radios with identical manufacturing processes that transmit the same baseband waveforms) labeled from 1 to n, (ii) transmitting waveforms over two environments. Base environment dataset employed to train a preliminary deep learning classifier, and (iii) \textit{SignCRF} reshapes the waveforms transmitted from the new environment to reform the unique hardware impairments associated with these waveforms to match the primary impairments associated with the base environment.} %\vspace{0.2cm}
    %\vspace{-20pt}
    \label{fig:cgan_intro}
\end{figure*}

In spite of the promising progress in this domain, recent studies have revealed a number of challenges in developing robust and reliable \gls{rffdl} models for practical applications. Key among them are the non-stationary characteristic of the wireless channel and the dynamic nature of the environmental actions in a mobile scenario, which significantly limit fingerprinting accuracy by obscuring the hardware impairments associated with the transmitted waveform~\cite{shawabka2020exposing, restuccia2019deepradioid,al2021deeplora}. In particular, without intervention of sophisticated \gls{dsp} algorithms, the existing \gls{rffdl} models suffer from significant performance degradation when training and testing of their deep learning classifier are done on separate segments of the data that are collected in different days/environments. This problem is introduced as the worst-case scenario in~\cite{restuccia2019deepradioid}, \cite{shawabka2020exposing}, \cite{al2021deeplora}, and \cite{soltanimore}, but to the best of our knowledge has not been resolved yet.

In an effort to address these unique challenges, the authors of~\cite{restuccia2019deepradioid} proposed the \textit{DeepRadioID} system that leverages a carefully optimized digital \gls{fir} filter at the transmitter’s side, which slightly modifies its base-band signal to compensate for current channel conditions; however, the system depends on an overhead \gls{dsp} at the transmitter with feedback required from the receiver. %In particular, DeepRadioID transmitter adaptively reshapes the newly transmitted signals based on a feedback received from the receiver side to compensate for the channel impact. 
Besides the need for additional computational resources, this approach can be challenging to implement as a practical radio fingerprinting approach %for applications such as radio fingerprinting
due to its dependency on the feedback signal and the manipulation of the transmitted signal.
%But RF Signature Emulator doesn't manipulate the transmitted signal to compensate for the channel impact.
In another work, the authors of~\cite{al2021deeplora} developed \textit{DeepLoRa}, a data augmentation mechanism to improve the \gls{rffdl} performance using LoRa technology datasets collected from 100 devices with \textit{bit-similar} radios (\textit{bit-similar} radios are devices with identical manufacturing processes that transmit almost the same packets over different days). However, the \textit{DeepLoRa} solution (i) lacks generalization because the solution is applicable only to LoRa technology; (ii) even though it enhances the classification accuracy of the dataset collected from LoRa radios spanning different days, its improvement is not enough for \gls{rffdl} practical implementations. The authors reported 22\% accuracy as their best fingerprinting result, and \textit{DeepLoRa} was only able to correctly recognize 7 devices out of the 20 devices under test, leaving it an impractical solution.

%\noindent \textbf{Unmet Need.}
Up to this date, the developed \gls{rffdl}-based models lack a generic and robust authentication process for a practical real-time radio fingerprinting system that is (i) independent of real-time feedback from the receiver, (ii) does not rely on overhead \gls{dsp} or artificial impairments, and (iii) preserves its high performance independent of the changes in a  mobile wireless environment. Most importantly, the dynamic behavior of the non-stationary wireless channel along with the environmental changes alter the transmitted signals in ways that makes it challenging even for the finest and most sophisticated classifiers to extract the unique impairments associated to each device. This significantly impacts the fingerprinting accuracy especially when the \gls{rffdl} \gls{nn} is trained on a dataset that is collected in a day or environment that is different from the day or environment that is used for classification inference --the main reason that none of the prior work were successful in achieving good performance in this worst-case scenario \cite{restuccia2019deepradioid,jian2020deep,soltanimore,shawabka2020exposing,al2021deeplora}.\\

Our proposed \gls{signcrf} brings to the radio fingerprinting landscape what it has been missing so far: a scalable, channel-agnostic data-driven radio authentication platform with unmatched precision in fingerprinting wireless devices based on their unique manufacturing impairments, and independent of the environment/channel irregularities. \textit{SignCRF} consists of three main components (Fig.~\ref{fig:cgan_intro}): (i) a baseline classifier that is developed in a base environment with minimum channel dynamics, and finely trained to authenticate devices with high accuracy and at scale; (ii) a domain/environment translator that is carefully designed and trained to remove the channel impairments from RF signals while it maintains and highlights the radio specific ``signature'' by leveraging the \gls{cgan}~\cite{zhu2017unpaired, pix2pix2017, goodfellow2014generative} concept from computer vision; and (iii) a Max Rule module that compares and selects the highest precision authentication technique between the baseline classifier and the environment translator output per radio. We design, train and validate the performance of \gls{signcrf} for multiple technologies in dynamic environments and at scale (100 LoRa and 20 WiFi devices, the largest datasets available in the literature). We demonstrate that \textit{SignCRF} can significantly improve the \gls{rffdl} performance by achieving as high as 100\% correct authentication for WiFi devices and 80\% correctly authenticate LoRa devices, a 5x and 8x improvement when compared to the state-of-the-art respectively. Furthermore, we show that \gls{signcrf} is resilient to adversarial actions by suppressing the device recognition accuracy from 73\% to 6\% which leads to zero mis-authentication of adversary radios who try to impersonate legitimate devices.\\

%Hence, neither of these techniques can practically resolve the issue since these techniques still do not know what characteristics are embedded in the old signal used to train the classifier and are not available in the new dataset.

% \begin{figure}[b!]
%         \centering
%         \vspace{-10pt}
%         \input{./Figures/mix_days_wifi_10radios}
%         \vspace{-5pt}
%         \caption{Classification accuracy of 10 WiFi radios. Training dataset created by mixing dataset collected over several days for three different Neural Networks, the testing dataset related to a day other than the days used to train the Neural Network models.}
%         \label{fig:mixed_days_10radios}
% \end{figure}

%\newpage
\noindent \textbf{Summary of Novel Contributions.} \gls{signcrf} is a novel \gls{rffdl} system that, for the first time in the literature, decouples the hardware impairments from channel and environmental conditions to achieve a channel-agnostic and practical data-driven authentication technique. A summary of our novel contributions are as follows:

{$\bullet$} We propose a framework for the real-time channel- and adversary-resilient approach of \gls{rffdl}. The key innovation behind \gls{signcrf}, as illustrated in Fig.~\ref{fig:cgan_intro}, is to \textbf{reshape the classifier input dataset} collected from a particular device at a time different from the time used to train that classifier, and \textbf{decouple the radio and channel impairments}. \gls{signcrf} leverages the \gls{gan} concept to apply controlled modifications to the IQ samples to decouple the environmental variations caused by data collection in different days while maintaining and highlighting the hardware impairments associated with a specific radio device.  %where both IQ samples transmitted from the same device and transmitting the same data structure.
%The RF Signature Emulator slightly modifies a baseband signal in one day/environment to compensate for the new wireless channel conditions and environmental changes emphasized in another day/environment's signal that is used to train the primary classifier.

{$\bullet$} We extensively evaluate the performance of the \gls{signcrf} system on experimental datasets (i) \textbf{20 WiFi} \textit{bit-similar} devices, i.e., transmitting the same baseband signal through nominally-identical RF interfaces and antennas,  introduced in \cite{shawabka2020exposing}; (ii) \textbf{100 LoRa} \textit{bit-similar} devices introduced in \cite{al2021deeplora}.

{$\bullet$} We demonstrate that \gls{signcrf} improves the classification accuracy of 5 and 20 WiFi \textit{bit-similar} devices (under the worst-case scenario) from 18\% to 83\% and from 9\% to 34\%, respectively, and the number of correctly classified devices from \textbf{20\% to 100\%} and from \textbf{30\% to 90\%} for the same radios dataset, respectively.
%, when compared to \hl{what reference are we benchmarking this against?}\textcolor{blue}{Amani. in \cite{shawabka2020exposing} Average TTOD of these 20 devices over10 training days using 10 different testing days was 5\%. The only problem is that in that paper we defined TTOD as train test one day and we used TDTA Train-One-Day-Test-Another}.
To the best of our knowledge, we are the first ever to achieve as high as 100\% precise authentication in the worst-case scenario of ``in the wild'' WiFi benchmark dataset when testing the pre-trained classifier with data collected on a day other than the day used to train the model.

{$\bullet$} We demonstrate that \gls{signcrf} outperforms the data augmentation technique proposed in \cite{al2021deeplora} for the LoRa technology. %where 22\% was reported as the best accuracy for fingerprinting 20 \textit{bit-similar} LoRa radios using payload dataset after applying DeepLoRa.
\gls{signcrf} improves the number of correctly classified/authenticated devices \textbf{from 20\% to 80\%} and \textbf{from 15\% to 75\%} for 5 and 20 LoRa radios using their payload datasets, respectively. Besides, to the best of our knowledge, we are the first to improve the fingerprinting performance of the LoRa device authentication based on ``preamble'' only datasets, which demonstrates the robustness of \gls{signcrf}.

%The results show improvements of (i) \textbf{from 18\% to 80\%}, \textbf{from 13\% to 73\%} and \textbf{from 6\% to 53\%} for 5, 20 and 100 \textit{bit-similar} LoRa radios using their payload datasets, respectively, and
%the number of correctly classified devices \textbf{from 20\% to 80\%}, \textbf{from 15\% to 75\%} and \textbf{from 9\% to 73\%} for the same radios' datasets respectively;
%(ii) \textbf{from 5\% to 39\%}, \textbf{from 4\% to 38\%} and \textbf{from 2\% to 9\%}, and the number of correctly classified devices from \textbf{5\% to 45\%}, \textbf{from 5\% to 60\%} and \textbf{from 1\% to 29\%} for 5, 20 and 100 \textit{bit-similar} LoRa radios using their preamble datasets, respectively.
%our system; the LoRa Preamble dataset is not the best candidate to use for practical RFFP. Even though our proposed system improved the fingerprinting performance significantly using this dataset. For practical use, we recommend counting on LoRa payload dataset. 

{$\bullet$} Lastly, we demonstrate the \textit{scalability} of the \gls{signcrf}, by showcasing the classification of \textbf{100 LoRa \textit{bit-similar} radios} using LoRa payload dataset. We report an improvement in device recognition at this scale \textbf{from 9\% to 73\%}, a \textbf{8x improvement} compared to the state-of-the-art~\cite{al2021deeplora}.

In addition to the above contributions, the codes used in this paper will be made publicly available for the researchers use.

\section{Related Work}\label{sec:related}

Shawabka \emph{et al.}~\cite{shawabka2020exposing} were the first to propose a systematic investigation of the role of the wireless channel in RFFDL algorithms' overall performance. This extensive evaluation determined a significant impact of the wireless channel actions on the accuracy of \gls{cnn}-based radio fingerprinting algorithms. Sankhe \emph{et al.}~\cite{sankhe2019oracle} introduced the ORACLE system to mitigate the channel impacts. However, it requires injecting pre-defined impairments to the transmitter, which is not a practical solution for existing/pre-deployed radio devices. Some other recent works~\cite{al2021deeplora,soltanimore,piva2021tags} propose data augmentation techniques to combat the wireless channel action in \gls{rffdl} applications. However, finding the perfect channel simulation that can enhance the fingerprinting for different technologies at scale is challenging and impractical. To the best of our knowledge, no existing work in the literature reported a practical radio fingerprinting solution to date that can perform reliably and independent of the environment and wireless channel dynamics.

Restuccia \emph{et al.}~\cite{restuccia2019deepradioid} proposed DeepRadioID, a system to improve the channel resilience of \gls{rffdl} algorithms through the application of carefully optimized \gls{fir} filters. However, our work is different as we propose a framework to enhance the fingerprinting accuracy based on deep learning processes, with no extra digital signaling processes, and without need to make any changes at the transmitter nor to incorporate the receiver feedback.

In another work~\cite{al2021deeplora}, authors have investigated data augmentation, and reported a marginal increase in \gls{rffdl} performance using the LoRa payload dataset. The best results after applying the data augmentation on the worst-case scenarios using 10 and 20 \textit{bit-similar} radios were an increase from 19\% to 36\% and from 13\% to 22\%, respectively. While the increased percentage reported in ~\cite{al2021deeplora} brings some hope in cracking the problem, it is not close to a practical solution that is commercially implementable at scale.

%about CycleGAN in Computer-vision
\gls{signcrf} brings to the radio fingerprinting landscape what it has been missing so far: a scalable, channel-agnostic data-driven radio authentication platform with unmatched precision in fingerprinting wireless devices based on their unique manufacturing impairments, and independent of the environment/channel irregularities. \textit{SignCRF} consists of: (i) a baseline classifier that is developed in a base environment with minimum channel dynamics, and finely trained to authenticate devices with high accuracy and at scale; (ii) a domain/environment translator that is carefully designed and trained to remove the channel impairments from RF signals while it maintains and highlights the radio specific ``signature'' by leveraging the \gls{cgan}~\cite{zhu2017unpaired, pix2pix2017, goodfellow2014generative} concept from computer vision; and (iii) a Max Rule module that compares and selects the highest precision authentication technique between the baseline classifier and the environment translator output per radio. We design, train and validate the performance of \gls{signcrf} for multiple technologies in dynamic environments and at scale (100 LoRa and 20 WiFi devices, the largest datasets available in the literature). Furthermore, we show that \gls{signcrf} is resilient to adversarial actions which cannot be achieved by the prior \gls{rffdl} techniques.

\section{\gls{signcrf}: Channel-Agnostic Radio \mbox{Authentication} Model}\label{sec:cgan}
\vspace{10pt}
\subsection{\gls{signcrf} Framework Design}\label{sec:cgan_framework}

%The discriminator and generator models are trained like standard GAN models. The generators learn to trick the discriminators better, and the discriminators learn to detect fake IQ slices better. Together, the models find equilibrium during the training process.
%Additionally, the generator models are regularized to create new IQ samples in the target domain, which is a  translated version of the input IQ slices from the source domain. This is achieved by using generated IQ slices as input to the corresponding generator model and comparing the output IQ slices to the original slices.

\begin{figure}[!t]
    \centering
    \includegraphics[width=1\columnwidth]{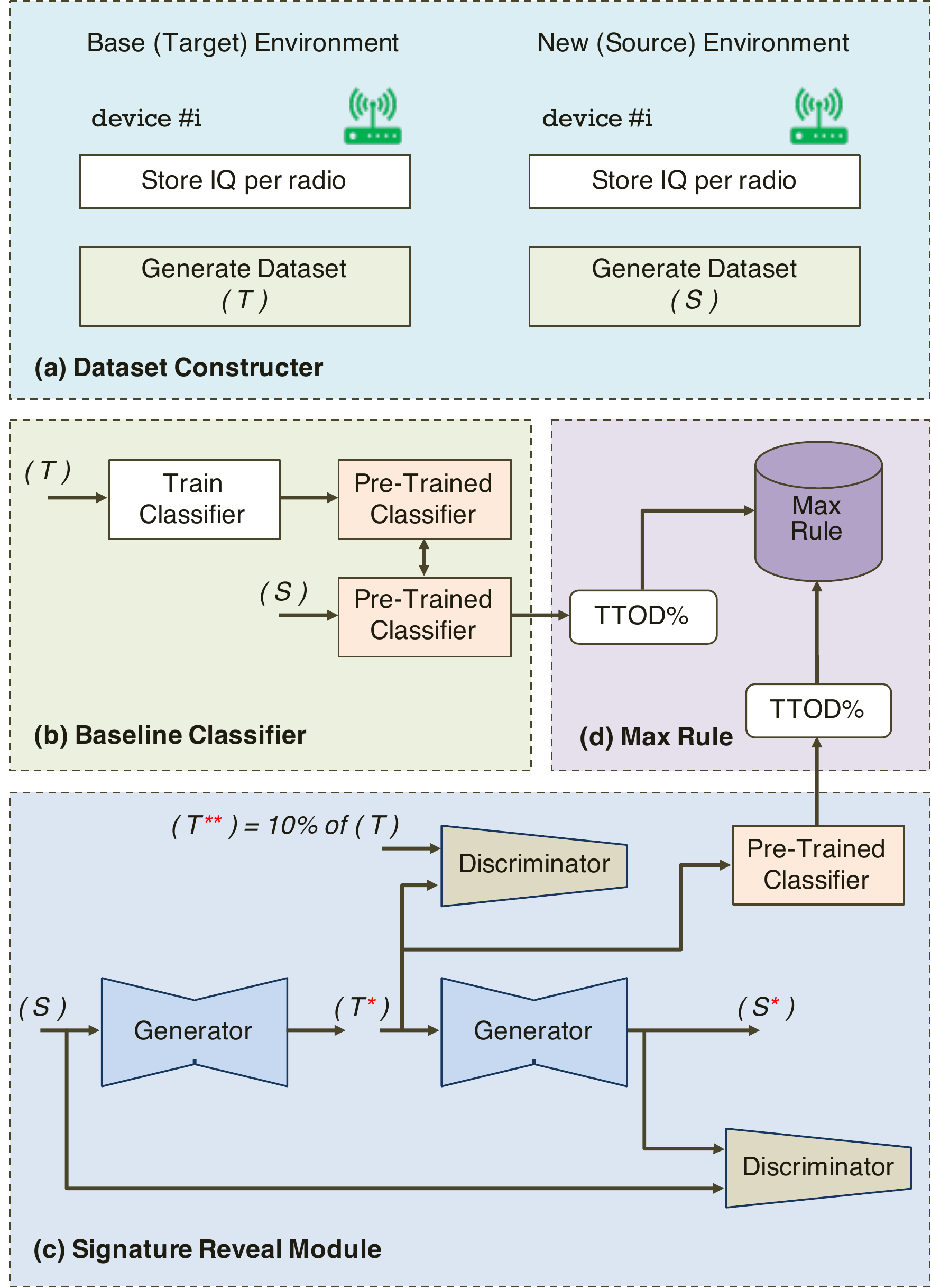}
    \caption{\gls{signcrf} framework: (a) Dataset constructer; (b) Baseline classifier using base (target) domain dataset; (c) Signature reveal module leveraging the \gls{cgan} concept; and (d) Apply Maximum Rule to (b) and (c) outputs.} %\vspace{0.2cm}
    %\vspace{-10pt}
    \label{fig:cgan_app}
\end{figure}

Fig.~\ref{fig:cgan_app} shows a high-level overview of the \gls{signcrf} architecture, which consists of 3 main modules:

$\bullet$ \textit{\underline{First}}. Figures~\ref{fig:cgan_app}(a) and ~\ref{fig:cgan_app}(b) show a baseline \gls{cnn} that is designed and optimized for best performance in a base environment with limited changes in the channel dynamics, i.e., trained and tested with a dataset collected in a short period of time in the same day and at the same environment. This classifier is then used as a baseline to classify any new dataset, including the data collected at different conditions, environments and/or timing (Fig.~\ref{fig:cgan_app}(a)). We employ the \gls{2d} baseline \gls{cnn} model proposed in~\cite{al2021deeplora} and illustrated later in Fig.~\ref{fig:cgan_arc}(d). This model consists of \gls{2d} ConvLayers with rectified linear units (ReLU) activation functions, each followed by a MaxPooling layer. This setup was repeated five times and then fed to three fully connected (FC) layers with ReLU activation. Finally, an FC layer with a Softmax activation function is used to generate the radio classification probabilities; 

$\bullet$ \textit{\underline{Second}}. A domain/environment translator that is carefully designed and trained to remove the channel impairments from RF signals while it maintains and highlights the radio specific ``signature''. We create and customize this RF ``signature'' reveal model per each device/radio by employing the methodology described later in detail (\ref{sec:cgan_aproach}). Fig.~\ref{fig:cgan_app}(c) shows the implementation of the RF ``signature-reveal'' model, which includes two generators and two discriminators employed to convert statistical representation of IQ sequences collected at the time, day, or environment $S$ (source domain $S$) from radio $i$ to IQ slices that represent the same statistical characteristic pertaining to the time, day, or environment $T$ (target domain $T$) collected from the same radio, where $S \neq T$. Our approach assumes that the unique radio impairments did not vanish over time or environment. However, they are altered by the channel and the environment. \gls{signcrf} maintains and highlights these device-specific impairments by removing the channel dynamics that are introduced in the new environment;

$\bullet$ \textit{\underline{Third}}. In the third step, as presented in Fig.~\ref{fig:cgan_app}(d), we compare the classification outputs of the first and second modules by using the ``TTOD'' metric, which represents the Train-and-Test-in-Other-Domain as explained later in Sec.~\ref{sec:cgan_met}. We apply a Max Rule criteria as described in equation~(\ref{eq:max_rule}) to pick the highest classification accuracy that determines whether the RF signature-reveal model is required for a specific radio or not. This step is essential, especially when the source domain dataset is not altered much during the transmission and the baseline classifier can still authenticate the device with high confidence. This step is considered as a precautionary step to avoid adding unnecessary complexity to the system and prevent potential accuracy degradation due to the signal manipulation caused by the signature-reveal module.
\begin{equation}\label{eq:max_rule}
     Max Rule = \arg \max_{i} Pr_{softmax_i} (S_i,  S'_i),
\end{equation}
\noindent where $i$ is the radio label, $Pr_{softmax_i}$ is the classification probability of radio $i$ after applying Softmax layer in the pre-trained classifier shown in Fig.~\ref{fig:cgan_app}, $S_i$ is the testing IQ slices collected in source domain $S$ before applying RF signature-reveal module, and $S'_i$ is the testing IQ slices collected in domain $S$ after applying RF signature-reveal, $F(S)$.   

%In this case, we note that any slight alteration can negatively influence the fingerprinting accuracy if the RF signature-reveal model customization is not precise enough.
%It is worth mentioning that our dataset was collected from \textit{bit-similar} devices transmitting almost the same dataset structure (which is considered the worst-case scenario in radio fingerprinting~\cite{shawabka2020exposing}); therefore, altering the characteristics of these radios (radios with high classification accuracy before RF Signature Emulator) may result in degradation of accuracy.

%The RF Signature Emulator creates deep learning models that can transform any sample related to that specific radio in a real-time manner before being fed to the Baseline classifier without the need to re-train a new RF Signature Emulator. Our approach has proved its capability in reshaping samples collected in a third domain, i.e., not related to $S$ nor $T$, and not seen by the RF Signature Emulator or the Baseline classifier during the training phase. The results show that the fingerprinting accuracy, in this case, is still improved by a factor of 2x when testing the Baseline classifier before and after RF Signature Emulator as explained in detail later in Sec. \ref{sec:wifi_res}.

\subsection{RF Signature-Reveal Module}\label{sec:cgan_aproach}

\begin{figure}[!t]
    \centering
    \includegraphics[width=1\columnwidth]{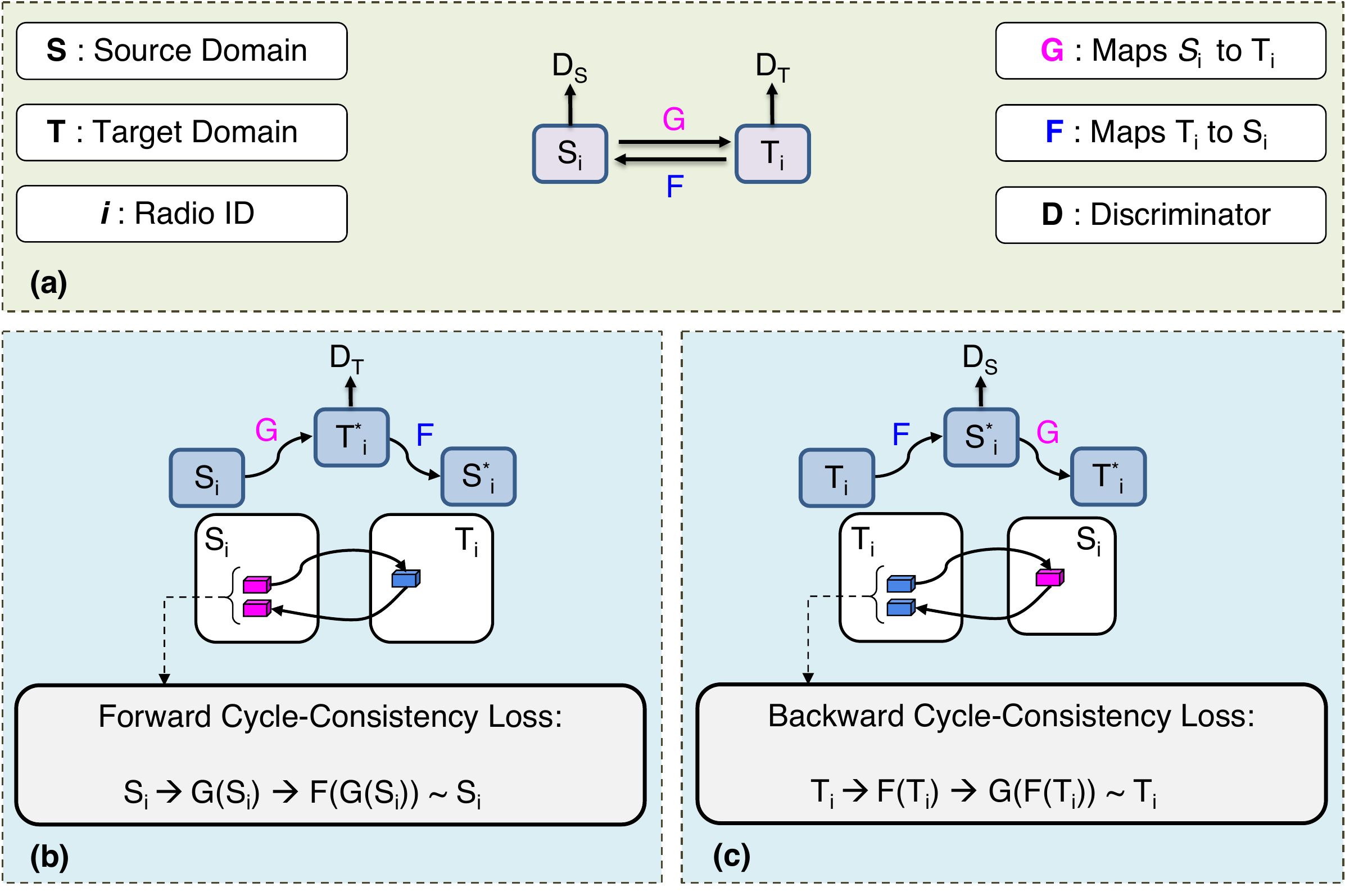}
    \caption{RF Signature-Reveal Module of \gls{signcrf} (a) The process of reshaping IQ slices collected in source domain $S$ to IQ slices collected in target domain $T$; (b) Forward cycle consistency; (c) Backward cycle consistency.} %\vspace{0.2cm}
    % \vspace{-20pt}
    \label{fig:cgan_methodology}
\end{figure}

At its core, \gls{signcrf} consists of an RF signature-reveal module (Fig.~\ref{fig:cgan_methodology}) that highlights the RF signature by eliminating the signal impairments caused by environmental changes. This module leverages the \gls{cgan} concept proposed in~\cite{zhu2017unpaired, pix2pix2017} to enhance the deep learning feature detection process in \gls{rffdl}. \gls{signcrf} employs two GANs to ensure that the neural network output will still reside in the same subspace as the designated data within a canonical dataset. As shown in Fig.~\ref{fig:cgan_methodology}, we intend to learn a mapping function $G$ to reform a dataset in one source domain, e.g., $S$, to a dataset in a target domain, e.g., $T$. Yet, to control this adaptation, we still need to learn an inverse mapping function, $F$, to reconstruct the source dataset, $S$, from the constructed target domain, $T$. In our case, the source, $S$, and the target, $T$, domains are IQ samples collected on different days and/or environments. \textbf{First, the source and the target domains should share common characteristics, i.e., the hardware's unique impairments as the received signal is transmitted from the same radio $i$}. The \gls{signcrf} relies on these common characteristics between the source and the target domains to improve the fingerprinting performance. The signature-reveal module employs the first \gls{gan} as indicated in Fig.~\ref{fig:cgan_methodology}(b) to create function $G$, which learns how to adjust the source domain characteristics that are altered during the wireless transmission to fit the target domain's prominent features. Then, the second \gls{gan} performs the same task while also providing an ``inverse function'' $F$ to learn mapping the dataset from target domain $T$ back to source domain $S$. $G(S)$ represents the output of the function that learns to map IQ slices in the source domain $S$ to IQ slices of the target domain $T$ using an adversarial loss. The adversarial loss aims to shape $G(S)$ distribution to be indistinguishable from $T$ slices distribution. As discussed before, an inverse mapping $F(T)$ is coupled with $G(S)$ to constrain the space of the constructed and reconstructed IQ slices using a cycle consistency loss that ensures $F(G(S))\approx S$ and $G(F(T))\approx T$. To achieve this goal, \gls{signcrf} uses two adversarial discriminators $D_T$  and $D_S$ to determine if the IQ slices generated using $G(S)$ and $F(T)$ match the original IQ slices from target domain $T$ and source domain $S$, respectively.

We customize \gls{signcrf} per device $i \in [1,2, ...N]$, where $N$ is the total number of devices we wish to authenticate. To do so, we collect IQ datasets from \textit{bit-similar} radios in different domains (each domain represented by a specific day, time, and environment). \gls{signcrf} reshapes the IQ dataset collected from radio $i$ in the source domain to a target domain space. This operation boosts the unique hardware impairments within the constructed dataset $G(S)$. The reshaped dataset will be then fed to the baseline classifier that is pre-trained using the target domain dataset (also referred to as ``base environment''). %The proposed RF Signature Emulator demonstrates that the fingerprinting accuracy of the worst-case scenario can be significantly improved.   
The mapping function $G$ is trained to reshape IQ slices to be similar to domain $T$ IQ slices. At the same time, the discriminator $D_T$ is trained to determine if the produced samples can successfully fit in the target domain space. The adversarial loss that expresses this $G$ mapping function and the $D_T$ discriminator is shown in equation~(\ref{eq:gan1_loss}). Similarly, adversarial loss in equation~(\ref{eq:gan2_loss}) is defined for the mapping function $F$ and its discriminator $D_S$. Note that instead of the log-likelihood loss proposed in~\cite{zhu2017unpaired}, we use the least-squares loss function presented in~\cite{mao2017least}, which demonstrated better performance and provided more stability during the training phase.

\begin{equation}\label{eq:gan1_loss}
    \begin{aligned}[b]
    & L_{GAN_1}(G, D_T, S, T) = \mathbb{E}_{t \sim P_t} [(D_T(T) -1)^2]\\
    &  \hspace{3cm}   +\mathbb{E}_{s \sim P_{s}} [D_T(G(S))^2],
    % & L_{GAN_1}(G, D_Y, X, Y) = \frac{1}{m} \sum\limits_{j=1}^m (1-D_Y(G(X_j))^2,
    \end{aligned}
\end{equation}
\begin{equation}\label{eq:gan2_loss}
    \begin{aligned}[b]
    % & L_{GAN_2}(F, D_X, Y, X) = \frac{1}{m} \sum\limits_{j=1}^m (1-D_X(F(Y_j))^2,
    & L_{GAN_2}(F, D_S, T, S) = \mathbb{E}_{s \sim P_s} [(D_S(S)-1)^2]\\
    &  \hspace{3cm}   +\mathbb{E}_{t \sim P_{t}} [D_T(F(T))^2],
    \end{aligned}
\end{equation}

%\noindent where, $j$ is IQ slice $j \in [1, 2, ...m]$.
\noindent where each $S$ and $T$ has $m$ number of IQ slices that are represented by $[s_i]_j^m  \in s$, and  $[t_i]_j^m \in T$, where ${s \sim P_s}$ and ${t \sim P_t}$ are the data distribution of each related domain.

\gls{signcrf} uses the cycle consistency loss approach in Machine Translation that requires a reverse translation, e.g., to translate a phrase from French to English, one should translate it back from English to French to enforce forward and backward consistency that should reduce the space of possible mapping functions. Applying this concept to \gls{signcrf} means that the IQ slices developed by the mapping function $G(S)$ in the first generator will be used as input to the mapping function $F(G(S))$ of the second generator, Thus, the resulted IQ slices should fit the original IQ slices, i.e., $F(G(S))\approx S$. This is called ``forward cycle consistency'' as illustrated in Fig.~\ref{fig:cgan_methodology}(b). The reverse operation is also true, called ``backward cycle consistency'', and leads to constructing $G(F(T))\approx T$, as shown in Fig.~\ref{fig:cgan_methodology}(c). Equation~(\ref{eq:cycle_loss}) represents the mathematical representation of the cycle consistency loss. This operation helps regularize the generators and guides the IQ slices construction process. The total objective loss is presented in equation~(\ref{eq:loss_full}). This loss sums up all discussed losses and introduces $\lambda$ as a new hyperparameter to weight the relative importance of the two objective functions. We consider $\lambda=10$ in our setup, as suggested in~\cite{zhu2017unpaired}.

\begin{equation}\label{eq:cycle_loss}
    \begin{aligned}[b]
    %  & L_{cyc}(G,F) = \frac{1}{m} \sum\limits_{j=1}^m [(F(G(S_j))-S_j )\\
    %  & \hspace{1.5cm} +(G(F(T_j))-T_j)]  
    & L_{cyc}(G,F) = \mathbb{E}_{s \sim P_s} [\| F(G(S))-S \|_1]\\
    &  \hspace{1.5cm}   +\mathbb{E}_{t \sim P_{t}} [\| G(F(T))-T \|_1]
    \end{aligned}
\end{equation}
\begin{equation}\label{eq:loss_full}
    \begin{aligned}[b]
     & L_{full} = L_{GAN_1} + L_{GAN_2} + \lambda L_{cyc}
    % & L_{cyc}(G,F) = \mathbb{E}_{X \sim P_X} [\| F(G(X))-x \|_1]\\
    % &  \hspace{1.5cm}   +\mathbb{E}_{Y \sim P_{Y}} [\| G(F(Y))-Y \|_1]
    \end{aligned}
\end{equation}

Note that $F$ and $G$ are autoencoders in nature~\cite{kingma2013auto}. We train them jointly to map IQ slices to themselves through an intermediate representation that is a translation of IQ slices into another domain. Adversarial losses are used to train these autoencoders to match the targeted distribution each time, which can be seen as a distinct case of adversarial autoencoders~\cite{makhzani2015adversarial,zhu2017unpaired}. In \gls{signcrf} framework the target distribution for the source domain $S$ is the target domain $T$ distribution. The final goal of \gls{signcrf} is to minimize the following objective function:
\begin{equation}\label{eq:full_obj}
     G, F = \arg \min_{G,F} \max_{D_S,D_T} L(G,F,D_S,D_T).
\end{equation}

%%%%%%%%%%%%%%%%%%%%%%%%%
%%%%%%%%%%%%%%%%%%%%%%%%%

\subsection{\gls{signcrf} Neural Network Architecture}\label{sec:cgan_arc}

\begin{figure}[!t]
    %\vspace{-10pt}
    \centering
    \includegraphics[width=1\columnwidth]{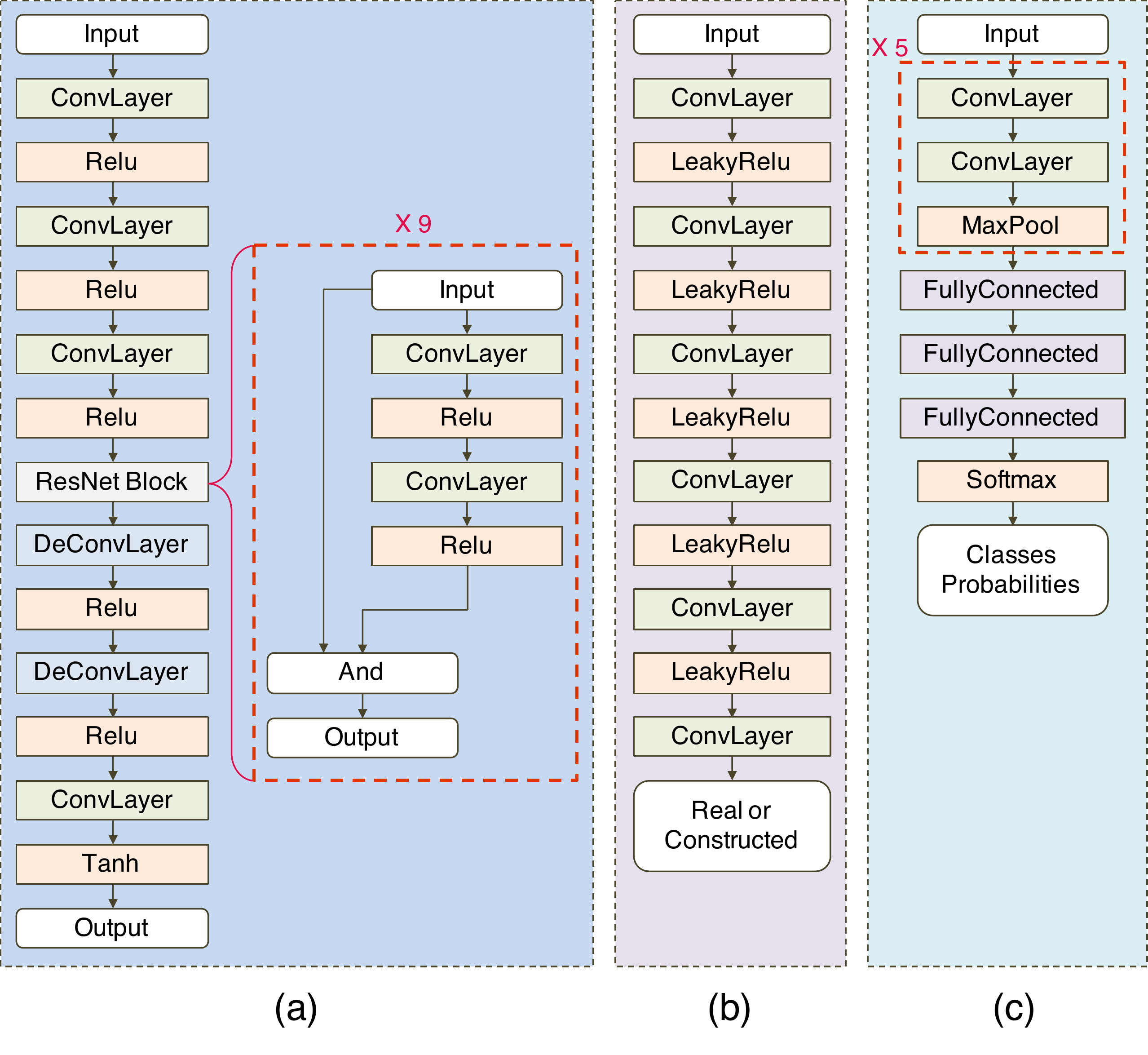}
    \caption{\gls{signcrf} Deep Learning \gls{nn} architecture (a) Generators $F$ and $G$, with a ResNet block; (b) Discriminators $D_S$ and $D_T$; (c) Baseline classifier.} %\vspace{0.2cm}
    \label{fig:cgan_arc}
\end{figure}

\begin{figure}[!b]
    %\vspace{-10pt}
    \centering
    \includegraphics[width=1\columnwidth]{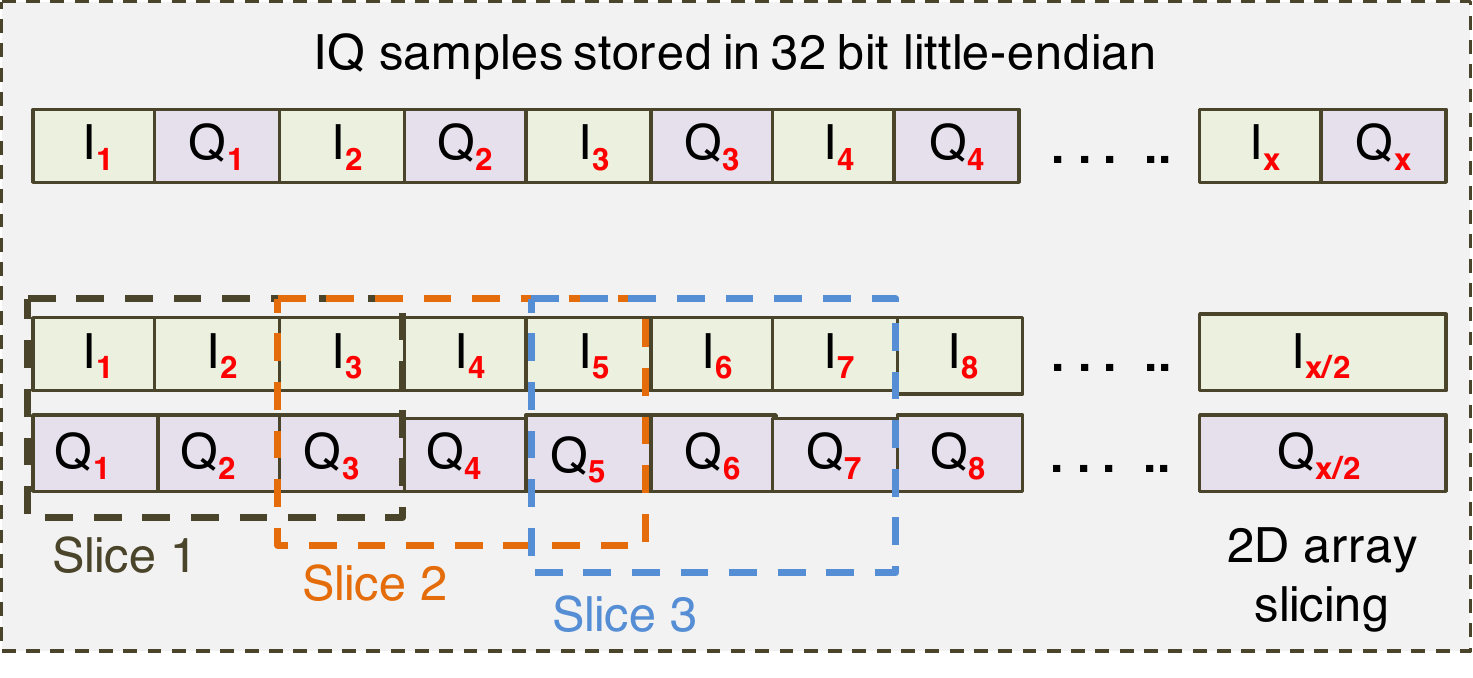}
    \caption{Sliding window approach to create 2D slices.} %\vspace{0.2cm}
    \label{fig:sliding_appr}
\end{figure}

Fig.~\ref{fig:cgan_arc} illustrates the deep learning model architectures used for (a) both generators to develop the mapping functions $G$ and $F$, (b) both discriminators $D_S$ and $D_T$, and (c) the baseline classifier that is trained in the base environment and used to authenticate the radios; we employ this classifier to authenticate all datasets determined in Sec.~\ref{sec:rf_cgan_data}. We use the sliding window approach to cut the received transmission to a predefined number of slices. Specifically, given an I/Q sequence in complex 32-bit floating-point samples stored in little-endian format as shown in Fig.~\ref{fig:sliding_appr}, we create a \gls{2d} array for I and Q values. To slice the \gls{2d} array, we need to specify the number of required slices, slice length, and stride. %In our case, we generate 12000 slices each with length 288 and stride 1 for the WiFi dataset as introduced in~\cite{shawabka2020exposing} and 9000 slices each slice with length 128 and stride 1 as specified in~\cite{al2021deeplora} for the LoRa dataset. The number of slices is constrained by the number of samples available in the collected datasets and the required slice length. We refer the readers to~\cite{shawabka2020exposing} for more insights on how to train a Baseline classifier.
For more details on how we slice our datasets please refer to Sec.~\ref{sec:train_NN_}.

\gls{signcrf} takes IQ slices as input and generates transposed IQ slices as output. The model, as shown in Fig.~\ref{fig:cgan_arc}(a) applies (i) a sequence of three downsampling convolutional blocks to encode the input IQ slices, and nine residual network (ResNet) convolutional blocks to transform the IQ slices; and (ii) two upsampling convolutional blocks to generate the output slices. We do not compile the generator models directly during the training. Instead, they are updated via composite models that are utilized to train each generator. It is also required to share the weights with the related discriminator and the other generator models. Both generators are updated to minimize the adversarial loss predicted by the discriminator for the generated IQ slices marked as ``real''. By this step, the generators are maintained to create IQ slices that better fit into the target domain. The discriminator model shown in Fig.~\ref{fig:cgan_arc}(b) predicts whether the IQ slices that are fed to this discriminator are ``real'' or ``fake''. We refer interested readers to~\cite{li2016precomputed, isola2017image} for more insights about \gls{gan} discriminator structure and to~\cite{zhu2017unpaired, cgan} for more detailed insights about the \gls{gan} generator and the composite models structures.

\noindent\textit{\underline{Baseline Classifier.}} Fig.~\ref{fig:cgan_arc}(c) illustrates the 1D CNN baseline model architecture inherited from AlexNet CNN~\cite{AlexNet}. This model is employed to classify the Generator $G$ output, explained earlier in Fig.~\ref{fig:cgan_app}(c). The model consists of two one-dimension (1D) ConvLayers with 128 filters each. Filter sizes of 1x7 and 1x5 are utilized, respectively, for the first and second ConvLayers. Each ConvLayer is followed by a rectified linear units (ReLU) activation function and 1x2 MaxPooling layer. This structure is repeated five times and then fed to three fully connected (FC) layers with 256, 256, and 128 neurons, respectively. A ReLU activation function follows each FC layer. Finally, an FC layer with a softmax activation function is used to determine the radios classification probabilities.

\noindent\textit{\underline{Per Device Customization Process.}} To further improve the performance of the \gls{signcrf}, we optimize the model to encounter the most promising combination of the hyperparameters and the number of epochs that can lead to the most willing feature detection for a specific radio. Upon accepting this work, we will share with the research community the related codes with full explanations of such optimizations.

\section{Experiment Results}\label{sec:exp_result}
% \subsection{RF Signature Emulator Performance Results.}\label{sec:res}
We analyze the performance results of the \gls{signcrf} with WiFi~\cite{shawabka2020exposing}, and LoRa (payload, and preamble) datasets~\cite{al2021deeplora}.

\subsection{Performance Metrics}\label{sec:cgan_met}

To assess the \gls{signcrf} performance, we consider the following performance metrics:

\noindent{\textit{Per-slice Testing Accuracy \textbf{(PSTest)}},} evaluates the performance of a pre-trained deep learning model (baseline model) using new datasets (not seen during the training process), and it measures the percentage of the correctly predicted slices out of the total testing slices;

\noindent{\textit{Train-and-Test-in-Other-Domain \textbf{(TTOD)} Accuracy},} represents the PSTest performance when the testing dataset is collected in a domain other than the domain used to train the pre-trained classifier. The new domain can be another day or environments; However, the other testing settings remain the same, e.g., radios, the technology used, and the structure of the packets;

\noindent{\textit{Radio Recognition Percentage \textbf{(RRP)}}.}  We consider radio $i$  to be correctly classified if its Softmax classification probability of being itself (actual radio) is the highest among all other radios. Based on this, the RRP represents the percentage of the correctly classified radios among all others. By this metric, we measure the capability of our \gls{signcrf} in improving the classification probability of the correct radios.

\noindent{\textit{Percentage Increase \textbf{(Improvement)}},} represents the amount of increase from the original \textit{TTOD or RRP} values before applying \gls{signcrf} to the final \textit{TTOD or RRP} values after implementing \gls{signcrf} in terms of 100 parts of the original values.

\subsection{Datasets}\label{sec:rf_cgan_data}

\subsubsection{The WiFi Dataset: An Overview}\label{sec:rf_cgan_data_wifi}
We employ I/Q samples transmitted using the IEEE 802.11a/g (WiFi) standard~\cite{IEEE-WiFi-Standard} that is publicly shared in~\cite{shawabka2020exposing}. This WiFi standard applies orthogonal frequency division multiplexing (OFDM) and thus multiple subcarriers to transmit each digital symbol using BPSK, QPSK, 16QAM or 64 QAM modulation, with different levels of convolutional coding (1/2 or 3/4). The used WiFi framework consists of a short training sequence (STS) for frame detection, and a long training sequence (LTS) for time synchronization. Then, to recover the transmitted digital symbols, the algorithm uses Fast Fourier Transform (FFT) to transfer the I/Q samples from the time domain to the frequency domain. Finally, before decoding the equalized IQ samples, the LTS is again employed to perform channel estimation and offset correction.  This dataset is collected in Arena~\cite{bertizzolo2019arena}, an ``in the wild''  testbed using 20 software-defined radios (SDRs), composed of thirteen N210 and seven X310 Universal Software Radio Peripherals (USRPs). %The receiver is an N210 SDR.
The testbed presents a rich and diverse environment to collect challenging datasets. To evaluate \gls{signcrf}, we use the equalized dataset collected from 20 USRPs in three different days to validation and testing in Real-time implementations.

\subsubsection{The LoRa Dataset: An Overview}\label{sec:rf_cgan_data_lora}
We employ the LoRa dataset introduced in~\cite{al2021deeplora} and publicly released by InterDigital\footnote{https://www.interdigital.com/data\_sets/lora\--radio\--data}. This dataset is collected in rich environments spanning different days with 100 \textit{bit-similar} Pysense sensors connected to 100 \textit{bit-similar} FiPy radios that operate as transmitters on a carrier of 902.3~MHz in the 915~MHz ISM band. Each radio transmits ten consecutive bursts of packets in each test environment. One second of silence separates each burst. The burst consists of 100 consecutive packets separated by 10 ms. Each packet contains the payload information carrying the temperature, the humidity, and the device voltage readings. We extract the preamble and the payload datasets collected from two different days in an outdoor environment based on the methodology described in~\cite{al2021deeplora}. Then we use these datasets to validate the performance of \gls{signcrf}. 

\subsubsection{Train \gls{nn} and Dataset Generation: }\label{sec:train_NN_}
\begin{figure}[!b]
    %\vspace{-10pt}
    \centering
    \includegraphics[width=1\columnwidth]{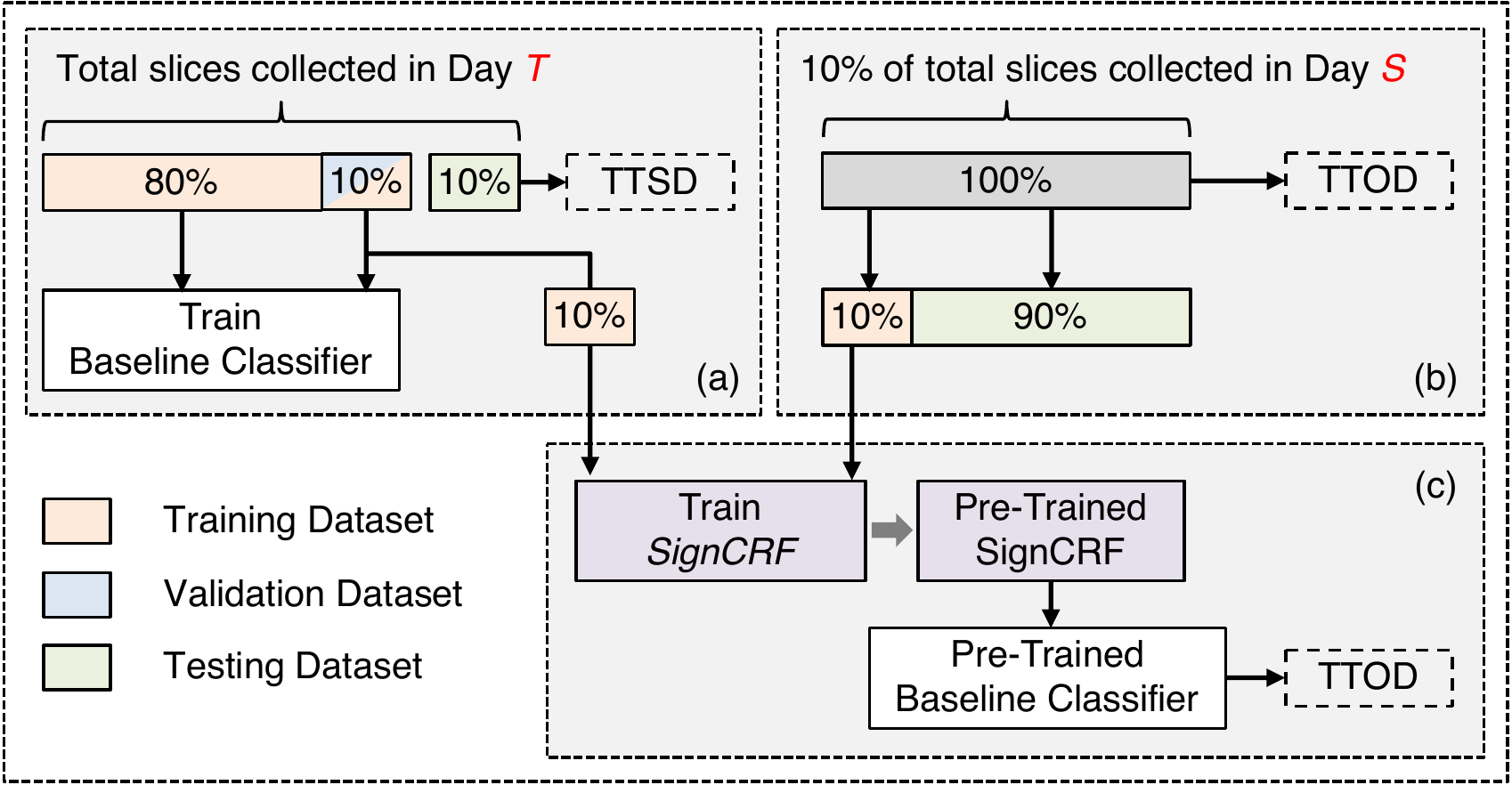}
    \caption{Dataset structure to (a) train the Baseline Classifier and evaluate the performance of TTSD; (b) evaluate the performance of TTOD; (c) train the \gls{signcrf} and evaluate the performance of TTOD.} %\vspace{0.2cm}
    \label{fig:dataset_split}
\end{figure}

We train our NNs using Keras running on top of TensorFlow over an NVIDIA Cuda-enabled Tesla K80m GPU system. We generate 12000 slices each with length 288 and stride 1 for the WiFi dataset as introduced  in~\cite{shawabka2020exposing} and 9000 slices each slice with length 128 and stride 1 as specified in~\cite{al2021deeplora} for the LoRa dataset. The number of samples available in the collected datasets and the required slice length constrain the number of slices. We employed 80\%, 10\%, and 10\% of the slices generated on Day \textit{T} to train, validate, and test (TTSD performance) the baseline model, respectively, as shown in Fig.~\ref{fig:dataset_split}(a). To get the TTOD results without \gls{signcrf}, we test the baseline model using 10\% of the slices generated on Day \textit{S}, as indicated in Fig.~\ref{fig:dataset_split}(b). 
To train \gls{signcrf} (i) we reserve 10\% of the validation slices generated on Day \textit{T}; and (ii) 10\% of the testing slices generated on Day \textit{S}, as displayed in Fig.~\ref{fig:dataset_split}(c). Note that we take 10\% of the data collected in Day \textit{S} that is consistent with the size of the data used for testing the Baseline Classifier. 
The remaining 90\% of the testing slices generated on Day S will pass through the pre-trained \gls{signcrf} system before being classified to create the new TTOD results. 

\subsection{WiFi Dataset Results}\label{sec:wifi_res}

% Fig.~\ref{fig:wifi_per_device_per_epoch} shows the per device per epoch TTOD accuracy of RF Signature Emulator created to convert IQ samples collected from five X310 radios transmitting the same WiFi packets from day/source $X$ to day/target $Y$. This figure indicates that the RF Signature Emulator successfully boosts the prominent impairments of 4 out of 5 of the tested radios. Nonetheless, in limited cases, e.g., one radio out of five (around 20\% of the overall radios), using RF Signature Emulator may not improve the results, which can be subject of future work. In addition a minor degradation can be observed while training the RF Signature Emulator with several epochs, but this behavior is expected as (i) the dataset is transferred from one domain to another, and slight drifts in the impairments is unavoidable and may mislead the overall classification, especially when we are investigating a very challenging \gls{rffdl} scenario (\textit{bit-similar} devices transmitting the same packets over and over); and (ii) we cannot guarantee the quality of the impairments associated with all samples used to train the model for any specific signature; however, new samples can be recollected for such devices for further validation. In any case, the degradation explained here can be avoided by using the Max Rule step explained earlier in Sec.~\ref{sec:cgan_framework}

\begin{figure}[!t]
\captionsetup{skip=5pt}
    %\vspace{-10pt}
    \centering
    \begin{subfigure}[t]{0.48\columnwidth}
       \captionsetup{skip=5pt}
        \centering
        \setlength\fwidth{0.7\columnwidth}
        \setlength\fheight{0.5\columnwidth}
        \input{./Figures/CM_wifi_5dev_without_cgan}
        \caption{ {Before} \gls{signcrf} implementations the TTOD\% performance = 18\%, and RRP\% performance = 20\%.(1 out of 5 radios classified correctly).}
        \label{fig:cm_5_wifi}%\vspace{-0.2cm}
    \end{subfigure} 
    \hfill
    \begin{subfigure}[t]{0.48\columnwidth}
       \captionsetup{skip=5pt}
       \centering
        \setlength\fwidth{0.7\columnwidth}
        \setlength\fheight{0.5\columnwidth}
        \input{./Figures/CM_wifi_5dev_with_cgan}
        \caption{ {After} \gls{signcrf} implementation the TTOD \% performance = 83\%, and RRP\% performance = 100\%.(5 out of 5 radios classified correctly).}
        \label{fig:cm_5_wifi_withcgan}%\vspace{-0.2cm}
    \end{subfigure} 
    \caption{Confusion matrices of equalized WiFi dataset collected from 5 \textit{bit-similar} radios transmitting the same data over WiFi technology; (i)  {before},  and (ii) {after} applying \gls{signcrf}.}
    \label{fig:cm_5_wifi_w_withcgan}%\vspace{-0.2cm}
\end{figure}
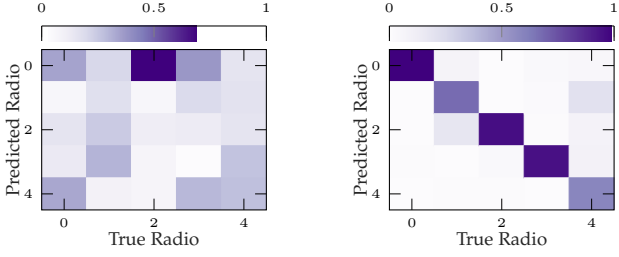

\noindent\textbf{Small-scale WiFi testbed:} We follow the three steps specified in Sec.~\ref{sec:cgan_framework} to assess the performance of applying \gls{signcrf} to a small WiFi testbed consisting of 5 \textit{bit-similar} X310 radios. Figures~\ref{fig:cm_5_wifi} and \ref{fig:cm_5_wifi_withcgan} show the TTOD and RRP performance before and after \gls{signcrf}, respectively. The figures indicate that \gls{signcrf} improves the \textit{TTOD results} \textbf{from 18\% to 83\%} and the \textit{RRP} \textbf{from 20\% to 100\%}, i.e., \gls{signcrf} is able to authenticate all the devices with 100\% confidence, a significant improvement compared to the state-of-the-art~\cite{shawabka2020exposing}.\\

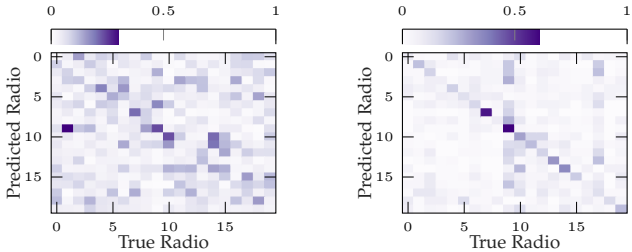
\begin{figure}[!b]
\captionsetup{skip=5pt}
    \centering
    %\vspace{-10pt}
    \begin{subfigure}[t]{0.48\columnwidth}
       \captionsetup{skip=5pt}
        \centering
        \setlength\fwidth{0.7\columnwidth}
        \setlength\fheight{0.5\columnwidth}
        \input{./Figures/CM_wif_20dev_without_cgan}
        \caption{ {Before} \gls{signcrf} implementation-TTOD\% performance = 9\%, and RRP\% performance = 30\%.(6 out of 20  radios classified correctly).}
        \label{fig:cm_20_wifi}%\vspace{-0.2cm}
    \end{subfigure} 
    \hfill
    \begin{subfigure}[t]{0.48\columnwidth}
       \captionsetup{skip=5pt}
       \centering
        \setlength\fwidth{0.7\columnwidth}
        \setlength\fheight{0.5\columnwidth}
        \input{./Figures/CM_wifi_20dev_with_cgan}
        \caption{ {After} \gls{signcrf} implementation the TTOD\% performance = 34\%, and RRP\% performance = 90\%.(18 out of 20 radios classified correctly.}
        \label{fig:cm_20_wifi_withcgan}%\vspace{-0.2cm}
    \end{subfigure} 
    \caption{Confusion matrices of equalized WiFi dataset collected from 20 \textit{bit-similar} radios transmitting the same data over WiFi technology; (i)  {before},  and (ii) {after} applying \gls{signcrf}.}
    \label{fig:cm_20_wifi_w_withcgan}%\vspace{-0.2cm}
    %\vspace{-10pt}
\end{figure}

\noindent\textbf{Medium-scale WiFi testbed:} \gls{signcrf} preserves promising results in improving the \gls{rffdl} performance for the WiFi dataset collected from 20 \textit{bit-similar} radios over different days. We use the same methodology described in Fig.~\ref{fig:cgan_app} to improve the TTOD accuracy of medium-size testbed with 20 \textit{bit-similar} radios transmitting the same WiFi packets repeatedly. Our proposed method improves the \textit{TTOD accuracy} from 9\% to 34\%. To the best of our knowledge, this paper, for the first time, improves the \gls{rffdl} of the WiFi technology by a significant factor of 4x in worst-case scenario, i.e., training in one domain and testing in another, using deep learning. Figures~\ref{fig:cm_20_wifi} and \ref{fig:cm_20_wifi_withcgan} show the classification results in confusion matrices before and after applying \gls{signcrf} to convert domain $S$ to domain $T$. Additionally, \gls{signcrf} boosted the \textit{RRP} \textbf{from 30\% to 90\%}, i.e., \gls{signcrf} is able to accurately authenticate 90\% of the devices, a significant improvement compared to the state-of-the-art which can only authenticate 30\% of the devices~\cite{shawabka2020exposing}. 

We benchmark our WiFi dataset results with DeepRadioID system proposed by Restuccia \emph{et al.}~\cite{restuccia2019deepradioid}. Authors of DeepRadioID have been able to improve the channel resilience of \gls{rffdl} algorithms by carefully applying optimized \gls{fir} filters. The best improvements of the \gls{rffdl} accuracy that DeepRadioID made were 1.5x and 1.7x using 5 and 20 WiFi radios, respectively. Our \gls{signcrf} outperforms DeepRadioID with accuracy improvements of 4.6x and 3.7x using 5 and 20 WiFi radios, respectively, achieving up to 100\% and 90\% \textit{RRP} as reported later in Sec.~\ref{sec:conclusion}, Table~\ref{table:summary_rrp}.

\subsection{LoRa Payload Dataset Results}\label{sec:lora_pay_res}

To further evaluate the capability of \gls{signcrf} in improving the data-driven authentication for diverse technologies, we use a dataset generated using LoRa technology transmission over the air. We extract the payload of the received LoRa packets to create our dataset using the methodology described in~\cite{al2021deeplora}. Then we assess the performance of \gls{signcrf} system in three testbed scales, namely, small, medium, and large with 5, 20, and 100 \textit{bit-similar} LoRa radios respectively. This further verifies the \gls{signcrf} proficiency in improving the \gls{rffdl} performance regardless of the (i) testbed scale, (ii) the adopted technology, and (iii) the type of IQ samples used in the classification process. i.e., the preamble or the payload part of the received packets.\\

% \begin{figure}[h!]
% \captionsetup{skip=5pt}

%     \vspace{-10pt}
%     \centering
%     \begin{subfigure}[t]{0.48\columnwidth}
%       \captionsetup{skip=5pt}
%         \centering
%         \setlength\fwidth{0.7\columnwidth}
%         \setlength\fheight{0.5\columnwidth}
%         \input{./Figures/100_lora_payload_nogan}
%         \caption{ {Before} xxx implementation the TTOD \% performance = 6\%, and RRP\% performance = 9\%.}
%         \label{fig:100_lora}%\vspace{-0.2cm}
%     \end{subfigure} 
%     \hfill
%     \begin{subfigure}[t]{0.48\columnwidth}
%       \captionsetup{skip=5pt}
%       \centering
%         \setlength\fwidth{0.7\columnwidth}        \setlength\fheight{0.5\columnwidth}
%         \input{./Figures/100_lora_payload_w_cgan}
%         \caption{ {After} xxx implementation the TTOD \% performance = 53\%, and RRP\% performance = 73\%.}
%         \label{fig:cm_100_lora_withcgan}%\vspace{-0.2cm}
%     \end{subfigure} 
%     \caption{Confusion matrices of LoRa payload dataset collected from 100\textit{bit-similar} LoRa radios (i) {before}, and (ii) {after} applying RF Signature Emulator to reshape day $X$ datasets.}
%     \label{fig:100_lora_payload}\vspace{-0.2cm}
% \end{figure}

\begin{figure}[t!]
\captionsetup{skip=5pt}

    %\vspace{-10pt}
    \centering
    \begin{subfigure}[t]{0.48\columnwidth}
      \captionsetup{skip=5pt}
        \centering
        \setlength\fwidth{0.7\columnwidth}
        \setlength\fheight{0.5\columnwidth}
        \input{./Figures/CM_lora_5_withot_cgan}
        \caption{ {Before} \gls{signcrf} implementations the TTOD\% performance = 18\%, and RRP\% performance = 20\%.(1 out of 5 radios classified correctly).}
        \label{fig:cm_5_lora}%\vspace{-0.2cm}
    \end{subfigure} 
    \hfill
    \begin{subfigure}[t]{0.48\columnwidth}
      \captionsetup{skip=5pt}
      \centering
        \setlength\fwidth{0.7\columnwidth}
        \setlength\fheight{0.5\columnwidth}
        \input{./Figures/CM_lora_5_with_cgan}
        \caption{ {After} \gls{signcrf} implementation the TTOD\% performance = 80\%, and RRP\% performance = 80\%.(4 out of 5 radios classified correctly).}
        \label{fig:cm_5_lora_withcgan}%\vspace{-0.2cm}
    \end{subfigure} 
    \caption{Confusion matrices of LoRa payload dataset collected from 5 \textit{bit-similar} LoRa radios transmitting the same data over LoRa technology; (i) {before}, and (ii) {after} applying \gls{signcrf}.}
    \label{fig:cm_5_lora_w_withcgan}%\vspace{-0.2cm}
    %\vspace{-5pt}
\end{figure}
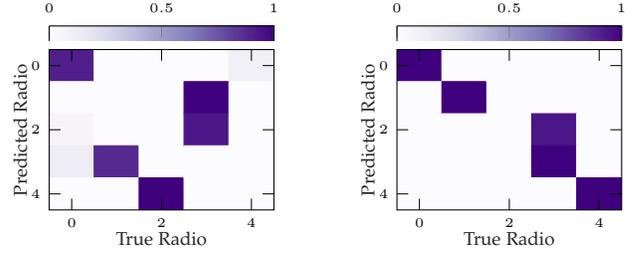

\noindent\textbf{Small-scale LoRa payload testbed:} Our small testbed consists of the payload dataset extracted from 5 \textit{bit-similar} LoRa radios. Each LoRa radio transmits the voltage, temperature, and humidity readings in its payload. The results are shown in Figures~\ref{fig:cm_5_lora} and \ref{fig:cm_5_lora_withcgan} for before and after applying our proposed \gls{signcrf} respectively, indicating that the \gls{signcrf} system succeeded in enhancing the \textit{TTOD results} \textbf{from 18\% to 80\%} and the \textit{RRP} \textbf{from 20\% to 80\%}.\\

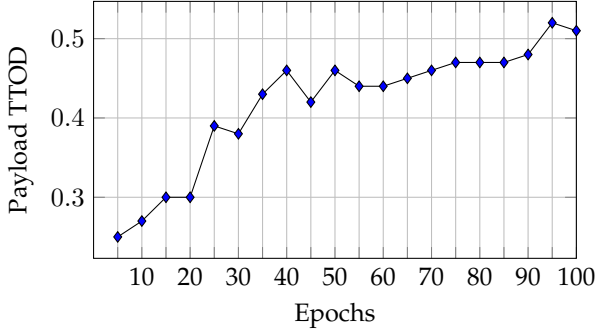
\begin{figure}[t!]
\captionsetup{skip=5pt}
    %\vspace{-10pt}
    \centering
    \input{./Figures/payload_acc_lora}
    \caption{TTOD\% performance per epoch after implementing \gls{signcrf} over LoRa payload dataset collected from 20 \textit{bit-similar} LoRa radios transmitting the same data over LoRa technology. \textbf{Note: this figure shows a smooth increase of TTOD\% after applying the general \gls{signcrf} solution without any further per device customization.}}
    \label{fig:TTOD_payload_per_epoch}%\vspace{-0.2cm}
    %\vspace{-10pt}
\end{figure}

\noindent\textbf{Medium-scale LoRa payload testbed:} Fig.~\ref{fig:TTOD_payload_per_epoch} illustrates the effectiveness of the \gls{signcrf} in increasing the TTOD accuracy per epoch during the classification process of 20 LoRa \textit{bit-similar} radios using their payload dataset without applying any per device customization. This figure shows a consistent trend in improving the fingerprinting performance while increasing the number of epochs. \gls{signcrf} works perfectly in polishing the impairments associated with 15 LoRa radios out of 20. The \textit{TTOD accuracy} of this scenario boosted \textbf{from 13\% to 52\%} after applying \gls{signcrf} without applying any further per device customization, and \textbf{from 13\% to 73\%} with the customized \gls{signcrf}, using the per device customization process described earlier in Sec.~\ref{sec:cgan}.

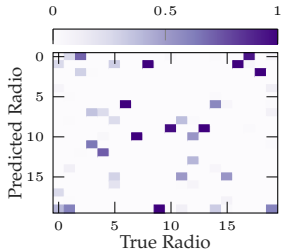
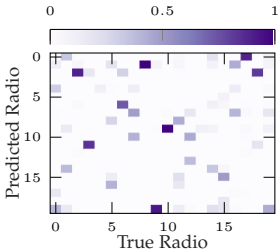
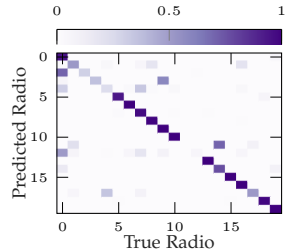
\begin{figure}[b!]
\captionsetup{skip=5pt}
    %\vspace{-10pt}
    \centering
    \begin{subfigure}[t]{0.48\columnwidth}
    \captionsetup{skip=5pt}
        \centering
        \setlength\fwidth{.7\columnwidth}
        \setlength\fheight{0.5\columnwidth}
        \input{./Figures/CM_lora_20_orig}
        \caption{{Before} both \textit{DeepLoRa} data augmentation and \gls{signcrf} the TTOD\% performance = 13\%, and RRP\% performance = 15\%. (3 of 20 radios classified correctly). }
        \label{fig:20_dev_no_aug}
    \end{subfigure}
    \vfill
    \begin{subfigure}[t]{0.48\columnwidth}
    \captionsetup{skip=5pt}
        \centering
        \setlength\fwidth{.7\columnwidth}
        \setlength\fheight{0.5\columnwidth}
        \input{./Figures/CM_lora_20_deeplora_pa}
        \caption{{After} \textit{DeepLoRa} data augmentation\cite{al2021deeplora} the TTOD\% performance = 22\%, and RRP\% performance = 35\%.(7 out of 20 radios classified correctly).}
        \label{fig:20_dev_PA}
    \end{subfigure}
     \hfill
    \begin{subfigure}[t]{0.48\columnwidth}
    \captionsetup{skip=5pt}
        \centering
        \setlength\fwidth{.7\columnwidth}
        \setlength\fheight{0.5\columnwidth}
        \input{./Figures/CM_lora_20_with_cgan}
        \caption{{After} applying customized \gls{signcrf} the TTOD\% performance = 73\%, and RRP\% performance = 75\%.(15 out of 20 radios classified correctly)}
        \label{fig:20_dev_lora_RF Signature Emulator}
    \end{subfigure}
    \caption{Confusion matrices of LoRa payload dataset collected from 20 \textit{bit-similar} LoRa radios transmitting the same data over LoRa technology in an outdoor environment. \ref{fig:20_dev_no_aug} {before} both \textit{DeepLoRa} and \gls{signcrf}; \ref{fig:20_dev_PA}, {after} \textit{DeepLoRa} outdoor-to-indoor pedestrian environments with channels A)\cite{al2021deeplora}; and \ref{fig:20_dev_lora_RF Signature Emulator} {after} \gls{signcrf}.}
    %\vspace{-5pt}
     \label{fig:TTOD_CM_with/out_Aug_cgan}%\vspace{-0.2cm}
\end{figure}

The authors in~\cite{al2021deeplora} applied \textit{DeepLoRa}, a data augmentation methodology to enhance the \textit{TTOD accuracy} of 20 \textit{bit-similar} LoRa radios using their payload dataset. The \textit{TTOD accuracy} before applying \textit{DeepLoRa}~\cite{al2021deeplora} is 13\% using a CNN-2D deep learning model proposed in~\cite{al2021deeplora}. \textit{DeepLoRa}~\cite{al2021deeplora} improves the \textit{TTOD accuracy} from 13\% to only 22\%, and the \textit{RRP performance} from 15\% to only 35\%. Fig.~\ref{fig:TTOD_CM_with/out_Aug_cgan} benchmarks \gls{signcrf} against \textit{DeepLoRa}~\cite{al2021deeplora}. It compares the performance (i) after applying the Baseline classification model, (ii) after applying \textit{DeepLoRa}, and (iii) after applying \gls{signcrf}. Our proposed method, the \gls{signcrf}, significantly outperforms \textit{DeepLoRa} by improving the \textit{TTOD accuracy} \textbf{from 13\% to 73\%} and the \textit{RRP performance} \textbf{from 15\% to 75\%}, a significant improvement compared to the state-of-the-art~\cite{al2021deeplora}.\\

\begin{figure}[!h]
    %\vspace{-5pt}
    \centering
    \includegraphics[width=1\columnwidth]{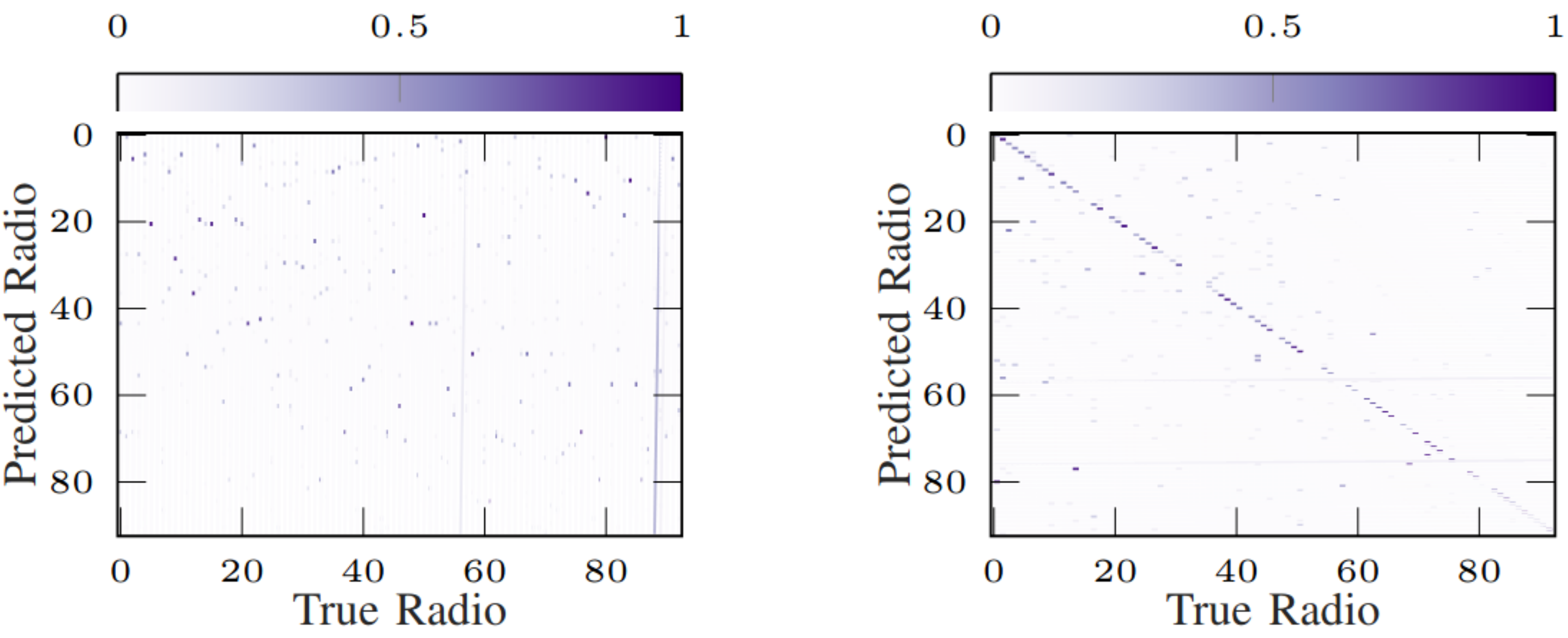}
    \caption{Confusion matrices of LoRa payload dataset collected from 100\textit{bit-similar} radios (i) {Before} \gls{signcrf} implementation the TTOD \% performance = 6\%, and RRP\% performance = 9\%, and (ii) {After} \gls{signcrf} implementation the TTOD \% performance = 53\%, and RRP\% performance = 73\%.} %\vspace{0.2cm}
    \label{fig:100_lora_payload}
\end{figure}

% To prove the effectiveness of our new approach over the previous one, we applied the RF Signature Emulator approach over this specific scenario to benchmark our results with DeepLoRa in \cite{}. Fig.~\ref{fig:benchmerck_payload_lora} compare’s the best TTOD classifications accuracy results after utilizing (i) DeepLoRa\cite{} methodology; (ii) RF Signature Emulator with a Max Rule methodology proposed in this paper. This Figure determines how RF Signature Emulator outperformed the DeepLoRa and increased the TTOD accuracy from 13\% to 71\% with the Max Rule technique; besides, the number of radios classified correctly rose from 3 to 15 radios out of 20 among the worst-case scenario as defined in \cite{}.
% Such improvements can serve the constraints associated with the practical implementation of Radio Fingerprinting based on deep learning algorithms. RF Signature Emulator plays a significant role in mapping the impairments related to the targeted radio under certain circumstances to the same radio in different conditions, opening the door for the practical implementations of Radio Fingerprinting based on deep learning algorithms.
%
%
% \begin{figure}[hbt!]
% \captionsetup{skip=5pt}
%     \centering
%     \input{./Figures/payload_acc_with_without_cgan}
%     \caption{{}.}
%       \label{fig:benchmerck_payload_lora}\vspace{-0.2cm}
% \end{figure} 
\noindent\textbf{Large-scale LoRa testbed:} To exhibit the ultimate effectiveness of \gls{signcrf} approach at scale, we evaluate its performance using the largest dataset in the literature collected from 100 \textit{bit-similar} LoRa radios. Fig.~\ref{fig:100_lora_payload} shows the effectiveness of \gls{signcrf} in increasing the \textit{TTOD accuracy} \textbf{from 6\% to 53\%} and \textit{RRP performance} \textbf{from 9\% to 73\%}. To the best of our knowledge, we are the first to develop a channel-agnostic \gls{rffdl} framework that improves the performance in the worst-case scenario using a large-scale dataset collected in the wild, unearthing a guaranteed way to authenticate the IoT radios based on \gls{rffdl} in the real world is necessary. As these results show, \gls{signcrf}, with its endowment to be customized per device, plays a significant role in opening the doors for practical real-time \gls{rffdl} implementations.

\subsection{LoRa Preamble Dataset Results}\label{sec:lora_prea_res}

We further validate \gls{signcrf} using preamble datasets collected from \textit{bit-similar} LoRa devices. Authenticating the radios based on the preamble dataset is in particular challenging and non-trivial as indicated in~\cite{al2021deeplora}; mainly due to the LoRa preamble characteristics, which make it resilient to noise, fading, and interference~\cite{sundaram2019survey, al2021deeplora}. However, our proposed solution boosts the \gls{rffdl} performance even for this complex case. Table~\ref{table:preamble_TTOD_rrp} summarizes the \gls{signcrf} effect on the radio fingerprinting performance using 5, 20, and 100 \textit{bit-similar} LoRa preamble datasets. The table shows an increase of \textbf{6.8x}, \textbf{8.5x} and \textbf{3.5x} on the \textit{TTOD accuracy} and an increase of \textbf{8x}, \textbf{11x} and \textbf{21x} on the \textit{RRP performance} of 5, 20, 100-bit-similar LoRa devices respectively, after implementing the \gls{signcrf}. This robustness of the proposed \gls{signcrf} in challenging conditions.

\begin{table}[h!]
\captionsetup{skip=5pt}
    %\vspace{-5pt}
    \centering
    % \ra{0.8}
    \begin{threeparttable}
    \caption{Preamble dataset Summary Table: \gls{rffdl} performance of LoRa preamble datasets per testbed size.}\label{table:preamble_TTOD_rrp}
    \begin{tabular}{M{0.9cm}M{0.6cm}M{0.5cm}M{0.5cm}M{0.5cm}M{0.5cm}M{0.5cm}M{0.5cm}M{0.5cm}M{0.5cm}}\toprule

    \multicolumn{2}{c}{\multirow{3}{*}{\textbf{Model}}} & \multicolumn{3}{c}{\textbf{TTOD}} & \phantom{abc}& \multicolumn{3}{c}{\textbf{RRP}} \\
\cmidrule{3-5} \cmidrule{7-9} && 5 & 20 & 100 &&  5 & 20 & 100\\ \cmidrule{3-5} \cmidrule{7-9}
       
        \multicolumn{2}{c}{Baseline} & 5 & 4  & 2  && 5 & 5 & 4 \\
        \multicolumn{2}{c}{\gls{signcrf}} & 39 & 38  & 9 && 45 & 60 & 29\\
        \\
        \multicolumn{2}{c}{\textbf{Improvement}}  & \textbf{6.8x} & \textbf{8.5x} & \textbf{3.5x} && \textbf{8x} & \textbf{11x} & \textbf{21x}
    \\\bottomrule
    \end{tabular}
    \end{threeparttable}
    %\vspace{-5pt}
\end{table}

% \begin{figure}[t!]
% \captionsetup{skip=5pt}

%     \centering
%     \begin{subfigure}[t]{0.48\columnwidth}
%       \captionsetup{skip=5pt}
%         \centering
%         \setlength\fwidth{0.7\columnwidth}
%         \setlength\fheight{0.5\columnwidth}
%         \input{./Figures/CM_lora_preamble_20_no_cgan}
%         \caption{{Before} RF Signature Emulator implementations the TTOD\% performance = 5\%, and RRP\% performance = 5\%.(1 out of 20 radios classified correctly).}
%         \label{fig:lora_pream_all_testbeds}%\vspace{-0.2cm}
%     \end{subfigure} 
%     \hfill
%     \begin{subfigure}[t]{0.48\columnwidth}
%       \captionsetup{skip=5pt}
%       \centering
%         \setlength\fwidth{0.7\columnwidth}
%         \setlength\fheight{0.5\columnwidth}
%         \input{./Figures/CM_lora_preamble_20_cgan}
%         \caption{{After} RF Signature Emulator implementations the TTOD\% performance = 39\%, and RRP\% performance = 45\%.(9 out of 20 radios classified correctly).}
%         \label{fig:cm_20_lora_pream_with_cgan}%\vspace{-0.2cm}
%     \end{subfigure} 
%     \caption{Confusion matrices of LoRa preamble dataset collected from 20 \textit{bit-similar} LoRa radios transmitting the same data over LoRa technology; (i) {before}, and (ii) {after} applying RF Signature Emulator to reshape day $X$ datasets.}
%     \label{fig:cm_20_lora_pream}%\vspace{-0.2cm}
%     %\vspace{-10pt}
% \end{figure}

\subsection{The Impact of Adversarial Action on \gls{signcrf}}\label{sec:adv_action}

\begin{figure}[t!]
    %\vspace{-5pt}
    \centering
    \includegraphics[width=1\columnwidth]{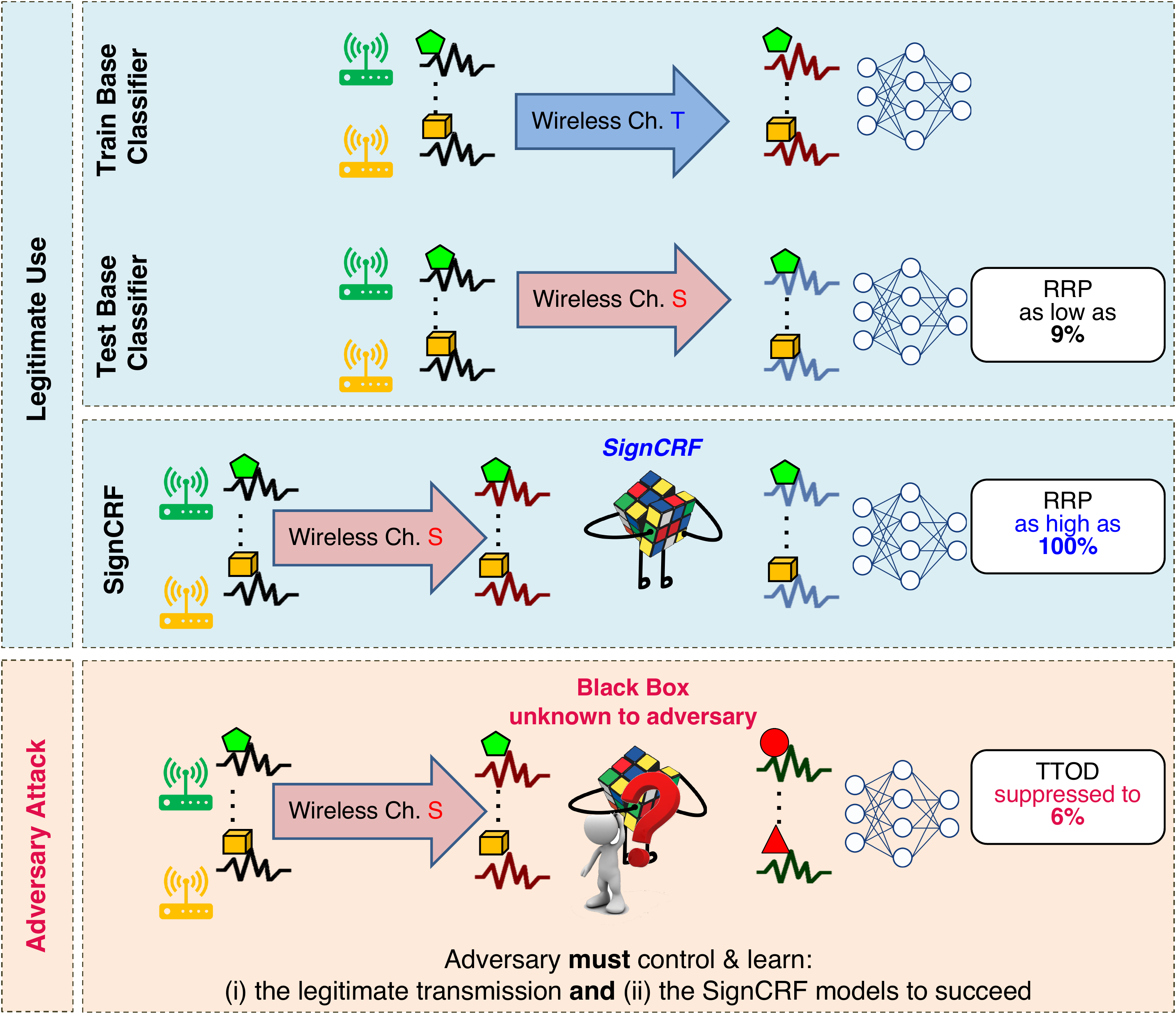}
    \caption{\textit{TTOD accuracy} of Adversarial scenario described in~\ref{sec:adv_action} for payload dataset extracted from 20\textit{bit-similar} LoRa radios: (top) before applying \gls{signcrf}; (middle) after \gls{signcrf}; and (bottom) After adversarial action.}
    %\vspace{-20pt}
    \label{fig:adv_results}
\end{figure}

Lastly, to test the robustness of \gls{signcrf}, we design an adversarial scenario in which we use the model created by \gls{signcrf} for radio $i$ to reshape the dataset collected by radio $j$, where $j$ $\neq$ $i$. We create and test this adversarial case with the payload dataset extracted from 20 LoRa \textit{bit-similar} radios. Fig.~\ref{fig:adv_results} summarizes the fingerprinting performance (top) without applying the proposed \gls{signcrf}; (middle) with \gls{signcrf}, and (bottom) with an adversary who managed to control the transmission of legitimate devices but stuck with the \gls{signcrf} puzzle. The \textit{TTOD accuracy} for this adversarial scenario is suppressed \textbf{from 73\% to 6\%}, which translates to zero mis-authentication of adversary radios who try to impersonate legitimate devices. Even if an adversary can control the legitimate radios, it still needs to know how to map the waveforms transmitted from this legitimate device on the \gls{signcrf} side. The \gls{signcrf} is designed to map the waveforms of one domain to another domain depending on common characteristics generated by radio $i$ and presented in both domains since it is a device-oriented approach. This explains why the adversary will not be able to break the \gls{signcrf} system. The adversary can succeed in hacking the system if, and only if, the hacker can access and control (i) the legitimate radio itself and (ii) the \gls{signcrf} structure which is completely unknown to any transmitter (including the adversary); therefore, \gls{signcrf} is immune from adversarial actions by adding an extra security layer.

% \begin{figure}[b!]
%     %\vspace{-5pt}
%     \centering
%     \includegraphics[width=1\columnwidth]{Figures/conc_table_9.pdf}
%     \caption{Summary table of the \textit{RRP performance} per technology dataset(WiFi, LoRa preamble and LoRa Payload) per testbed size(5, 20 and 100 radios.}
%     %\vspace{-5pt}
%     \label{fig:summary_rrp}
% \end{figure}

\section{Conclusions}\label{sec:conclusion}

This work proposes \gls{signcrf}, a novel and data-driven framework for real-time, diverse, and channel-agnostic \gls{rffdl} applications. To the best of our knowledge, we are the first to: (i) leverage the \gls{gan} concept to fingerprint wireless radios independent of the channel/environment dynamics and in a real-time and practical manner in the worst-case scenario, i.e., when several radios with identical manufacturing processes transmit the same baseband signal over different days and/or in diverse environments; (ii) improve the fingerprinting accuracy of WiFi technology up to 4x (100\% correct authentication) without applying any \gls{dsp} techniques and at scale; (iii) improve the fingerprinting accuracy for LoRa payload up to 6.3x (80\% correct authentication) at scale using 100 \textit{bit-similar} radios; (iv) prove that our technique works even with only preamble datasets in \gls{rffdl} applications to fingerprint LoRa radios; and (v) introducing a data-driven authentication approach that is robust and resilient to adversarial actions. In addition, our results show that \gls{signcrf} outperforms previous research for small- and medium-scale testbeds, and is the first successfully tested data-driven authentication method in a large-scale testbed. Our TTOD\% and RRP\% results are summarized in Tables \ref{table:summary_ttod} and \ref{table:summary_rrp}.

\begin{table}[hbt!]
 \begin{adjustbox}{width=\columnwidth,center}
    \centering
    \begin{tabular}{ccccccccc}\toprule
        \centering
        &\multicolumn{6}{c}{\textbf{TTOD\% per testbed scale per dataset}} & \phantom{abc}
        \\
        \cmidrule{2-7}
        &\multicolumn{3}{c}{\textbf{LoRa}} & \phantom{abc}& \multicolumn{2}{c}{\textbf{WiFi}}
        \\
        \multirow{3}{*}{Model} & \multicolumn{3}{c}{\textbf{Payload}} & \phantom{abc}& \multicolumn{2}{c}{\textbf{Equalized}}
        \\
        \cmidrule{2-4} \cmidrule{6-6} \cmidrule{6-7} & 5& 20 & 100 && 5& 20 \\ \cmidrule{2-4} \cmidrule{6-6} \cmidrule{6-7}
        Baseline &18\%&13\%&6\%&&18\%&9\%  \\
        DeeploRa\cite{al2021deeplora} &-&22\%& - && - & - \\
        \gls{signcrf} &80\%&73\%&53\%&&83\%&34\% \\
        \\
        {\textbf{Improvement}} & \textbf{4.5x} & \textbf{5.6x} & \textbf{8.8x} && \textbf{4.6x}& \textbf{3.7x}
        \\\bottomrule
    \end{tabular}%\vspace{-0.3cm}
    \end{adjustbox}
    \captionsetup{skip=5pt}
    \caption{Summary table for TTOD\% per technology per testbed size.}\label{table:summary_ttod}
\end{table}

\begin{table}[hbt!]
 \begin{adjustbox}{width=\columnwidth,center}

    \centering
    \begin{tabular}{ccccccccc}\toprule
        \centering
        &\multicolumn{6}{c}{\textbf{RRP\% per testbed scale and type}} & \phantom{abc}
        \\
        \cmidrule{2-7}
        &\multicolumn{3}{c}{\textbf{LoRa}} & \phantom{abc}& \multicolumn{2}{c}{\textbf{WiFi}}
        \\
        \multirow{3}{*}{Model} & \multicolumn{3}{c}{\textbf{Payload}} & \phantom{abc}& \multicolumn{2}{c}{\textbf{Equalized}}
        \\
        \cmidrule{2-4} \cmidrule{6-6} \cmidrule{6-7} & 5& 20 & 100 && 5& 20 \\ \cmidrule{2-4} \cmidrule{6-6} \cmidrule{6-7}
        Baseline &20\%&15\%&9\%&&20\%&30\%  \\
        DeeploRa\cite{al2021deeplora} &-&35\%& - && - & - \\
        \gls{signcrf} &80\%&75\%&73\%&&100\%&90\% \\
        \\
        {\textbf{Improvement}} & \textbf{4x} & \textbf{5x} & \textbf{8x} && \textbf{5x}& \textbf{3x}

        \\\bottomrule
    \end{tabular}%\vspace{-0.3cm}
      \end{adjustbox}
      \captionsetup{skip=5pt}
    \caption{Summary table for RRP\% per technology per testbed size.}\label{table:summary_rrp}
\end{table}
%\section*{Acknowledgments}
%\footnotesize
\bibliographystyle{IEEEtran}
\bibliography{mybib}
\vskip 0pt plus -1fil

\begin{IEEEbiography}
[{\includegraphics[width=1in,height=1.25in,keepaspectratio]{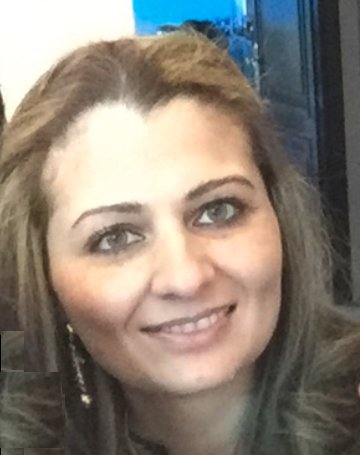}}]{Amani Al-Shawabka} is currently a Ph.D. candidate at Northeastern University in the Electrical and Computer Engineering department. She received her Master's degree in Computer Engineering from Northeastern University in 2019. She started her professional career in mobile network companies after completing her Bachelor's degree in Communication Engineering from Yarmouk University in Jordan. Her research interests are wireless communications through deep learning, embedded systems, and wireless network security.
\end{IEEEbiography}

\vskip 0pt plus -1fil

\begin{IEEEbiography}
[{\includegraphics[width=1in,height=1.25in,clip,keepaspectratio]{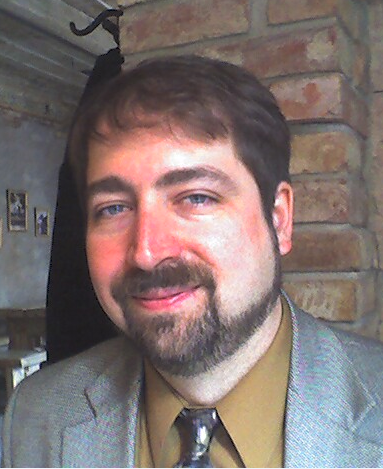}}]{Philip Pietraski} received the PhD, EE degree from Polytechnic University, now NYU (2000). Phil’s background is in signal processing and communication. Phil has been with InterDigital since 2001 and has a diverse history spanning modem algorithm design \& simulation, radio propagation \& impairments studies, and air interface R\&D spanning CDMA, HSPA, LTE, 5G, mmW, WiFi. He was the algorithm lead for a HSDPA coprocessor and facilitated tech-transfer to customers.  He was a 3GPP WG4 and WG1 delegate and hold over 100 patents. He spent some time with the Government Solutions team where he designed hybrid beamforming algorithms and related protocols for interference suppression and signature minimization. He is currently the lead of the AI-native air interface program at InterDigital – exploring the exciting intersection of Machine Learning and Wireless Communications. Prior to joining InterDigital, he worked as a Research Engineer at the DoE lab National Synchrotron Light Source (NSLS) at Brookhaven National Laboratory, working in diverse areas including pulsed-power RF systems, electron-beam control and diagnostics, and high-rate X-ray detectors; ultimately becoming Lead Engineer of the X-ray detector development project at the NSLS.  He also does volunteer work by serving on the board of trustees until 2019 and currently on the advisory committee for DeVry University, NJ; and coaching/mentoring for FIRST Robotics.
\end{IEEEbiography}

\vskip 0pt plus -1fil

\begin{IEEEbiography}
[{\includegraphics[width=1in,height=1.25in,clip,keepaspectratio]{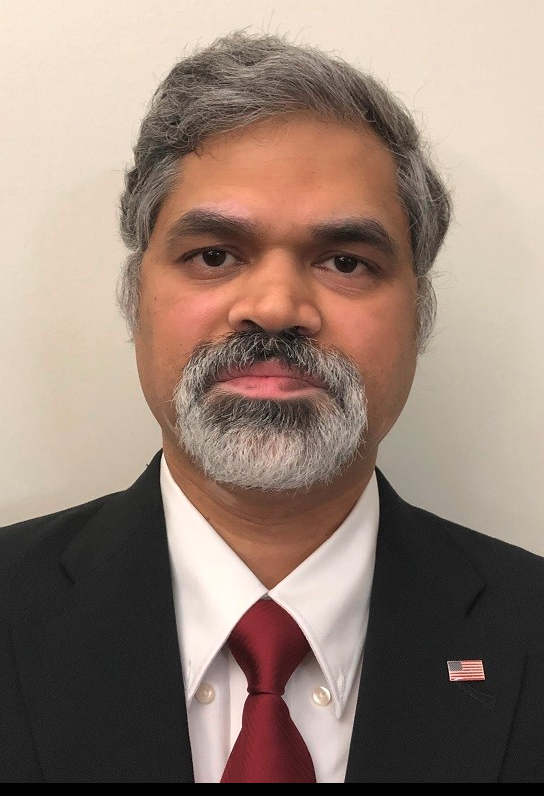}}]{Sudhir B Pattar} received  MSEE degree from Indian Institute of Science, Bangalore (1997). Sudhir’s background and interests are in wireless communications, signal processing, system engineering and application of machine learning for wireless communications. Sudhir has been with InterDigital since 2002 and is currently leading the Government Solutions program at Interdigital. In past, Sudhir has worked on modem algorithms, and protocols. Sudhir is currently Principal Investigator for multiple government funded projects. Sudhir has represented InterDigital at multiple cellular standards organizations. Sudhir worked at Broadcom as a Principal Engineer and worked on LTE modem development. Prior to joining InterDigital, he worked as a Research Engineer at NEC labs, working on coexistence of Wireless and cellular wireless systems.
\end{IEEEbiography}

\vskip 0pt plus -1fil

\begin{IEEEbiography}
[{\includegraphics[width=1in,height=1.25in,clip,keepaspectratio]{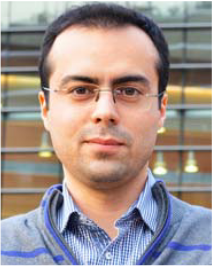}}]{Pedram Johari} is a Principal Research Scientist at the Institute for Wireless Internet of Things at Northeastern University, Boston MA, since January 2020. Pedram has received his Ph.D. in Electrical Engineering from the University at Buffalo, the State University of New York in 2018. He was a Research Assistant Professor and Adjunct Lecturer at the University at Buffalo in 2018 and 2019. Pedram was also the CTO of NanoThings Inc., New York City NY, in 2018 and 2019. His research interests are in the fusion of AI and future generation of cellular networks (5G and beyond), in particular focused on spectrum sharing, vehicular communications, full protocol wireless network emulators enabling wireless digital twins, and Internet of wireless medical things. Pedram has collaborated with several academic, industrial, and governmental research collaborators, including University at Buffalo, Georgia Tech, Qualcomm, MathWorks, InterDigital, MITRE, and VIAVI. Pedram is also the managing editor of Elsevier journal on Software Impacts.
\end{IEEEbiography}
\vskip 0pt plus -1fil

\begin{IEEEbiography}[{\includegraphics[width=1in,height=1.25in,clip,keepaspectratio]{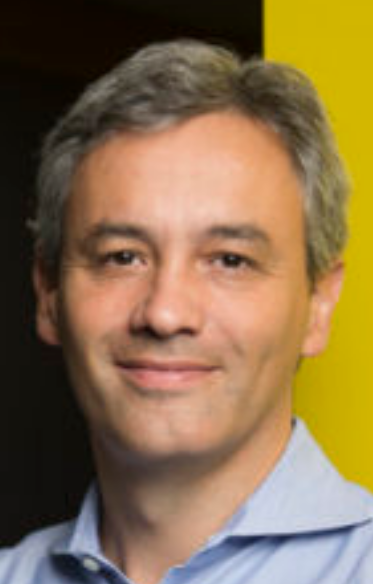}}]{Tommaso Melodia } (M'07, SM'16, F'18) is the William Lincoln Smith Chair Professor with the Department of Electrical and Computer Engineering at Northeastern University in Boston. He is also the Founding Director of the Institute for the Wireless Internet of Things and the Director of Research for the PAWR Project Office. He received his Ph.D. in Electrical and Computer Engineering from the Georgia Institute of Technology in 2007. He is a recipient of the National Science Foundation CAREER award. Prof. Melodia has served as Associate Editor fo IEEE Transactions on Wireless Communications, IEEE Transactions on Mobile Computing, Elsevier Computer Networks, among others. He has served as Technical Program Committee Chair for IEEE Infocom 2018, General Chair for IEEE SECON 2019, ACM Nanocom 2019, and ACM WUWnet 2014. Prof. Melodia is the Director of Research for the Platforms for Advanced Wireless Research (PAWR) Project Office, a \$100M public-private partnership to establish 4 city-scale platforms for wireless research to advance the US wireless ecosystem in years to come. Prof. Melodia's research on modeling, optimization, and experimental evaluation of Internet-of-Things and wireless networked systems has been funded by the National Science Foundation, the Air Force Research Laboratory the Office of Naval Research, DARPA, and the Army Research Laboratory.  Prof. Melodia is a Fellow of the IEEE and a Senior Member of the ACM.
 
\end{IEEEbiography}

\end{document}

%% file: acronyms.tex
\newacronym{6g}{6G}{sixth generation}
\newacronym{3gpp}{3GPP}{3rd Generation Partnership Project}
\newacronym{adc}{ADC}{Analog to Digital Converter}
\newacronym{dac}{DAC}{Digital to Analog Converter}
\newacronym{5g}{5G}{5th generation}
\newacronym{aimd}{AIMD}{Additive Increase Multiplicative Decrease}
\newacronym{am}{AM}{Acknowledged Mode}
\newacronym{amc}{AMC}{Adaptive Modulation and Coding}
\newacronym{aoa}{AoA}{Angle of Arrival}
\newacronym{aod}{AoD}{Angle of Departure}
\newacronym{aqm}{AQM}{Active Queue Management}
\newacronym{awgn}{AGWN}{Additive White Gaussian Noise}
\newacronym{balia}{BALIA}{Balanced Link Adaptation}
\newacronym{bdp}{BDP}{Bandwidth-Delay Product}
\newacronym{bf}{BF}{Beamforming}
\newacronym{fpga}{FPGA}{field-programmable gate array}
\newacronym{cc}{CC}{Congestion Control}
\newacronym{cdf}{CDF}{Cumulative Distribution Function}
\newacronym{cn}{CN}{Core Network}
\newacronym{cm}{CM}{confusion matrix}
\newacronym[plural=\gls{cnn}s,firstplural=convolutional neural networks (CNNs)]{cnn}{CNN}{convolutional neural network}
\newacronym{cqi}{CQI}{Channel Quality Information}
\newacronym{cp}{CP}{Control Plane}
\newacronym{csirs}{CSI-RS}{Channel State Information - Reference Signal}
\newacronym{dc}{DC}{Dual Connectivity}
\newacronym{dce}{DCE}{Direct Code Execution}
\newacronym{dci}{DCI}{Downlink Control Information}
\newacronym{dmr}{DMR}{Deadline Miss Ratio}
\newacronym{dmrs}{DMRS}{DeModulation Reference Signal}
\newacronym{e2e}{E2E}{End-to-End}
\newacronym{ecn}{ECN}{Explicit Congestion Notification}
\newacronym{ebs}{EBS}{exhaustive beam sweep}
\newacronym{edf}{EDF}{Earliest Deadline First}
\newacronym{enb}{eNB}{evolved Node Base}
\newacronym{epc}{EPC}{Evolved Packet Core}
\newacronym{es}{ES}{Edge Server}
\newacronym{fdma}{FDMA}{Frequency Division Multiple Access}
\newacronym{fdd}{FDD}{Frequency Division Duplexing}
\newacronym[firstplural=Radio Access Technologies (RATs)]{rat}{RAT}{Radio Access Technology}
\newacronym{fs}{FS}{Fast Switching}
\newacronym{txer}{TX}{transmitter}
\newacronym{rxer}{RX}{receiver}
\newacronym{bt}{BT}{beam tracking}
\newacronym{ftp}{FTP}{File Transfer Protocol}
\newacronym{gnb}{gNB}{Next Generation Node Base}
\newacronym{bs}{BS}{Base Station}
\newacronym{harq}{HARQ}{Hybrid Automatic Repeat reQuest}
\newacronym{hetnet}{HetNet}{Heterogeneous Network}
\newacronym{hh}{HH}{Hard Handover}
\newacronym{hol}{HOL}{Head-of-Line}
\newacronym{ia}{IA}{initial access}
\newacronym{imt}{IMT}{International Mobile Telecommunication}
\newacronym{iot}{IoT}{Internet of Things}
\newacronym{ioimt}{IoIMT}{Internet of Implantable Medical Things}
\newacronym{los}{LOS}{Line-of-Sight}
\newacronym{lte}{LTE}{Long Term Evolution}
\newacronym{m2m}{M2M}{Machine to Machine}
\newacronym{ml}{ML}{machine learning}
\newacronym{dl}{DL}{deep learning}
\newacronym{mac}{MAC}{Medium Access Control}
\newacronym{mc}{MC}{Multi-Connectivity}
\newacronym{mcs}{MCS}{Modulation and Coding Scheme}
\newacronym{mec}{MEC}{Mobile Edge Cloud}
\newacronym{mi}{MI}{Mutual Information}
\newacronym{mimo}{MIMO}{Multiple Input, Multiple Output}
\newacronym{mmwave}{mmWave}{millimeter wave}
\newacronym{mmWave}{mmWave}{Millimeter wave}
\newacronym{mptcp}{MPTCP}{Multipath TCP}
\newacronym{mr}{MR}{Maximum Rate}
\newacronym{mss}{MSS}{Maximum Segment Size}
\newacronym{mtd}{MTD}{Machine-Type Device}
\newacronym{mtu}{MTU}{Maximum Transmission Unit}
\newacronym{nfv}{NFV}{Network Function Virtualization}
\newacronym{nlos}{NLOS}{Non-Line-of-Sight}
\newacronym{nr}{NR}{New Radio}
\newacronym{ofdm}{OFDM}{Orthogonal Frequency Division Multiplexing}
\newacronym{pdcch}{PDCCH}{Physical Downlink Control Channel}
\newacronym{pdcp}{PDCP}{Packet Data Convergence Protocol}
\newacronym{pdsch}{PDSCH}{Physical Downlink Shared Channel}
\newacronym{pdu}{PDU}{Packet Data Unit}
\newacronym{pf}{PF}{Proportional Fair}
\newacronym{pgw}{PGW}{Packet Gateway}
\newacronym{phy}{PHY}{Physical}
\newacronym{pbch}{PBCH}{Physical Broadcast Channel}
\newacronym[plural=\gls{mme}s,firstplural=Mobility Management Entities (MMEs)]{mme}{MME}{Mobility Management Entity}
\newacronym{prb}{PRB}{Physical Resource Block}
\newacronym{pss}{PSS}{Primary Synchronization Signal}
\newacronym{pucch}{PUCCH}{Physical Uplink Control Channel}
\newacronym{pusch}{PUSCH}{Physical Uplink Shared Channel}
\newacronym{rach}{RACH}{Random Access Channel}
\newacronym{ran}{RAN}{Radio Access Network}
\newacronym{red}{RED}{Random Early Detection}
\newacronym{rf}{RF}{Radio Frequency}
\newacronym{rlc}{RLC}{Radio Link Control}
\newacronym{rlf}{RLF}{Radio Link Failure}
\newacronym{rrc}{RRC}{Radio Resource Control}
\newacronym{rrm}{RRM}{Radio Resource Management}
\newacronym{rr}{RR}{Round Robin}
\newacronym{rs}{RS}{Remote Server}
\newacronym{rsrp}{RSRP}{Reference Signal Received Power}
\newacronym{rss}{RSS}{Received Signal Strength}
\newacronym{rtt}{RTT}{Round Trip Time}
\newacronym{rw}{RW}{Receive Window}
\newacronym{rx}{RX}{Receiver}
\newacronym{sa}{SA}{standalone}
\newacronym{sack}{SACK}{Selective Acknowledgment}
\newacronym{sap}{SAP}{Service Access Point}
\newacronym{ap}{AP}{Access Point}
\newacronym{sch}{SCH}{Secondary Cell Handover}
\newacronym{scoot}{SCOOT}{Split Cycle Offset Optimization Technique}
\newacronym{sdma}{SDMA}{Spatial Division Multiple Access}
\newacronym{sinr}{SINR}{Signal to Interference plus Noise Ratio}
\newacronym{sm}{SM}{Saturation Mode}
\newacronym{snr}{SNR}{Signal-to-Noise-Ratio}
\newacronym{son}{SON}{Self-Organizing Network}
\newacronym{ss}{SS}{Synchronization Signal}
\newacronym{ssbs}{SSBs}{synchronization signal blocks}
\newacronym{ssb}{SSB}{synchronization signal block}
\newacronym{srs}{SRS}{Sounding Reference Signal}
\newacronym{sss}{SSS}{Secondary Synchronization Signal}
\newacronym{tb}{TB}{Transport Block}
\newacronym{tcp}{TCP}{Transmission Control Protocol}
\newacronym{tdd}{TDD}{Time Division Duplexing}
\newacronym{tdma}{TDMA}{Time Division Multiple Access}
\newacronym{tfl}{TfL}{Transport for London}
\newacronym{tm}{TM}{Transparent Mode}
\newacronym{trp}{TRP}{Transmitter Receiver Pair}
\newacronym{tti}{TTI}{Transmission Time Interval}
\newacronym{ttt}{TTT}{Time-to-Trigger}
\newacronym{tx}{TX}{Transmitter}
\newacronym{ue}{UE}{User Equipment}
\newacronym{ul}{UL}{Uplink}
\newacronym{uml}{UML}{Unified Modeling Language}
\newacronym{um}{UM}{Unacknowledged Mode}
\newacronym{utc}{UTC}{Urban Traffic Control}
\newacronym{vm}{VM}{Virtual Machine}
\newacronym{rsrq}{RSRQ}{Reference Signal Received Quality}
\newacronym{rssi}{RSSI}{Received Signal Strength Indicator}
\newacronym{crs}{CRS}{Cell Reference Signal}
\newacronym{nsa}{NSA}{Non Stand Alone}
\newacronym{mrdc}{MR-DC}{Multi \gls{rat} \gls{dc}}
\newacronym{endc}{EN-DC}{E-UTRAN-\gls{nr} \gls{dc}}
\newacronym{5gc}{5GC}{5G Core}
\newacronym{si}{SI}{Study Item}
\newacronym{iab}{IAB}{Integrated Access and Backhaul}
\newacronym{wf}{WF}{Wired-first}
\newacronym{hqf}{HQF}{Highest-quality-first}
\newacronym{pa}{PA}{Position-aware}
\newacronym{mlr}{MLR}{Maximum-local-rate}
\newacronym{wbf}{WBF}{Wired Bias Function}
\newacronym{mib}{MIB}{Master Information Block}
\newacronym{sib}{SIB}{Secondary Information Block}
\newacronym{kpi}{KPI}{Key Performance Indicator}
\newacronym{ppp}{PPP}{Poisson Point Process}
\newacronym{gtp}{GTP}{GPRS Tunneling Protocol}
\newacronym{amf}{AMF}{Access and Mobility Management Function}
\newacronym{dash}{DASH}{Dynamic Adaptive Streaming over HTTP}
\newacronym{http}{HTTP}{HyperText Transfer Protocol}
\newacronym{qos}{QoS}{Quality of Service}
\newacronym{udp}{UDP}{User Datagram Protocol}
\newacronym{cu}{CU}{Central Unit}
\newacronym{du}{DU}{Distributed Unit}
\newacronym{mt}{MT}{Mobile Termination}
\newacronym{sdap}{SDAP}{Service Data Adaptation Protocol}
\newacronym{tdm}{TDM}{Time Division Multiplexing}
\newacronym{fdm}{FDM}{Frequency Division Multiplexing}
\newacronym{sdm}{SDM}{Space Division Multiplexing}
\newacronym{dag}{DAG}{Directed Acyclic Graph}
\newacronym{st}{ST}{Spanning Tree}
\newacronym{ummimo}{UM-MIMO}{Ultra-massive Multiple Input, Multiple Output}
\newacronym{uavs}{UAVs}{Unmanned Aerial Vehicles}
\newacronym{wlan}{WLAN}{Wireless LAN}
\newacronym{rlnc}{RLNC}{Random Linear Network Coding}
\newacronym{drx}{DRX}{Discontinuous Reception}
\newacronym{cpu}{CPU}{Central Processing Unit}
\newacronym{txb}{TXB}{transmitter's beam}
\newacronym{rxb}{RXB}{receiver's beam}
\newacronym{sifs}{SIFS}{Short Interframe Space}
\newacronym{difs}{DIFS}{DCF Interframe Space}

\newacronym{fph}{FPH}{false positives per hour}
\newacronym{fp}{FP}{false positive}

\newacronym{dsp}{DSP}{digital signal processing}
\newacronym{rfid}{RFID}{Radio Frequency Identification}
\newacronym{rfp}{RFP}{radio fingerprinting}
\newacronym{sdr}{SDR}{software-defined radio}
\newacronym{dnn}{DNN}{deep neural network}
\newacronym{ieeg}{iEEG}{intracranial electroencephalogram}
\newacronym{auc}{AUC}{area under the curve}
\newacronym{soc}{SoC}{system on chip}
\newacronym[plural=\gls{imd}s,firstplural=implantable medical devices (IMDs)]{imd}{IMD}{implantable medical device}

\newacronym{rffdl}{RFFDL}{Radio Frequency Fingerprinting through Deep Learning}
\newacronym{fir}{FIR}{Finite Impulse Response}
\newacronym{nn}{NN}{Neural Network}
\newacronym{gan}{GAN}{Generative Adversarial Network}
\newacronym{cgan}{CycleGAN}{Cycle-Consistent Generative Adversarial Networks}
\newacronym{2d}{2D}{two dimension}
\newacronym{signcrf}{\textit{SignCRF}}{Scalable Intelligent GAN Network-based Channel-agnostic Radio Fingerprinting system}

%% file: Figures/CM_wifi_5dev_without_cgan.tex
\begin{tikzpicture}
\pgfplotsset{every tick label/.append style={font=\tiny}}

\begin{axis}[
enlargelimits=false,
colorbar,
colormap/Purples,
width=\fwidth,
height=\fheight,
at={(0\fwidth,0\fheight)},
scale only axis,
tick align=inside,
ylabel={Predicted Radio},
xmin=-0.5,
xmax=4.5,
xtick style={color=black},
xlabel style={font=\scriptsize\color{white!15!black}},
ylabel style={font=\scriptsize\color{white!15!black}},
xlabel={True Radio},
ymin=-0.5,
ymax=4.5,
xlabel shift=-5pt,
ylabel shift=-5pt,
ytick style={color=black},
axis background/.style={fill=white},
colorbar horizontal,
colorbar style={
at={(0,1.05)},               % <-- (changed)
anchor=below south west,    % <-- (changed)
% change the width of the colorbar relative to the main `axis' environment
width=\pgfkeysvalueof{/pgfplots/parent axis width},
xtick={0, 0.5, 1},
xmin=0,
xmax=1,
axis x line*=top,
xticklabel shift=2pt,
},
colorbar/width=2mm,
]
\addplot [matrix plot,point meta=explicit]
 coordinates {
 
% [[3356.  582. 1521. 1272. 3269.]
%  [2009. 1708. 2418. 2998.  867.]
%  [ 704. 5486. 1123.  864.  462.]
%  [3689. 1956. 1251.  276. 2828.]
%  [1555. 1674. 1519. 2595. 2657.]]
% original updated

%  (0,0) [0.34]   (1,0) [0.06]    (2,0) [0.15]   (3,0) [0.13]    (4,0) [0.33]	
 
%  (0,1) [0.20]	(1,1) [0.17]    (2,1) [0.24]    (3,1) [0.30]    (4,1) [0.09]				
 
%  (0,2) [0.69]   (1,2) [0.06]	(2,2) [0.11]    (2,3) [0.07]	(4,2) [0.07]		
 
%  (0,3) [0.37]	(1,3) [0.19]  (2,3) [0.12]  (3,3) [0.03]    (4,3) [0.29]			
 
%  (0,4) [0.15]	(1,4) [0.16]	(2,4) [0.15]    (4,3) [0.26]    (4,4) [0.27]	

 (0,0) [0.34]   (0,1) [0.06]    (0,2) [0.15]   (0,3) [0.13]    (0,4) [0.33]	
 
 (1,0) [0.20]	(1,1) [0.17]    (1,2) [0.24]    (1,3) [0.30]    (1,4) [0.09]				
 
 (2,0) [0.69]   (2,1) [0.06]	(2,2) [0.11]    (2,3) [0.07]	(2,4) [0.07]		
 
 (3,0) [0.37]	(3,1) [0.19]  (3,2) [0.12]  (3,3) [0.03]    (3,4) [0.29]			
 
 (4,0) [0.15]	(4,1) [0.16]	(4,2) [0.15]    (4,3) [0.26]    (4,4) [0.27]	

};
\end{axis}
\end{tikzpicture}

%% file: Figures/CM_wifi_5dev_with_cgan.tex
\begin{tikzpicture}
\pgfplotsset{every tick label/.append style={font=\tiny}}

\begin{axis}[
enlargelimits=false,
colorbar,
colormap/Purples,
width=\fwidth,
height=\fheight,
at={(0\fwidth,0\fheight)},
scale only axis,
tick align=inside,
ylabel={Predicted Radio},
xmin=-0.5,
xmax=4.5,
xtick style={color=black},
xlabel style={font=\scriptsize\color{white!15!black}},
ylabel style={font=\scriptsize\color{white!15!black}},
xlabel={True Radio},
ymin=-0.5,
ymax=4.5,
xlabel shift=-5pt,
ylabel shift=-5pt,
ytick style={color=black},
axis background/.style={fill=white},
colorbar horizontal,
colorbar style={
at={(0,1.05)},               % <-- (changed)
anchor=below south west,    % <-- (changed)
% change the width of the colorbar relative to the main `axis' environment
width=\pgfkeysvalueof{/pgfplots/parent axis width},
xtick={0, 0.5, 1},
xmin=0,
xmax=1,
axis x line*=top,
xticklabel shift=2pt,
},
colorbar/width=2mm,
]
\addplot [matrix plot,point meta=explicit]
 coordinates {
 
 %[[0.3356 0.0582 0.1521 0.1272 0.3269]
 %[0.0034 0.5093 0.1937 0.115  0.1786] after 60 epochs
 %[0  9.670e-02 7.234e-01 2.850e-02 1.479e-01] after 90 epochs
 %[0.081  0.3825 0.2114 0.1674 0.1577] after 60 epochs
 %[0.0033 0.1725 0.3653 0.0121 0.4468] after 60
%  original
(0,0) [0.99]	(0,1) [0]	(0,2) [0]	(0,3) [0.01]	(0,4) [0]
% [9925.    0.    0.   75.    0.]
 
(1,0) [0.07]	(1,1) [0.67]	(1,2) [0.17]	(1,3) [0]	(1,4) [0.01]

% [ 692. 6655. 1688.   61.  904.]
 
%  (2,0) [15]	(2,1) [19]	(2,2) [72]	(2,3) [21]	(2,4) [37]

(2,0) [0.001]	(2,1) [0.001]	(2,2) [0.95]	(2,3) [0.03]	(2,4) [0.01]

% 1.000e+00 1.000e+00 9.548e+03 3.380e+02 1.120e+02
 
%  (3,0) [13]	(3,1) [11]	(3,2) [3]	(3,3) [17]	(3,4) [1]

 (3,0) [0.03]	(3,1) [0.02]	(3,2) [0.01]	(3,3) [0.94]	(3,4) [0.002] 
 
% [311.  175.   80. 9416.   18.]
 
%  (4,0) [33]	(4,1) [18]	(4,2) [15]	(4,3) [16]	(4,4) [45]	
% 
(4,0) [0.04]	(4,1) [0.2]	(4,2) [0.08]	(4,3) [0.09]	(4,4) [0.58]	

% [ 390. 2047.  792.  930. 5841.]

};
\end{axis}
\end{tikzpicture}

%% file: Figures/CM_wif_20dev_without_cgan.tex
\begin{tikzpicture}
\pgfplotsset{every tick label/.append style={font=\tiny}}

\begin{axis}[
enlargelimits=false,
colorbar,
colormap/Purples,
width=\fwidth,
height=\fheight,
at={(0\fwidth,0\fheight)},
scale only axis,
tick align=inside,
ylabel={Predicted Radio},
xmin=-0.5,
xmax=19.5,
xtick style={color=black},
xlabel style={font=\scriptsize\color{white!15!black}},
ylabel style={font=\scriptsize\color{white!15!black}},
xlabel={True Radio},
ymin=-0.5,
ymax=19.5,
xlabel shift=-5pt,
ylabel shift=-5pt,
ytick style={color=black},
axis background/.style={fill=white},
colorbar horizontal,
colorbar style={
at={(0,1.05)},               % <-- (changed)
anchor=below south west,    % <-- (changed)
% change the width of the colorbar relative to the main `axis' environment
width=\pgfkeysvalueof{/pgfplots/parent axis width},
xtick={0, 0.5, 1},
xmin=0,
xmax=1,
axis x line*=top,
xticklabel shift=2pt,
},
colorbar/width=2mm,
]
\addplot [matrix plot,point meta=explicit]
 coordinates {

(0 , 0)  	[0.0128]
(0 , 1)  	[0.0866]
(0 , 2)  	[0.0464]
(0 , 3)  	[0.0203]
(0 , 4)  	[0.0285]
(0 , 5)  	[0.0534]
(0 , 6)  	[0.0415]
(0 , 7)  	[0.0188]
(0 , 8)  	[0.0204]
(0 , 9)  	[0.0623]
(0 , 10)  	[0.0356]
(0 , 11)  	[0.049]
(0 , 12)  	[0.0359]
(0 , 13)  	[0.0431]
(0 , 14)  	[0.0883]
(0 , 15)  	[0.0616]
(0 , 16)  	[0.0522]
(0 , 17)  	[0.0954]
(0 , 18)  	[0.1124]
(0 , 19)  	[0.0355]

(1 , 0)  	[0.0366]
(1 , 1)  	[0.0163]
(1 , 2)  	[0.0847]
(1 , 3)  	[0.09]
(1 , 4)  	[0.0385]
(1 , 5)  	[0.0026]
(1 , 6)  	[0.0607]
(1 , 7)  	[0.0435]
(1 , 8)  	[0.0097]
(1 , 9)  	[0.3017]
(1 , 10)  	[0.0453]
(1 , 11)  	[0.0187]
(1 , 12)  	[0.0263]
(1 , 13)  	[0.0319]
(1 , 14)  	[0.0209]
(1 , 15)  	[0.0695]
(1 , 16)  	[0.054]
(1 , 17)  	[0.0007]
(1 , 18)  	[0.016]
(1 , 19)  	[0.0324]

(2 , 0)  	[0.1463]
(2 , 1)  	[0.0401]
(2 , 2)  	[0.02]
(2 , 3)  	[0.0973]
(2 , 4)  	[0.0111]
(2 , 5)  	[0.0459]
(2 , 6)  	[0.0818]
(2 , 7)  	[0.011]
(2 , 8)  	[0.0191]
(2 , 9)  	[0.0999]
(2 , 10)  	[0.0158]
(2 , 11)  	[0.0146]
(2 , 12)  	[0.0295]
(2 , 13)  	[0.082]
(2 , 14)  	[0.0245]
(2 , 15)  	[0.0568]
(2 , 16)  	[0.0815]
(2 , 17)  	[0.0304]
(2 , 18)  	[0.0653]
(2 , 19)  	[0.0271]

(3 , 0)  	[0.0465]
(3 , 1)  	[0.0945]
(3 , 2)  	[0.0376]
(3 , 3)  	[0.0134]
(3 , 4)  	[0.0908]
(3 , 5)  	[0.0248]
(3 , 6)  	[0.0038]
(3 , 7)  	[0.0714]
(3 , 8)  	[0.0389]
(3 , 9)  	[0.1167]
(3 , 10)  	[0.0239]
(3 , 11)  	[0.0437]
(3 , 12)  	[0.018]
(3 , 13)  	[0.0038]
(3 , 14)  	[0.0267]
(3 , 15)  	[0.0671]
(3 , 16)  	[0.0308]
(3 , 17)  	[0.0421]
(3 , 18)  	[0.1224]
(3 , 19)  	[0.0831]

(4 , 0)  	[0.084]
(4 , 1)  	[0.0549]
(4 , 2)  	[0.0661]
(4 , 3)  	[0.0377]
(4 , 4)  	[0.1804]
(4 , 5)  	[0.0427]
(4 , 6)  	[0.0949]
(4 , 7)  	[0.0286]
(4 , 8)  	[0.0153]
(4 , 9)  	[0.0251]
(4 , 10)  	[0.0565]
(4 , 11)  	[0.0207]
(4 , 12)  	[0.0515]
(4 , 13)  	[0.0256]
(4 , 14)  	[0.0191]
(4 , 15)  	[0.0224]
(4 , 16)  	[0.0456]
(4 , 17)  	[0.0233]
(4 , 18)  	[0.0545]
(4 , 19)  	[0.0511]

(5 , 0)  	[0.0695]
(5 , 1)  	[0.007]
(5 , 2)  	[0.021]
(5 , 3)  	[0.073]
(5 , 4)  	[0.0573]
(5 , 5)  	[0.1305]
(5 , 6)  	[0.0833]
(5 , 7)  	[0.0662]
(5 , 8)  	[0.0792]
(5 , 9)  	[0.0066]
(5 , 10)  	[0.017]
(5 , 11)  	[0.0224]
(5 , 12)  	[0.026]
(5 , 13)  	[0.0249]
(5 , 14)  	[0.0385]
(5 , 15)  	[0.0322]
(5 , 16)  	[0.0477]
(5 , 17)  	[0.1058]
(5 , 18)  	[0.0476]
(5 , 19)  	[0.0443]

(6 , 0)  	[0.0495]
(6 , 1)  	[0.0539]
(6 , 2)  	[0.01]
(6 , 3)  	[0.117]
(6 , 4)  	[0.1548]
(6 , 5)  	[0.0979]
(6 , 6)  	[0.0191]
(6 , 7)  	[0.0109]
(6 , 8)  	[0.027]
(6 , 9)  	[0.0422]
(6 , 10)  	[0.024]
(6 , 11)  	[0.1348]
(6 , 12)  	[0.0117]
(6 , 13)  	[0.0226]
(6 , 14)  	[0.0881]
(6 , 15)  	[0.0115]
(6 , 16)  	[0.0096]
(6 , 17)  	[0.0064]
(6 , 18)  	[0.071]
(6 , 19)  	[0.038]

(7 , 0)  	[0.0246]
(7 , 1)  	[0.0379]
(7 , 2)  	[0.022]
(7 , 3)  	[0.02]
(7 , 4)  	[0.0214]
(7 , 5)  	[0.0503]
(7 , 6)  	[0.0618]
(7 , 7)  	[0.2032]
(7 , 8)  	[0.0469]
(7 , 9)  	[0.0439]
(7 , 10)  	[0.0552]
(7 , 11)  	[0.0397]
(7 , 12)  	[0.0516]
(7 , 13)  	[0.0109]
(7 , 14)  	[0.0353]
(7 , 15)  	[0.0142]
(7 , 16)  	[0.0638]
(7 , 17)  	[0.1303]
(7 , 18)  	[0.0329]
(7 , 19)  	[0.0341]

(8 , 0)  	[0.0913]
(8 , 1)  	[0.0165]
(8 , 2)  	[0.0315]
(8 , 3)  	[0.0154]
(8 , 4)  	[0.0701]
(8 , 5)  	[0.0742]
(8 , 6)  	[0.051]
(8 , 7)  	[0.0853]
(8 , 8)  	[0.0844]
(8 , 9)  	[0.1465]
(8 , 10)  	[0.0213]
(8 , 11)  	[0.076]
(8 , 12)  	[0.0241]
(8 , 13)  	[0.0257]
(8 , 14)  	[0.055]
(8 , 15)  	[0.0443]
(8 , 16)  	[0.0026]
(8 , 17)  	[0.0316]
(8 , 18)  	[0.0257]
(8 , 19)  	[0.0275]

(9 , 0)  	[0.0115]
(9 , 1)  	[0.0362]
(9 , 2)  	[0.1286]
(9 , 3)  	[0.0479]
(9 , 4)  	[0.0571]
(9 , 5)  	[0.039]
(9 , 6)  	[0.1735]
(9 , 7)  	[0.0343]
(9 , 8)  	[0.0277]
(9 , 9)  	[0.2331]
(9 , 10)  	[0.027]
(9 , 11)  	[0.0812]
(9 , 12)  	[0.0057]
(9 , 13)  	[0.0047]
(9 , 14)  	[0.055]
(9 , 15)  	[0.0057]
(9 , 16)  	[0.0007]
(9 , 17)  	[0.017]
(9 , 18)  	[0.0118]
(9 , 19)  	[0.0023]

(10 , 0)  	[0.0081]
(10 , 1)  	[0.0094]
(10 , 2)  	[0.0005]
(10 , 3)  	[0.0068]
(10 , 4)  	[0.0747]
(10 , 5)  	[0.0085]
(10 , 6)  	[0.0943]
(10 , 7)  	[0.0031]
(10 , 8)  	[0.0044]
(10 , 9)  	[0.023]
(10 , 10)  	[0.2287]
(10 , 11)  	[0.1532]
(10 , 12)  	[0.0343]
(10 , 13)  	[0.0196]
(10 , 14)  	[0.1248]
(10 , 15)  	[0.0627]
(10 , 16)  	[0.0006]
(10 , 17)  	[0.0863]
(10 , 18)  	[0.0105]
(10 , 19)  	[0.0465]

(11 , 0)  	[0.0699]
(11 , 1)  	[0.0019]
(11 , 2)  	[0.0194]
(11 , 3)  	[0.1394]
(11 , 4)  	[0.061]
(11 , 5)  	[0.0382]
(11 , 6)  	[0.0136]
(11 , 7)  	[0.0597]
(11 , 8)  	[0.0009]
(11 , 9)  	[0.0514]
(11 , 10)  	[0.0929]
(11 , 11)  	[0.0278]
(11 , 12)  	[0.0149]
(11 , 13)  	[0.0384]
(11 , 14)  	[0.1151]
(11 , 15)  	[0.0449]
(11 , 16)  	[0.1038]
(11 , 17)  	[0.0658]
(11 , 18)  	[0.0154]
(11 , 19)  	[0.0256]

(12 , 0)  	[0.0284]
(12 , 1)  	[0.0634]
(12 , 2)  	[0.0187]
(12 , 3)  	[0.1113]
(12 , 4)  	[0.0259]
(12 , 5)  	[0.0328]
(12 , 6)  	[0.046]
(12 , 7)  	[0.0032]
(12 , 8)  	[0.0406]
(12 , 9)  	[0.0385]
(12 , 10)  	[0.0171]
(12 , 11)  	[0.0285]
(12 , 12)  	[0.0394]
(12 , 13)  	[0.0288]
(12 , 14)  	[0.1203]
(12 , 15)  	[0.0669]
(12 , 16)  	[0.063]
(12 , 17)  	[0.1266]
(12 , 18)  	[0.0507]
(12 , 19)  	[0.0499]

(13 , 0)  	[0.0143]
(13 , 1)  	[0.0205]
(13 , 2)  	[0.069]
(13 , 3)  	[0.1054]
(13 , 4)  	[0.1064]
(13 , 5)  	[0.0399]
(13 , 6)  	[0.0185]
(13 , 7)  	[0.0088]
(13 , 8)  	[0.0581]
(13 , 9)  	[0.0208]
(13 , 10)  	[0.0142]
(13 , 11)  	[0.0449]
(13 , 12)  	[0.0989]
(13 , 13)  	[0.1107]
(13 , 14)  	[0.0394]
(13 , 15)  	[0.0175]
(13 , 16)  	[0.0538]
(13 , 17)  	[0.0316]
(13 , 18)  	[0.0708]
(13 , 19)  	[0.0565]

(14 , 0)  	[0.0149]
(14 , 1)  	[0.0504]
(14 , 2)  	[0.0049]
(14 , 3)  	[0.0204]
(14 , 4)  	[0.0368]
(14 , 5)  	[0.011]
(14 , 6)  	[0.0227]
(14 , 7)  	[0.005]
(14 , 8)  	[0.0109]
(14 , 9)  	[0.007]
(14 , 10)  	[0.2049]
(14 , 11)  	[0.1952]
(14 , 12)  	[0.0211]
(14 , 13)  	[0.0321]
(14 , 14)  	[0.1426]
(14 , 15)  	[0.0422]
(14 , 16)  	[0.064]
(14 , 17)  	[0.0585]
(14 , 18)  	[0.0393]
(14 , 19)  	[0.0161]

(15 , 0)  	[0.028]
(15 , 1)  	[0.0202]
(15 , 2)  	[0.0227]
(15 , 3)  	[0.0145]
(15 , 4)  	[0.0274]
(15 , 5)  	[0.0653]
(15 , 6)  	[0.0559]
(15 , 7)  	[0.0125]
(15 , 8)  	[0.0188]
(15 , 9)  	[0.0393]
(15 , 10)  	[0.1124]
(15 , 11)  	[0.1184]
(15 , 12)  	[0.1379]
(15 , 13)  	[0.0151]
(15 , 14)  	[0.1267]
(15 , 15)  	[0.0951]
(15 , 16)  	[0.0044]
(15 , 17)  	[0.0278]
(15 , 18)  	[0.0286]
(15 , 19)  	[0.029]

(16 , 0)  	[0.075]
(16 , 1)  	[0.0748]
(16 , 2)  	[0.0468]
(16 , 3)  	[0.1346]
(16 , 4)  	[0.0352]
(16 , 5)  	[0.1023]
(16 , 6)  	[0.0373]
(16 , 7)  	[0.0471]
(16 , 8)  	[0.0106]
(16 , 9)  	[0.0391]
(16 , 10)  	[0.0087]
(16 , 11)  	[0.0881]
(16 , 12)  	[0.026]
(16 , 13)  	[0.0106]
(16 , 14)  	[0.0161]
(16 , 15)  	[0.0107]
(16 , 16)  	[0.073]
(16 , 17)  	[0.0329]
(16 , 18)  	[0.1059]
(16 , 19)  	[0.0252]

(17 , 0)  	[0.0942]
(17 , 1)  	[0.0764]
(17 , 2)  	[0.0547]
(17 , 3)  	[0.0028]
(17 , 4)  	[0.0531]
(17 , 5)  	[0.0242]
(17 , 6)  	[0.05]
(17 , 7)  	[0.0056]
(17 , 8)  	[0.0235]
(17 , 9)  	[0.0042]
(17 , 10)  	[0.0258]
(17 , 11)  	[0.0362]
(17 , 12)  	[0.0906]
(17 , 13)  	[0.0379]
(17 , 14)  	[0.0144]
(17 , 15)  	[0.1499]
(17 , 16)  	[0.0938]
(17 , 17)  	[0.0836]
(17 , 18)  	[0.0493]
(17 , 19)  	[0.0298]

(18 , 0)  	[0.0922]
(18 , 1)  	[0.0082]
(18 , 2)  	[0.0735]
(18 , 3)  	[0.0719]
(18 , 4)  	[0.033]
(18 , 5)  	[0.1411]
(18 , 6)  	[0.0199]
(18 , 7)  	[0.0196]
(18 , 8)  	[0.0686]
(18 , 9)  	[0]
(18 , 10)  	[0.0021]
(18 , 11)  	[0.0223]
(18 , 12)  	[0.0637]
(18 , 13)  	[0.0044]
(18 , 14)  	[0.0354]
(18 , 15)  	[0.0445]
(18 , 16)  	[0.0796]
(18 , 17)  	[0.0271]
(18 , 18)  	[0.1527]
(18 , 19)  	[0.0402]

(19 , 0)  	[0.0613]
(19 , 1)  	[0.0058]
(19 , 2)  	[0.098]
(19 , 3)  	[0.01]
(19 , 4)  	[0.0049]
(19 , 5)  	[0.0494]
(19 , 6)  	[0.0474]
(19 , 7)  	[0.0741]
(19 , 8)  	[0.0163]
(19 , 9)  	[0.0637]
(19 , 10)  	[0.0929]
(19 , 11)  	[0.0482]
(19 , 12)  	[0.0643]
(19 , 13)  	[0.018]
(19 , 14)  	[0.0174]
(19 , 15)  	[0.116]
(19 , 16)  	[0.1063]
(19 , 17)  	[0.054]
(19 , 18)  	[0.03]
(19 , 19)  	[0.022]

};
\end{axis}
\end{tikzpicture}

%% file: Figures/CM_wifi_20dev_with_cgan.tex
\begin{tikzpicture}
\pgfplotsset{every tick label/.append style={font=\tiny}}

\begin{axis}[
enlargelimits=false,
colorbar,
colormap/Purples,
width=\fwidth,
height=\fheight,
at={(0\fwidth,0\fheight)},
scale only axis,
tick align=inside,
ylabel={Predicted Radio},
xmin=-0.5,
xmax=19.5,
xtick style={color=black},
xlabel style={font=\scriptsize\color{white!15!black}},
ylabel style={font=\scriptsize\color{white!15!black}},
xlabel={True Radio},
ymin=-0.5,
ymax=19.5,
xlabel shift=-5pt,
ylabel shift=-5pt,
ytick style={color=black},
axis background/.style={fill=white},
colorbar horizontal,
colorbar style={
at={(0,1.05)},               % <-- (changed)
anchor=below south west,    % <-- (changed)
% change the width of the colorbar relative to the main `axis' environment
width=\pgfkeysvalueof{/pgfplots/parent axis width},
xtick={0, 0.5, 1},
xmin=0,
xmax=1,
axis x line*=top,
xticklabel shift=2pt,
},
colorbar/width=2mm,
]
\addplot [matrix plot,point meta=explicit]
 coordinates {

(0,0)	 [0.09]	(0,1)	 [0.02]	(0,2)	 [0.02]	(0,3)	 [0.05]	(0,4)	 [0.08]	(0,5)	 [0.07]	(0,6)	 [0]	(0,7)	 [0]	(0,8)	 [0]	(0,9)	 [0]	(0,10)	 [0.02]	(0,11)	 [0.02]	(0,12)	 [0.04]	(0,13)	 [0]	(0,14)	 [0.05]	(0,15)	 [0.01]	(0,16)	 [0.03]	(0,17)	 [0.01]	(0,18)	 [0.09]	(0,19)	 [0.07]

(1,0)	 [0.02]	(1,1)	 [0.22]	(1,2)	 [0.02]	(1,3)	 [0.02]	(1,4)	 [0.05]	(1,5)	 [0.01]	(1,6)	 [0.05]	(1,7)	 [0.01]	(1,8)	 [0.03]	(1,9)	 [0.04]	(1,10)	 [0.01]	(1,11)	 [0.01]	(1,12)	 [0.02]	(1,13)	 [0.01]	(1,14)	 [0]	(1,15)	 [0.03]	(1,16)	 [0.03]	(1,17)	 [0.10]	(1,18)	 [0.01]	(1,19)	 [0.01]						
(2,0)	 [0.01]	(2,1)	 [0.09]	(2,2)	 [0.17]	(2,3)	 [0.07]	(2,4)	 [0.07]	(2,5)	 [0.02]	(2,6)	 [0.07]	(2,7)	 [0.01]	(2,8)	 [0.06]	(2,9)	 [0.01]	(2,10)	 [0]	(2,11)	 [0.02]	(2,12)	 [0.08]	(2,13)	 [0.04]	(2,14)	 [0]	(2,15)	 [0.09]	(2,16)	 [0.09]	(2,17)	 [0.10]	(2,18)	 [0.07]	(2,19)	 [0.04]																				
(3,0)	 [0.01]	(3,1)	 [0.03]	(3,2)	 [0]	(3,3)	 [0.07]	(3,4)	 [0.04]	(3,5)	 [0.07]	(3,6)	 [0.03]	(3,7)	 [0]	(3,8)	 [0.03]	(3,9)	 [0.01]	(3,10)	 [0.01]	(3,11)	 [0]	(3,12)	 [0]	(3,13)	 [0.01]	(3,14)	 [0.01]	(3,15)	 [0.01]	(3,16)	 [0.02]	(3,17)	 [0.05]	(3,18)	 [0.07]	(3,19)	 [0.03]						
(4,0)	 [0.12]	(4,1)	 [0.02]	(4,2)	 [0]	(4,3)	 [0.01]	(4,4)	 [0.18]	(4,5)	 [0.06]	(4,6)	 [0.05]	(4,7)	 [0]	(4,8)	 [0.06]	(4,9)	 [0.01]	(4,10)	 [0.05]	(4,11)	 [0]	(4,12)	 [0]	(4,13)	 [0.02]	(4,14)	 [0.01]	(4,15)	 [0.01]	(4,16)	 [0.01]	(4,17)	 [0.03]	(4,18)	 [0.03]	(4,19)	 [0]						
(5,0)	 [0.05]	(5,1)	 [0.03]	(5,2)	 [0.01]	(5,3)	 [0.04]	(5,4)	 [0.04]	(5,5)	 [0.13]	(5,6)	 [0.07]	(5,7)	 [0.01]	(5,8)	 [0.01]	(5,9)	 [0.01]	(5,10)	 [0.01]	(5,11)	 [0.01]	(5,12)	 [0.02]	(5,13)	 [0.0]	(5,14)	 [0]	(5,15)	 [0.02]	(5,16)	 [0.01]	(5,17)	 [0.02]	(5,18)	 [0.14]	(5,19)	 [0.01]						
(6,0)	 [0.08]	(6,1)	 [0.01]	(6,2)	 [0.01]	(6,3)	 [0.08]	(6,4)	 [0.09]	(6,5)	 [0.08]	(6,6)	 [0.16]	(6,7)	 [0.07]	(6,8)	 [0.05]	(6,9)	 [0.06]	(6,10)	 [0.03]	(6,11)	 [0.02]	(6,12)	 [0.02]	(6,13)	 [0.01]	(6,14)	 [0.03]	(6,15)	 [0.07]	(6,16)	 [0.02]	(6,17)	 [0.04]	(6,18)	 [0.02]	(6,19)	 [0.06]						
(7,0)	 [0.04]	(7,1)	 [0.04]	(7,2)	 [0.04]	(7,3)	 [0.04]	(7,4)	 [0.03]	(7,5)	 [0.07]	(7,6)	 [0.01]	(7,7)	 [0.57]	(7,8)	 [0.02]	(7,9)	 [0.05]	(7,10)	 [0]	(7,11)	 [0.01]	(7,12)	 [0.01]	(7,13)	 [0]	(7,14)	 [0]	(7,15)	 [0.02]	(7,16)	 [0.03]	(7,17)	 [0.02]	(7,18)	 [0.02]	(7,19)	 [0.03]						
(8,0)	 [0.01]	(8,1)	 [0.03]	(8,2)	 [0.04]	(8,3)	 [0.04]	(8,4)	 [0.02]	(8,5)	 [0.08]	(8,6)	 [0.03]	(8,7)	 [0.02]	(8,8)	 [0.08]	(8,9)	 [0.02]	(8,10)	 [0]	(8,11)	 [0]	(8,12)	 [0.01]	(8,13)	 [0]	(8,14)	 [0]	(8,15)	 [0]	(8,16)	 [0.01]	(8,17)	 [0.01]	(8,18)	 [0.07]	(8,19)	 [0]																	
(9,0)	 [0.07]	(9,1)	 [0.20]	(9,2)	 [0.20]	(9,3)	 [0.26]	(9,4)	 [0.03]	(9,5)	 [0.01]	(9,6)	 [0.15]	(9,7)	 [0.17]	(9,8)	 [0.26]	(9,9)	 [0.61]	(9,10)	 [0.03]	(9,11)	 [0.14]	(9,12)	 [0.17]	(9,13)	 [0.15]	(9,14)	 [0.05]	(9,15)	 [0.12]	(9,16)	 [0.21]	(9,17)	 [0.06]	(9,18)	 [0]	(9,19)	 [0.08]						
(10,0)	 [0.13]	(10,1)	 [0.02]	(10,2)	 [0.05]	(10,3)	 [0.08]	(10,4)	 [0.06]	(10,5)	 [0.02]	(10,6)	 [0.05]	(10,7)	 [0.03]	(10,8)	 [0.10]	(10,9)	 [0.03]	(10,10)	 [0.29]	(10,11)	 [0.15]	(10,12)	 [0.13]	(10,13)	 [0.02]	(10,14)	 [0.24]	(10,15)	 [0.07]	(10,16)	 [0.10]	(10,17)	 [0.13]	(10,18)	 [0]	(10,19)	 [0.14]						
(11,0)	 [0.01]	(11,1)	 [0.01]	(11,2)	 [0.02]	(11,3)	 [0]	(11,4)	 [0.02]	(11,5)	 [0.02]	(11,6)	 [0.02]	(11,7)	 [0]	(11,8)	 [0.01]	(11,9)	 [0.02]	(11,10)	 [0.16]	(11,11)	 [0.23]	(11,12)	 [0.03]	(11,13)	 [0.04]	(11,14)	 [0.02]	(11,15)	 [0.03]	(11,16)	 [0.03]	(11,17)	 [0.01]	(11,18)	 [0.02]	(11,19)	 [0.04]																			
(12,0)	 [0.04]	(12,1)	 [0.04]	(12,2)	 [0.07]	(12,3)	 [0.05]	(12,4)	 [0.05]	(12,5)	 [0.03]	(12,6)	 [0.05]	(12,7)	 [0]	(12,8)	 [0.06]	(12,9)	 [0.02]	(12,10)	 [0.16]	(12,11)	 [0.08]	(12,12)	 [0.17]	(12,13)	 [0.07]	(12,14)	 [0.05]	(12,15)	 [0.03]	(12,16)	 [0.07]	(12,17)	 [0.10]	(12,18)	 [0.06]	(12,19)	 [0.03]							
(13,0)	 [0]	(13,1)	 [0.05]	(13,2)	 [0.06]	(13,3)	 [0.02]	(13,4)	 [0.03]	(13,5)	 [0.02]	(13,6)	 [0.05]	(13,7)	 [0.04]	(13,8)	 [0.03]	(13,9)	 [0]	(13,10)	 [0.02]	(13,11)	 [0.06]	(13,12)	 [0.01]	(13,13)	 [0.32]	(13,14)	 [0]	(13,15)	 [0.09]	(13,16)	 [0.06]	(13,17)	 [0.04]	(13,18)	 [0]	(13,19)	 [0.01]																				
(14,0)	 [0.08]	(14,1)	 [0.01]	(14,2)	 [0.02]	(14,3)	 [0.02]	(14,4)	 [0.02]	(14,5)	 [0.04]	(14,6)	 [0.04]	(14,7)	 [0]	(14,8)	 [0.04]	(14,9)	 [0.02]	(14,10)	 [0.03]	(14,11)	 [0.09]	(14,12)	 [0.06]	(14,13)	 [0.02]	(14,14)	 [0.35]	(14,15)	 [0.04]	(14,16)	 [0.02]	(14,17)	 [0.07]	(14,18)	 [0.04]	(14,19)	 [0.10]						

(15,0)	 [0.03]	(15,1)	 [0]	(15,2)	 [0.01]	(15,3)	 [0.01]	(15,4)	 [0.02]	(15,5)	 [0.03]	(15,6)	 [0]	(15,7)	 [0]	(15,8)	 [0.01]	(15,9)	 [0.01]	(15,10)	 [0.05]	(15,11)	 [0.01]	(15,12)	 [0.02]	(15,13)	 [0.04]	(15,14)	 [0.01]	(15,15)	 [00.19]	(15,16)	 [0.02]	(15,17)	 [0.01]	(15,18)	 [0.04]	(15,19)	 [0.05]	

(16,0)	 [0.09]	(16,1)	 [0]	(16,2)	 [0.01]	(16,3)	 [0.01]	(16,4)	 [0.05]	(16,5)	 [0.05]	(16,6)	 [0.04]	(16,7)	 [0.01]	(16,8)	 [0.02]	(16,9)	 [0.01]	(16,10)	 [0]	(16,11)	 [0.01]	(16,12)	 [0.02]	(16,13)	 [0.01]	(16,14)	 [0.06]	(16,15)	 [0.01]	(16,16)	 [0.07]	(16,17)	 [0.06]	(16,18)	 [0.08]	(16,19)	 [0.01]

(17,0)	 [0.03]	(17,1)	 [0.19]	(17,2)	 [0.22]	(17,3)	 [0.06]	(17,4)	 [0.02]	(17,5)	 [0.11]	(17,6)	 [0.14]	(17,7)	 [0.03]	(17,8)	 [0.07]	(17,9)	 [0.06]	(17,10)	 [0.04]	(17,11)	 [0.08]	(17,12)	 [0.15]	(17,13)	 [0.22]	(17,14)	 [0]	(17,15)	 [0.07]	(17,16)	 [0.11]	(17,17)	 [0.13]	(17,18)	 [0.03]	(17,19)	 [0.04]

(18,0)	 [0.04]	(18,1)	 [0]	(18,2)	 [0]	(18,3)	 [0.02]	(18,4)	 [0.05]	(18,5)	 [0.05]	(18,6)	 [0]	(18,7)	 [0]	(18,8)	 [0]	(18,9)	 [0]	(18,10)	 [0]	(18,11)	 [0.02]	(18,12)	 [0]	(18,13)	 [0.01]	(18,14)	 [0.06]	(18,15)	 [0]	(18,16)	 [0]	(18,17)	 [0.02]	(18,18)	 [0.15]	(18,19)	 [0.02]

(19,0)	 [0.04]	(19,1)	 [0.01]	(19,2)	 [0.02]	(19,3)	 [0.06]	(19,4)	 [0.05]	(19,5)	 [0.04]	(19,6)	 [0]	(19,7)	 [0]	(19,8)	 [0.06]	(19,9)	 [0.01]	(19,10)	 [0.08]	(19,11)	 [0.03]	(19,12)	 [0.04]	(19,13)	 [0.02]	(19,14)	 [0.06]	(19,15)	 [0.09]	(19,16)	 [0.06]	(19,17)	 [0.02]	(19,18)	 [0.04]	(19,19)	 [0.23]

};
\end{axis}
\end{tikzpicture}

%% file: Figures/CM_lora_5_withot_cgan.tex
\begin{tikzpicture}
\pgfplotsset{every tick label/.append style={font=\tiny}}

\begin{axis}[
enlargelimits=false,
colorbar,
colormap/Purples,
width=\fwidth,
height=\fheight,
at={(0\fwidth,0\fheight)},
scale only axis,
tick align=inside,
ylabel={Predicted Radio},
xmin=-0.5,
xmax=4.5,
xtick style={color=black},
xlabel style={font=\scriptsize\color{white!15!black}},
ylabel style={font=\scriptsize\color{white!15!black}},
xlabel={True Radio},
ymin=-0.5,
ymax=4.5,
xlabel shift=-5pt,
ylabel shift=-5pt,
ytick style={color=black},
axis background/.style={fill=white},
colorbar horizontal,
colorbar style={
at={(0,1.05)},               % <-- (changed)
anchor=below south west,    % <-- (changed)
% change the width of the colorbar relative to the main `axis' environment
width=\pgfkeysvalueof{/pgfplots/parent axis width},
xtick={0, 0.5, 1},
xmin=0,
xmax=1,
axis x line*=top,
xticklabel shift=2pt,
},
colorbar/width=2mm,
]
\addplot [matrix plot,point meta=explicit]
 coordinates {

% Original
% [0.42333333 0.52111111 0.05555556 0.         0.        ]
%  [0.         0.85222222 0.         0.14777778 0.        ]
%  [0.10555556 0.11666667 0.77333333 0.         0.00444444]
%  [0.         0.12666667 0.20777778 0.         0.66555556]
%  [0.26555556 0.21       0.51555556 0.00555556 0.00333333]]
% Overall Accuracy:  0.41044444444444445

% (0,0)	[0.4]	(0,1)	[0.5]	(0,2)	[0.055]	(0,3)	[0]	(0,4)	[0]

% (1,0)	[0]	(1,1)	[0.85]	(1,2)	[0]	(1,3)	[0.15]	(1,4)	[0]

% (2,0)	[0.1]	(2,1)	[0.12]	(2,2)	[0.77]	(2,3)	[0]	(2,4)	[0]

% (3,0)	[0]	(3,1)	[0.13]	(3,2)	[0.2]	(3,3)	[0]	(3,4)	[0.67]

% (4,0)	[0.27]	(4,1)	[0.21]	(4,2)	[0.52]	(4,3)	[0]	(4,4)	[0]

(0,0)	[0.90]	(0,1)	[0]	(0,2)	[0.06]	(0,3)	[0.13]	(0,4)	[0]

(1,0)	[0]	(1,1)	[0]	(1,2)	[0]	(1,3)	[0.87]	(1,4)	[0]

(2,0)	[0]	(2,1)	[0]	(2,2)	[0]	(2,3)	[0]	(2,4)	[1]

(3,0)	[0]	(3,1)	[1]	(3,2)	[0.93]	(3,3)	[0]	(3,4)	[0]

(4,0)	[0.10]	(4,1)	[0]	(4,2)	[0]	(4,3)	[0]	(4,4)	[0]

};
\end{axis}
\end{tikzpicture}

%% file: Figures/CM_lora_5_with_cgan.tex
\begin{tikzpicture}
\pgfplotsset{every tick label/.append style={font=\tiny}}

\begin{axis}[
enlargelimits=false,
colorbar,
colormap/Purples,
width=\fwidth,
height=\fheight,
at={(0\fwidth,0\fheight)},
scale only axis,
tick align=inside,
ylabel={Predicted Radio},
xmin=-0.5,
xmax=4.5,
xtick style={color=black},
xlabel style={font=\scriptsize\color{white!15!black}},
ylabel style={font=\scriptsize\color{white!15!black}},
xlabel={True Radio},
ymin=-0.5,
ymax=4.5,
xlabel shift=-5pt,
ylabel shift=-5pt,
ytick style={color=black},
axis background/.style={fill=white},
colorbar horizontal,
colorbar style={
at={(0,1.05)},               % <-- (changed)
anchor=below south west,    % <-- (changed)
% change the width of the colorbar relative to the main `axis' environment
width=\pgfkeysvalueof{/pgfplots/parent axis width},
xtick={0, 0.5, 1},
xmin=0,
xmax=1,
axis x line*=top,
xticklabel shift=2pt,
},
colorbar/width=2mm,
]
\addplot [matrix plot,point meta=explicit]
 coordinates {

%original

% (0,0) [0.99] (0,1)	 [0]	(0,2)	 [0.01]	(0,3)	 [0]	(0,4)	 [0]

% (1,0)	[0]	(1,1)	[0.85]	(1,2)	[0]	(1,3)	[0.15]	(1,4)	[0]

% (2,0)	[0]	(2,1)	[0]	(2,2)	[1]	(2,3)	[0]	(2,4)	[0]

% (3,0)	[0]	(3,1)	[0.13]	(3,2)	[0.2]	(3,3)	[0]	(3,4)	[0.67]

% (4,0)	 [0.02]	(4,1)	 [0.03]	(4,2)	 [0.07]	(4,3)	 [0]	(4,4)	 [0.89]

(0,0)	 [1]	(0,1)	 [0]	(0,2)	 [0]	(0,3)	 [0]	(0,4)	 [0]

(1,0)	 [0]	(1,1)	 [1]	(1,2)	 [0]	(1,3)	 [0]	(1,4)	 [0]

(2,0)	 [0]	(2,1)	 [0]	(2,2)	 [0]	(2,3)	 [0]	(2,4)	 [0]

(3,0)	 [0]	(3,1)	 [0]	(3,2)	 [0.93]	(3,3)	 [1]	(3,4)	 [0]

(4,0)	 [0]	(4,1)	 [0]	(4,2)	 [0]	(4,3)	 [0]	(4,4)	 [1]

};
\end{axis}
\end{tikzpicture}

%% file: Figures/payload_acc_lora.tex
\usetikzlibrary{patterns}
\pgfplotsset{width=10cm, height=5cm}

\pgfplotsset{
    compat=1.3,
    % define your own legend style here
    my ybar legend/.style={
        legend image code/.code={
            \draw [##1] (0cm,-0.6ex) rectangle +(3em,1.5ex);
        },
    },
}

\pgfplotscreateplotcyclelist{line_shape}{%
solid, every mark/.append style={solid, fill=blue}, mark=diamond*\\%
dotted, every mark/.append style={solid,  fill=gray}, mark=square*\\%
densely dotted, every mark/.append style={solid, fill=gray}, mark=otimes*\\%
loosely dotted, every mark/.append style={solid, fill=gray}, mark=triangle*\\%
dashed, every mark/.append style={solid, fill=gray},mark=halfdiamond*\\%
loosely dashed, every mark/.append style={solid, fill=gray},mark=*\\%
densely dashed, every mark/.append style={solid, fill=gray},mark=square*\\%
dashdotted, every mark/.append style={solid, fill=gray},mark=otimes*\\%
dashdotdotted, every mark/.append style={solid},mark=star\\%
densely dashdotted,every mark/.append style={solid, fill=gray},mark=diamond*\\%
}

\begin{tikzpicture}
% \selectcolormodel{gray}
\begin{axis}[
cycle list name = line_shape,
ylabel style={align=center}, 
ylabel={Payload TTOD},
xlabel={Epochs},
xticklabels={ ,10, , 20, ,30,, 40, ,50, ,60, ,70, ,80, ,90,,100},
legend style={at={(0.7,0.3)},
anchor=north,legend columns=2},
xtick=data,
mark size=2pt,
xmin =0,
xmax = 100,
width=8cm,
grid=major,
]
\addplot coordinates {(5, 0.25 ) (10, 0.27 )  (15, 0.30 ) (20, 0.30 ) (25, 0.39 )  (30, 0.38 ) (35, 0.43 ) (40, 0.46 )  (45, 0.42) (50, 0.46) (55, 0.44)  (60, 0.44) (65, 0.45) (70, 0.46 )  (75, 0.47 ) (80, 0.47 ) (85, 0.47 )  (90, 0.48 ) (95, 0.52 ) (100, 0.51)};
%(300, 60) (Max Rule, 71)
%\addplot coordinates {(5, 13 ) (10,13 )  (15, 13 ) (20, 13 ) (25, 13 )  (30, 13 ) (35, 13 ) (40,13 )  (45, 13) (50, 13) (55, 13)  (60, 13) (65, 13) (70,13 )  (75, 13 ) (80, 13 ) (85, 13 )  (90, 13 ) (95, 13 ) (100, 13) }; % (300, 13) (Max Rule, 13)
%\addplot coordinates {(5, 22 ) (10,22 )  (15, 22 ) (20, 22 ) (25, 22 )  (30, 22 ) (35, 22 ) (40,22 )  (45, 22) (50, 22) (55, 22)  (60, 22) (65, 22) (70,22)  (75, 22 ) (80, 22 ) (85, 22 )  (90, 22 ) (95, 22 ) (100, 22)  }; %(300, 22) (Max Rule, 22)

% \legend{$RFCycleGAN$}
\end{axis}
\end{tikzpicture}

%% file: Figures/CM_lora_20_orig.tex
\begin{tikzpicture}
\pgfplotsset{every tick label/.append style={font=\tiny}}

\begin{axis}[
enlargelimits=false,
colorbar,
colormap/Purples,
width=\fwidth,
height=\fheight,
at={(0\fwidth,0\fheight)},
scale only axis,
tick align=inside,
ylabel={Predicted Radio},
xmin=-0.5,
xmax=19.5,
xtick style={color=black},
xlabel style={font=\scriptsize\color{white!15!black}},
ylabel style={font=\scriptsize\color{white!15!black}},
xlabel={True Radio},
ymin=-0.5,
ymax=19.5,
xlabel shift=-5pt,
ylabel shift=-5pt,
ytick style={color=black},
axis background/.style={fill=white},
colorbar horizontal,
colorbar style={
at={(0,1.05)},               % <-- (changed)
anchor=below south west,    % <-- (changed)
% change the width of the colorbar relative to the main `axis' environment
width=\pgfkeysvalueof{/pgfplots/parent axis width},
xtick={0, 0.5, 1},
xmin=0,
xmax=1,
axis x line*=top,
xticklabel shift=2pt,
},
colorbar/width=2mm,
]
\addplot [matrix plot,point meta=explicit]
 coordinates {

(0,0) [0.03]	(0,1) [0.25]	(0,2) [0]	(0,3) [0]	(0,4) [0.06]	(0,5) [0]	(0,6) [0]	(0,7) [0]	(0,8) [0]	(0,9) [0]	(0,10) [0]	(0,11) [0]	(0,12) [0]	(0,13) [0]	(0,14) [0]	(0,15) [0]	(0,16) [0]	(0,17) [0.25]	(0,18) [0]	(0,19) [0.40]	

(1,0) [0.25]	(1,1) [0]	(1,2) [0]	(1,3) [0]	(1,4) [0]	(1,5) [0.01]	(1,6) [0]	(1,7) [0]	(1,8) [0.01]	(1,9) [0]	(1,10) [0]	(1,11) [0]	(1,12) [0]	(1,13) [0]	(1,14) [0.11]	(1,15) [0]	(1,16) [0]	(1,17) [0]	(1,18) [0.01]	(1,19) [0.61]	

(2,0) [0.70]	(2,1) [0]	(2,2) [0.29]	(2,3) [0]	(2,4) [0]	(2,5) [0]	(2,6) [0]	(2,7) [0]	(2,8) [0]	(2,9) [0]	(2,10) [0]	(2,11) [0]	(2,12) [0.01]	(2,13) [0]	(2,14) [0]	(2,15) [0]	(2,16) [0]	(2,17) [0]	(2,18) [0]	(2,19) [0]

(3,0) [0]	(3,1) [0]	(3,2) [0.01]	(3,3) [0]	(3,4) [0]	(3,5) [0]	(3,6) [0]	(3,7) [0.35]	(3,8) [0]	(3,9) [0]	(3,10) [0]	(3,11) [0.64]	(3,12) [0]	(3,13) [0]	(3,14) [0]	(3,15) [0]	(3,16) [0]	(3,17) [0]	(3,18) [0]	(3,19) [0]	

(4,0) [0]	(4,1) [0]	(4,2) [0]	(4,3) [0]	(4,4) [0]	(4,5) [0]	(4,6) [0]	(4,7) [0.24]	(4,8) [0]	(4,9) [0]	(4,10) [0.04]	(4,11) [0]	(4,12) [0.72]	(4,13) [0]	(4,14) [0]	(4,15) [0]	(4,16) [0]	(4,17) [0]	(4,18) [0]	(4,19) [0]	

(5,0) [0]	(5,1) [0.30]	(5,2) [0]	(5,3) [0]	(5,4) [0]	(5,5) [0]	(5,6) [0]	(5,7) [0]	(5,8) [0.24]	(5,9) [0]	(5,10) [0]	(5,11) [0]	(5,12) [0]	(5,13) [0]	(5,14) [0]	(5,15) [0.28]	(5,16) [0.19]	(5,17) [0]	(5,18) [0]	(5,19) [0]	

(6,0) [0]	(6,1) [0]	(6,2) [0]	(6,3) [0]	(6,4) [0]	(6,5) [0]	(6,6) [1]	(6,7) [0]	(6,8) [0]	(6,9) [0]	(6,10) [0]	(6,11) [0]	(6,12) [0]	(6,13) [0]	(6,14) [0]	(6,15) [0]	(6,16) [0]	(6,17) [0]	(6,18) [0]	(6,19) [0]	

(7,0) [0]	(7,1) [0]	(7,2) [0]	(7,3) [0]	(7,4) [0]	(7,5) [0]	(7,6) [0]	(7,7) [0]	(7,8) [0]	(7,9) [0]	(7,10) [1]	(7,11) [0]	(7,12) [0]	(7,13) [0]	(7,14) [0]	(7,15) [0]	(7,16) [0]	(7,17) [0]	(7,18) [0]	(7,19) [0]	

(8,0) [0]	(8,1) [1]	(8,2) [0]	(8,3) [0]	(8,4) [0]	(8,5) [0]	(8,6) [0]	(8,7) [0]	(8,8) [0]	(8,9) [0]	(8,10) [0]	(8,11) [0]	(8,12) [0]	(8,13) [0]	(8,14) [0]	(8,15) [0]	(8,16) [0]	(8,17) [0]	(8,18) [0]	(8,19) [0]	

(9,0) [0]	(9,1) [0]	(9,2) [0]	(9,3) [0]	(9,4) [0]	(9,5) [0]	(9,6) [0]	(9,7) [0]	(9,8) [0]	(9,9) [0]	(9,10) [0]	(9,11) [0]	(9,12) [0]	(9,13) [0]	(9,14) [0]	(9,15) [0]	(9,16) [0]	(9,17) [0]	(9,18) [0]	(9,19) [1]

(10,0) [0.03]	(10,1) [0]	(10,2) [0]	(10,3) [0]	(10,4) [0]	(10,5) [0]	(10,6) [0]	(10,7) [0]	(10,8) [0]	(10,9) [0.97]	(10,10) [0]	(10,11) [0]	(10,12) [0]	(10,13) [0]	(10,14) [0]	(10,15) [0]	(10,16) [0]	(10,17) [0]	(10,18) [0]	(10,19) [0]	

(11,0) [0]	(11,1) [0]	(11,2) [0]	(11,3) [0]	(11,4) [0]	(11,5) [0]	(11,6) [0]	(11,7) [0]	(11,8) [0.37]	(11,9) [0]	(11,10) [0]	(11,11) [0]	(11,12) [0]	(11,13) [0]	(11,14) [0]	(11,15) [0.51]	(11,16) [0.01]	(11,17) [0]	(11,18) [0]	(11,19) [0.11]	

(12,0) [0]	(12,1) [0]	(12,2) [0]	(12,3) [0]	(12,4) [0]	(12,5) [0]	(12,6) [0]	(12,7) [0]	(12,8) [0.02]	(12,9) [0]	(12,10) [0.52]	(12,11) [0]	(12,12) [0]	(12,13) [0.40]	(12,14) [0]	(12,15) [0]	(12,16) [0.06]	(12,17) [0]	(12,18) [0]	(12,19) [0]	

(13,0) [0]	(13,1) [0]	(13,2) [0]	(13,3) [0]	(13,4) [0]	(13,5) [0]	(13,6) [0]	(13,7) [0]	(13,8) [0]	(13,9) [1]	(13,10) [0]	(13,11) [0]	(13,12) [0]	(13,13) [0]	(13,14) [0]	(13,15) [0]	(13,16) [0]	(13,17) [0]	(13,18) [0]	(13,19) [0]	

(14,0) [0.03]	(14,1) [0]	(14,2) [0]	(14,3) [0]	(14,4) [0]	(14,5) [0.02]	(14,6) [0.58]	(14,7) [0]	(14,8) [0]	(14,9) [0]	(14,10) [0]	(14,11) [0]	(14,12) [0]	(14,13) [0]	(14,14) [0.05]	(14,15) [0]	(14,16) [0]	(14,17) [0]	(14,18) [0]	(14,19) [0.31]

(15,0) [0]	(15,1) [0]	(15,2) [0]	(15,3) [0]	(15,4) [0]	(15,5) [0]	(15,6) [0.05]	(15,7) [0]	(15,8) [0]	(15,9) [0]	(15,10) [0]	(15,11) [0]	(15,12) [0]	(15,13) [0]	(15,14) [0]	(15,15) [0.52]	(15,16) [0]	(15,17) [0]	(15,18) [0.08]	(15,19) [0.03]	

(16,0) [0]	(16,1) [1]	(16,2) [0]	(16,3) [0]	(16,4) [0]	(16,5) [0]	(16,6) [0]	(16,7) [0]	(16,8) [0]	(16,9) [0]	(16,10) [0]	(16,11) [0]	(16,12) [0]	(16,13) [0]	(16,14) [0]	(16,15) [0]	(16,16) [0]	(16,17) [0]	(16,18) [0]	(16,19) [0]	

(17,0) [0.94]	(17,1) [0.05]	(17,2) [0]	(17,3) [0]	(17,4) [0]	(17,5) [0]	(17,6) [0]	(17,7) [0]	(17,8) [0]	(17,9) [0.02]	(17,10) [0]	(17,11) [0]	(17,12) [0]	(17,13) [0]	(17,14) [0]	(17,15) [0]	(17,16) [0]	(17,17) [0]	(17,18) [0]	(17,19) [0]	

(18,0) [0]	(18,1) [0]	(18,2) [1]	(18,3) [0]	(18,4) [0]	(18,5) [0]	(18,6) [0]	(18,7) [0]	(18,8) [0]	(18,9) [0]	(18,10) [0]	(18,11) [0]	(18,12) [0]	(18,13) [0]	(18,14) [0]	(18,15) [0]	(18,16) [0]	(18,17) [0]	(18,18) [0]	(18,19) [0]	

(19,0) [0.07]	(19,1) [0]	(19,2) [0]	(19,3) [0]	(19,4) [0.12]	(19,5) [0]	(19,6) [0.02]	(19,7) [0.09]	(19,8) [0]	(19,9) [0]	(19,10) [0]	(19,11) [0.02]	(19,12) [0]	(19,13) [0]	(19,14) [0]	(19,15) [0]	(19,16) [0]	(19,17) [0]	(19,18) [0]	(19,19) [0.60]

};
\end{axis}
\end{tikzpicture}

%% file: Figures/CM_lora_20_deeplora_pa.tex
\begin{tikzpicture}
\pgfplotsset{every tick label/.append style={font=\tiny}}

\begin{axis}[
enlargelimits=false,
colorbar,
colormap/Purples,
width=\fwidth,
height=\fheight,
at={(0\fwidth,0\fheight)},
scale only axis,
tick align=inside,
ylabel={Predicted Radio},
xmin=-0.5,
xmax=19.5,
xtick style={color=black},
xlabel style={font=\scriptsize\color{white!15!black}},
ylabel style={font=\scriptsize\color{white!15!black}},
xlabel={True Radio},
ymin=-0.5,
ymax=19.5,
xlabel shift=-5pt,
ylabel shift=-5pt,
ytick style={color=black},
axis background/.style={fill=white},
colorbar horizontal,
colorbar style={
at={(0,1.05)},               % <-- (changed)
anchor=below south west,    % <-- (changed)
% change the width of the colorbar relative to the main `axis' environment
width=\pgfkeysvalueof{/pgfplots/parent axis width},
xtick={0, 0.5, 1},
xmin=0,
xmax=1,
axis x line*=top,
xticklabel shift=2pt,
},
colorbar/width=2mm,
]
\addplot [matrix plot,point meta=explicit]
 coordinates {

(0,0) [0.01]	(0,1) [0.12]	(0,2) [0]	(0,3) [0]	(0,4) [0.30]	(0,5) [0]	(0,6) [0]	(0,7) [0.01]	(0,8) [0]	(0,9) [0]	(0,10) [0]	(0,11) [0]	(0,12) [0]	(0,13) [0]	(0,14) [0]	(0,15) [0]	(0,16) [0]	(0,17) [0.16]	(0,18) [0]	(0,19) [0.39]	

(1,0) [0.33]	(1,1) [0]	(1,2) [0]	(1,3) [0]	(1,4) [0]	(1,5) [0]	(1,6) [0]	(1,7) [0]	(1,8) [0]	(1,9) [0.14]	(1,10) [0]	(1,11) [0]	(1,12) [0]	(1,13) [0]	(1,14) [0.38]	(1,15) [0]	(1,16) [0]	(1,17) [0]	(1,18) [0]	(1,19) [0.15]	

(2,0) [0.01]	(2,1) [0.05]	(2,2) [0.89]	(2,3) [0]	(2,4) [0]	(2,5) [0]	(2,6) [0.03]	(2,7) [0]	(2,8) [0]	(2,9) [0]	(2,10) [0]	(2,11) [0]	(2,12) [0.02]	(2,13) [0]	(2,14) [0]	(2,15) [0]	(2,16) [0]	(2,17) [0]	(2,18) [0]	(2,19) [0]	

(3,0) [0]	(3,1) [0]	(3,2) [0.17]	(3,3) [0]	(3,4) [0]	(3,5) [0]	(3,6) [0]	(3,7) [0]	(3,8) [0]	(3,9) [0]	(3,10) [0]	(3,11) [0.83]	(3,12) [0]	(3,13) [0]	(3,14) [0]	(3,15) [0]	(3,16) [0]	(3,17) [0]	(3,18) [0]	(3,19) [0]

(4,0) [0]	(4,1) [0]	(4,2) [0]	(4,3) [0]	(4,4) [0.100]	(4,5) [0]	(4,6) [0]	(4,7) [0]	(4,8) [0]	(4,9) [0]	(4,10) [0]	(4,11) [0]	(4,12) [0]	(4,13) [0]	(4,14) [0]	(4,15) [0]	(4,16) [0]	(4,17) [0]	(4,18) [0]	(4,19) [0]	

(5,0) [0]	(5,1) [0.18]	(5,2) [0]	(5,3) [0]	(5,4) [0.01]	(5,5) [0]	(5,6) [0]	(5,7) [0]	(5,8) [0.27]	(5,9) [0]	(5,10) [0]	(5,11) [0]	(5,12) [0]	(5,13) [0]	(5,14) [0]	(5,15) [0.13]	(5,16) [0.41]	(5,17) [0]	(5,18) [0]	(5,19) [0]	

(6,0) [0]	(6,1) [0]	(6,2) [0.28]	(6,3) [0]	(6,4) [0]	(6,5) [0]	(6,6) [0.72]	(6,7) [0]	(6,8) [0]	(6,9) [0]	(6,10) [0]	(6,11) [0]	(6,12) [0]	(6,13) [0]	(6,14) [0]	(6,15) [0]	(6,16) [0]	(6,17) [0]	(6,18) [0]	(6,19) [0]

(7,0) [0]	(7,1) [0]	(7,2) [0]	(7,3) [0]	(7,4) [0.12]	(7,5) [0]	(7,6) [0]	(7,7) [0.48]	(7,8) [0]	(7,9) [0]	(7,10) [0.41]	(7,11) [0]	(7,12) [0]	(7,13) [0]	(7,14) [0]	(7,15) [0]	(7,16) [0]	(7,17) [0]	(7,18) [0]	(7,19) [0]	

(8,0) [0]	(8,1) [0.99]	(8,2) [0]	(8,3) [0]	(8,4) [0]	(8,5) [0]	(8,6) [0]	(8,7) [0]	(8,8) [0]	(8,9) [0]	(8,10) [0]	(8,11) [0]	(8,12) [0]	(8,13) [0]	(8,14) [0]	(8,15) [0]	(8,16) [0]	(8,17) [0]	(8,18) [0]	(8,19) [0]	

(9,0) [0]	(9,1) [0.07]	(9,2) [0]	(9,3) [0]	(9,4) [0]	(9,5) [0]	(9,6) [0]	(9,7) [0]	(9,8) [0]	(9,9) [0.01]	(9,10) [0]	(9,11) [0]	(9,12) [0]	(9,13) [0]	(9,14) [0]	(9,15) [0]	(9,16) [0]	(9,17) [0]	(9,18) [0]	(9,19) [0.92]

(10,0) [0]	(10,1) [0]	(10,2) [0]	(10,3) [0]	(10,4) [0]	(10,5) [0]	(10,6) [0]	(10,7) [0.02]	(10,8) [0]	(10,9) [0.98]	(10,10) [0]	(10,11) [0]	(10,12) [0]	(10,13) [0]	(10,14) [0]	(10,15) [0]	(10,16) [0]	(10,17) [0]	(10,18) [0]	(10,19) [0]

(11,0) [0.03]	(11,1) [0.04]	(11,2) [0]	(11,3) [0]	(11,4) [0]	(11,5) [0]	(11,6) [0]	(11,7) [0]	(11,8) [0.43]	(11,9) [0]	(11,10) [0]	(11,11) [0]	(11,12) [0]	(11,13) [0]	(11,14) [0]	(11,15) [0.21]	(11,16) [0.16]	(11,17) [0]	(11,18) [0]	(11,19) [0.13]	

(12,0) [0]	(12,1) [0]	(12,2) [0]	(12,3) [0]	(12,4) [0]	(12,5) [0]	(12,6) [0]	(12,7) [0.16]	(12,8) [0]	(12,9) [0]	(12,10) [0.44]	(12,11) [0]	(12,12) [0]	(12,13) [0.37]	(12,14) [0]	(12,15) [0]	(12,16) [0]	(12,17) [0]	(12,18) [0]	(12,19) [0.02]	

(13,0) [0]	(13,1) [0]	(13,2) [0]	(13,3) [0]	(13,4) [0]	(13,5) [0]	(13,6) [0]	(13,7) [0]	(13,8) [0]	(13,9) [0.100]	(13,10) [0]	(13,11) [0]	(13,12) [0]	(13,13) [0]	(13,14) [0]	(13,15) [0]	(13,16) [0]	(13,17) [0]	(13,18) [0]	(13,19) [0]	

(14,0) [0.19]	(14,1) [0.08]	(14,2) [0]	(14,3) [0]	(14,4) [0]	(14,5) [0]	(14,6) [0.10]	(14,7) [0]	(14,8) [0]	(14,9) [0.07]	(14,10) [0]	(14,11) [0]	(14,12) [0]	(14,13) [0]	(14,14) [0.33]	(14,15) [0]	(14,16) [0]	(14,17) [0]	(14,18) [0]	(14,19) [0.21]	

(15,0) [0]	(15,1) [0.01]	(15,2) [0]	(15,3) [0]	(15,4) [0]	(15,5) [0.08]	(15,6) [0.04]	(15,7) [0]	(15,8) [0]	(15,9) [0]	(15,10) [0]	(15,11) [0]	(15,12) [0]	(15,13) [0]	(15,14) [0.04]	(15,15) [0.49]	(15,16) [0]	(15,17) [0]	(15,18) [0.16]	(15,19) [0.17]	

(16,0) [0]	(16,1) [0.43]	(16,2) [0]	(16,3) [0]	(16,4) [0.13]	(16,5) [0]	(16,6) [0]	(16,7) [0.40]	(16,8) [0]	(16,9) [0]	(16,10) [0]	(16,11) [0]	(16,12) [0]	(16,13) [0]	(16,14) [0]	(16,15) [0]	(16,16) [0]	(16,17) [0.02]	(16,18) [0]	(16,19) [0.02]	

(17,0) [0.88]	(17,1) [0]	(17,2) [0]	(17,3) [0]	(17,4) [0]	(17,5) [0]	(17,6) [0.01]	(17,7) [0.03]	(17,8) [0]	(17,9) [0.03]	(17,10) [0.01]	(17,11) [0]	(17,12) [0.02]	(17,13) [0]	(17,14) [0]	(17,15) [0]	(17,16) [0]	(17,17) [0.01]	(17,18) [0]	(17,19) [0]	

(18,0) [0]	(18,1) [0]	(18,2) [0.81]	(18,3) [0]	(18,4) [0]	(18,5) [0]	(18,6) [0]	(18,7) [0]	(18,8) [0]	(18,9) [0]	(18,10) [0]	(18,11) [0.18]	(18,12) [0]	(18,13) [0]	(18,14) [0]	(18,15) [0]	(18,16) [0]	(18,17) [0]	(18,18) [0]	(18,19) [0]	

(19,0) [0.10]	(19,1) [0.07]	(19,2) [0]	(19,3) [0]	(19,4) [0.03]	(19,5) [0]	(19,6) [0.01]	(19,7) [0.17]	(19,8) [0.01]	(19,9) [0.09]	(19,10) [0]	(19,11) [0]	(19,12) [0]	(19,13) [0]	(19,14) [0]	(19,15) [0]	(19,16) [0]	(19,17) [0]	(19,18) [0]	(19,19) [0.43]

};
\end{axis}
\end{tikzpicture}

%% file: Figures/CM_lora_20_with_cgan.tex
\begin{tikzpicture}
\pgfplotsset{every tick label/.append style={font=\tiny}}

\begin{axis}[
enlargelimits=false,
colorbar,
colormap/Purples,
width=\fwidth,
height=\fheight,
at={(0\fwidth,0\fheight)},
scale only axis,
tick align=inside,
ylabel={Predicted Radio},
xmin=-0.5,
xmax=19.5,
xtick style={color=black},
xlabel style={font=\scriptsize\color{white!15!black}},
ylabel style={font=\scriptsize\color{white!15!black}},
xlabel={True Radio},
ymin=-0.5,
ymax=19.5,
xlabel shift=-5pt,
ylabel shift=-5pt,
ytick style={color=black},
axis background/.style={fill=white},
colorbar horizontal,
colorbar style={
at={(0,1.05)},               % <-- (changed)
anchor=below south west,    % <-- (changed)
% change the width of the colorbar relative to the main `axis' environment
width=\pgfkeysvalueof{/pgfplots/parent axis width},
xtick={0, 0.5, 1},
xmin=0,
xmax=1,
axis x line*=top,
xticklabel shift=2pt,
},
colorbar/width=2mm,
]
\addplot [matrix plot,point meta=explicit]
 coordinates {

(0,0)	 [0.95]	(0,1)	 [0.03]	(0,2)	 [0.70]	(0,3)	 [0.06]	(0,4)	 [0.30]	(0,5)	 [0]	(0,6)	 [0]	(0,7)	 [0]	(0,8)	 [0]	(0,9)	 [0]	(0,10)	 [0]	(0,11)	 [0]	(0,12)	 [0.50]	(0,13)	 [0]	(0,14)	 [0.15]	(0,15)	 [0]	(0,16)	 [0]	(0,17)	 [0]	(0,18)	 [0]	(0,19)	 [0]

(1,0)	 [0]	(1,1)	 [0.51]	(1,2)	 [0]	(1,3)	 [0]	(1,4)	 [0]	(1,5)	 [0]	(1,6)	 [0]	(1,7)	 [0]	(1,8)	 [0]	(1,9)	 [0]	(1,10)	 [0]	(1,11)	 [0.21]	(1,12)	 [0.06]	(1,13)	 [0]	(1,14)	 [0]	(1,15)	 [0]	(1,16)	 [0]	(1,17)	 [0.05]	(1,18)	 [0]	(1,19)	 [0]

(2,0)	 [0]	(2,1)	 [0]	(2,2)	 [0.29]	(2,3)	 [0]	(2,4)	 [0]	(2,5)	 [0]	(2,6)	 [0]	(2,7)	 [0]	(2,8)	 [0]	(2,9)	 [0]	(2,10)	 [0]	(2,11)	 [0]	(2,12)	 [0]	(2,13)	 [0]	(2,14)	 [0]	(2,15)	 [0]	(2,16)	 [0]	(2,17)	 [0]	(2,18)	 [0]	(2,19)	 [0]	

(3,0)	 [0]	(3,1)	 [0]	(3,2)	 [0]	(3,3)	 [0.36]	(3,4)	 [0]	(3,5)	 [0]	(3,6)	 [0]	(3,7)	 [0]	(3,8)	 [0]	(3,9)	 [0]	(3,10)	 [0]	(3,11)	 [0]	(3,12)	 [0]	(3,13)	 [0]	(3,14)	 [0]	(3,15)	 [0]	(3,16)	 [0]	(3,17)	 [0]	(3,18)	 [0]	(3,19)	 [0]

(4,0)	 [0]	(4,1)	 [0.04]	(4,2)	 [0]	(4,3)	 [0]	(4,4)	 [0.23]	(4,5)	 [0]	(4,6)	 [0]	(4,7)	 [0]	(4,8)	 [0]	(4,9)	 [0]	(4,10)	 [0]	(4,11)	 [0]	(4,12)	 [0.01]	(4,13)	 [0]	(4,14)	 [0]	(4,15)	 [0]	(4,16)	 [0]	(4,17)	 [0.34]	(4,18)	 [0]	(4,19)	 [0]	

(5,0)	 [0]	(5,1)	 [0]	(5,2)	 [0]	(5,3)	 [0]	(5,4)	 [0]	(5,5)	 [0.93]	(5,6)	 [0]	(5,7)	 [0.02]	(5,8)	 [0]	(5,9)	 [0]	(5,10)	 [0]	(5,11)	 [0]	(5,12)	 [0]	(5,13)	 [0]	(5,14)	 [0]	(5,15)	 [0]	(5,16)	 [0]	(5,17)	 [0]	(5,18)	 [0.01]	(5,19)	 [0]	

(6,0)	 [0]	(6,1)	 [0]	(6,2)	 [0]	(6,3)	 [0]	(6,4)	 [0.03]	(6,5)	 [0]	(6,6)	 [1]	(6,7)	 [0]	(6,8)	 [0]	(6,9)	 [0]	(6,10)	 [0]	(6,11)	 [0]	(6,12)	 [0.05]	(6,13)	 [0]	(6,14)	 [0]	(6,15)	 [0]	(6,16)	 [0]	(6,17)	 [0]	(6,18)	 [0]	(6,19)	 [0]	

(7,0)	 [0]	(7,1)	 [0.05]	(7,2)	 [0]	(7,3)	 [0]	(7,4)	 [0.33]	(7,5)	 [0]	(7,6)	 [0]	(7,7)	 [0.97]	(7,8)	 [0]	(7,9)	 [0]	(7,10)	 [0]	(7,11)	 [0]	(7,12)	 [0.18]	(7,13)	 [0]	(7,14)	 [0]	(7,15)	 [0]	(7,16)	 [0]	(7,17)	 [0.08]	(7,18)	 [0]	(7,19)	 [0]	

(8,0)	 [0]	(8,1)	 [0.18]	(8,2)	 [0]	(8,3)	 [0]	(8,4)	 [0.02]	(8,5)	 [0]	(8,6)	 [0]	(8,7)	 [0]	(8,8)	 [0.99]	(8,9)	 [0]	(8,10)	 [0]	(8,11)	 [0]	(8,12)	 [0]	(8,13)	 [0]	(8,14)	 [0]	(8,15)	 [0]	(8,16)	 [0]	(8,17)	 [0]	(8,18)	 [0]	(8,19)	 [0]	

(9,0)	 [0]	(9,1)	 [0]	(9,2)	 [0]	(9,3)	 [0.58]	(9,4)	 [0]	(9,5)	 [0]	(9,6)	 [0]	(9,7)	 [0]	(9,8)	 [0]	(9,9)	 [1]	(9,10)	 [0]	(9,11)	 [0]	(9,12)	 [0.08]	(9,13)	 [0]	(9,14)	 [0]	(9,15)	 [0]	(9,16)	 [0]	(9,17)	 [0]	(9,18)	 [0]	(9,19)	 [0]	

(10,0)	 [0]	(10,1)	 [0]	(10,2)	 [0]	(10,3)	 [0]	(10,4)	 [0.04]	(10,5)	 [0]	(10,6)	 [0]	(10,7)	 [0.01]	(10,8)	 [0]	(10,9)	 [0]	(10,10)	 [1]	(10,11)	 [0]	(10,12)	 [0]	(10,13)	 [0]	(10,14)	 [0]	(10,15)	 [0]	(10,16)	 [0]	(10,17)	 [0]	(10,18)	 [0]	(10,19)	 [0]	

(11,0)	 [0]	(11,1)	 [0]	(11,2)	 [0]	(11,3)	 [0]	(11,4)	 [0]	(11,5)	 [0]	(11,6)	 [0]	(11,7)	 [0]	(11,8)	 [0]	(11,9)	 [0]	(11,10)	 [0]	(11,11)	 [0]	(11,12)	 [0]	(11,13)	 [0]	(11,14)	 [0]	(11,15)	 [0]	(11,16)	 [0]	(11,17)	 [0]	(11,18)	 [0]	(11,19)	 [0]	

(12,0)	 [0]	(12,1)	 [0]	(12,2)	 [0.01]	(12,3)	 [0]	(12,4)	 [0]	(12,5)	 [0]	(12,6)	 [0]	(12,7)	 [0]	(12,8)	 [0]	(12,9)	 [0]	(12,10)	 [0]	(12,11)	 [0]	(12,12)	 [0.05]	(12,13)	 [0]	(12,14)	 [0]	(12,15)	 [0]	(12,16)	 [0]	(12,17)	 [0]	(12,18)	 [0]	(12,19)	 [0]	

(13,0)	 [0]	(13,1)	 [0]	(13,2)	 [0]	(13,3)	 [0]	(13,4)	 [0]	(13,5)	 [0]	(13,6)	 [0]	(13,7)	 [0.01]	(13,8)	 [0]	(13,9)	 [0]	(13,10)	 [0]	(13,11)	 [0]	(13,12)	 [0]	(13,13)	 [1]	(13,14)	 [0]	(13,15)	 [0]	(13,16)	 [0]	(13,17)	 [0]	(13,18)	 [0]	(13,19)	 [0]	

(14,0)	 [0]	(14,1)	 [0]	(14,2)	 [0]	(14,3)	 [0]	(14,4)	 [0]	(14,5)	 [0]	(14,6)	 [0]	(14,7)	 [0]	(14,8)	 [0]	(14,9)	 [0]	(14,10)	 [0]	(14,11)	 [0.70]	(14,12)	 [0]	(14,13)	 [0]	(14,14)	 [0.76]	(14,15)	 [0]	(14,16)	 [0]	(14,17)	 [0]	(14,18)	 [0]	(14,19)	 [0]	

(15,0)	 [0]	(15,1)	 [0]	(15,2)	 [0]	(15,3)	 [0]	(15,4)	 [0]	(15,5)	 [0]	(15,6)	 [0]	(15,7)	 [0]	(15,8)	 [0]	(15,9)	 [0]	(15,10)	 [0]	(15,11)	 [0]	(15,12)	 [0]	(15,13)	 [0]	(15,14)	 [0]	(15,15)	 [1]	(15,16)	 [0]	(15,17)	 [0]	(15,18)	 [0.01]	(15,19)	 [0]

(16,0)	 [0]	(16,1)	 [0.19]	(16,2)	 [0]	(16,3)	 [0]	(16,4)	 [0.04]	(16,5)	 [0]	(16,6)	 [0]	(16,7)	 [0]	(16,8)	 [0.01]	(16,9)	 [0]	(16,10)	 [0]	(16,11)	 [0]	(16,12)	 [0.05]	(16,13)	 [0]	(16,14)	 [0]	(16,15)	 [0]	(16,16)	 [1]	(16,17)	 [0]	(16,18)	 [0]	(16,19)	 [0]	

(17,0)	 [0]	(17,1)	 [0]	(17,2)	 [0]	(17,3)	 [0]	(17,4)	 [0]	(17,5)	 [0]	(17,6)	 [0]	(17,7)	 [0]	(17,8)	 [0]	(17,9)	 [0]	(17,10)	 [0]	(17,11)	 [0.09]	(17,12)	 [0]	(17,13)	 [0]	(17,14)	 [0]	(17,15)	 [0]	(17,16)	 [0]	(17,17)	 [0.51]	(17,18)	 [0]	(17,19)	 [0]

(18,0)	 [0]	(18,1)	 [0]	(18,2)	 [0]	(18,3)	 [0]	(18,4)	 [0]	(18,5)	 [0.06]	(18,6)	 [0]	(18,7)	 [0]	(18,8)	 [0]	(18,9)	 [0]	(18,10)	 [0]	(18,11)	 [0]	(18,12)	 [0]	(18,13)	 [0]	(18,14)	 [0]	(18,15)	 [0]	(18,16)	 [0]	(18,17)	 [0]	(18,18)	 [0.98]	(18,19)	 [0]

(19,0)	 [0.05]	(19,1)	 [0]	(19,2)	 [0]	(19,3)	 [0]	(19,4)	 [0]	(19,5)	 [0]	(19,6)	 [0]	(19,7)	 [0]	(19,8)	 [0]	(19,9)	 [0]	(19,10)	 [0]	(19,11)	 [0]	(19,12)	 [0]	(19,13)	 [0]	(19,14)	 [0.09]	(19,15)	 [0]	(19,16)	 [0]	(19,17)	 [0]	(19,18)	 [0]	(19,19)	 [1]

};
\end{axis}
\end{tikzpicture}

%% file: main.bbl
\begin{thebibliography}{10}
\providecommand{\url}[1]{#1}
\csname url@samestyle\endcsname
\providecommand{\newblock}{\relax}
\providecommand{\bibinfo}[2]{#2}
\providecommand{\BIBentrySTDinterwordspacing}{\spaceskip=0pt\relax}
\providecommand{\BIBentryALTinterwordstretchfactor}{4}
\providecommand{\BIBentryALTinterwordspacing}{\spaceskip=\fontdimen2\font plus
\BIBentryALTinterwordstretchfactor\fontdimen3\font minus
  \fontdimen4\font\relax}
\providecommand{\BIBforeignlanguage}[2]{{%
\expandafter\ifx\csname l@#1\endcsname\relax
\typeout{** WARNING: IEEEtran.bst: No hyphenation pattern has been}%
\typeout{** loaded for the language `#1'. Using the pattern for}%
\typeout{** the default language instead.}%
\else
\language=\csname l@#1\endcsname
\fi
#2}}
\providecommand{\BIBdecl}{\relax}
\BIBdecl

\bibitem{yang2017survey}
Y.~Yang, L.~Wu, G.~Yin, L.~Li, and H.~Zhao, ``A survey on security and privacy
  issues in internet-of-things,'' \emph{IEEE Internet of Things Journal},
  vol.~4, no.~5, pp. 1250--1258, 2017.

\bibitem{shawabka2020exposing}
A.~Al-Shawabka, F.~Restuccia, S.~D'Oro, T.~Jian, B.~C. Rendon, N.~Soltani,
  J.~Dy, K.~Chowdhury, S.~Ioannidis, and T.~Melodia, ``{Exposing the
  Fingerprint: Dissecting the Impact of the Wireless Channel on Radio
  Fingerprinting},'' \emph{Proc. of INFOCOM}, 2020.

\bibitem{restuccia2019deepradioid}
F.~Restuccia, S.~D'Oro, A.~Al-Shawabka, M.~Belgiovine, L.~Angioloni,
  S.~Ioannidis, K.~Chowdhury, and T.~Melodia, ``{DeepRadioID: Real-Time
  Channel-Resilient Optimization of Deep Learning-based Radio Fingerprinting
  Algorithms},'' in \emph{Proc. of the ACM Intl. Symposium on Mobile Ad Hoc
  Networking and Computing}, 2019.

\bibitem{DARPA}
{Defense Advanced Research Projects Agency (DARPA)}, ``{The Radio Frequency
  Spectrum + Machine Learning = A New Wave in Radio Technology},''
  \url{https://www.darpa.mil/news-events/2017-08-11a}, 2017.

\bibitem{sankhe2019oracle}
K.~Sankhe, M.~Belgiovine, F.~Zhou, S.~Riyaz, S.~Ioannidis, and K.~Chowdhury,
  ``{ORACLE: Optimized Radio clAssification through Convolutional neuraL
  nEtworks},'' in \emph{IEEE INFOCOM 2019}.\hskip 1em plus 0.5em minus
  0.4em\relax IEEE, 2019.

\bibitem{merchant2018deep}
K.~Merchant, S.~Revay, G.~Stantchev, and B.~Nousain, ``{Deep Learning for RF
  Device Fingerprinting in Cognitive Communication Networks},'' \emph{IEEE
  Journal of Selected Topics in Signal Processing}, vol.~12, no.~1, pp.
  160--167, 2018.

\bibitem{al2021deeplora}
A.~Al-Shawabka, P.~Pietraski, S.~B. Pattar, F.~Restuccia, and T.~Melodia,
  ``Deeplora: Fingerprinting lora devices at scale through deep learning and
  data augmentation,'' in \emph{Proceedings of the Twenty-second International
  Symposium on Theory, Algorithmic Foundations, and Protocol Design for Mobile
  Networks and Mobile Computing}, 2021, pp. 251--260.

\bibitem{soltanimore}
N.~Soltani, K.~Sankhe, J.~Dy, S.~Ioannidis, and K.~Chowdhury, ``More is better:
  Data augmentation for channel-resilient rf fingerprinting.''

\bibitem{jian2020deep}
T.~Jian, B.~C. Rendon, E.~Ojuba, N.~Soltani, Z.~Wang, K.~Sankhe, A.~Gritsenko,
  J.~Dy, K.~Chowdhury, and S.~Ioannidis, ``Deep learning for rf fingerprinting:
  A massive experimental study,'' \emph{IEEE Internet of Things Magazine},
  vol.~3, no.~1, pp. 50--57, 2020.

\bibitem{zhu2017unpaired}
J.-Y. Zhu, T.~Park, P.~Isola, and A.~A. Efros, ``Unpaired image-to-image
  translation using cycle-consistent adversarial networks,'' in
  \emph{Proceedings of the IEEE international conference on computer vision},
  2017, pp. 2223--2232.

\bibitem{pix2pix2017}
P.~Isola, J.-Y. Zhu, T.~Zhou, and A.~A. Efros, ``Image-to-image translation
  with conditional adversarial networks,'' \emph{CVPR}, 2017.

\bibitem{goodfellow2014generative}
I.~Goodfellow, J.~Pouget-Abadie, M.~Mirza, B.~Xu, D.~Warde-Farley, S.~Ozair,
  A.~Courville, and Y.~Bengio, ``Generative adversarial nets,'' \emph{Advances
  in neural information processing systems}, vol.~27, 2014.

\bibitem{piva2021tags}
M.~Piva, G.~Maselli, and F.~Restuccia, ``The tags are alright: Robust
  large-scale rfid clone detection through federated data-augmented radio
  fingerprinting,'' \emph{arXiv preprint arXiv:2105.03671}, 2021.

\bibitem{mao2017least}
X.~Mao, Q.~Li, H.~Xie, R.~Y. Lau, Z.~Wang, and S.~Paul~Smolley, ``Least squares
  generative adversarial networks,'' in \emph{Proceedings of the IEEE
  international conference on computer vision}, 2017, pp. 2794--2802.

\bibitem{kingma2013auto}
D.~P. Kingma and M.~Welling, ``Auto-encoding variational bayes,'' \emph{arXiv
  preprint arXiv:1312.6114}, 2013.

\bibitem{makhzani2015adversarial}
A.~Makhzani, J.~Shlens, N.~Jaitly, I.~Goodfellow, and B.~Frey, ``Adversarial
  autoencoders,'' \emph{arXiv preprint arXiv:1511.05644}, 2015.

\bibitem{li2016precomputed}
C.~Li and M.~Wand, ``Precomputed real-time texture synthesis with markovian
  generative adversarial networks,'' in \emph{European conference on computer
  vision}.\hskip 1em plus 0.5em minus 0.4em\relax Springer, 2016, pp. 702--716.

\bibitem{isola2017image}
P.~Isola, J.-Y. Zhu, T.~Zhou, and A.~A. Efros, ``Image-to-image translation
  with conditional adversarial networks,'' in \emph{Proceedings of the IEEE
  conference on computer vision and pattern recognition}, 2017, pp. 1125--1134.

\bibitem{cgan}
\BIBentryALTinterwordspacing
J.~Brownlee. (2019) How to implement cyclegan models from scratch with keras.
  [Online]. Available:
  \url{https://machinelearningmastery.com/how-to-develop-cyclegan-models-from-scratch-with-keras/}
\BIBentrySTDinterwordspacing

\bibitem{AlexNet}
A.~Krizhevsky, I.~Sutskever, and G.~E. Hinton, ``{ImageNet Classification with
  Deep Convolutional Neural Networks}.''

\bibitem{IEEE-WiFi-Standard}
IEEE, ``{IEEE Standard for Information Technology--Telecommunications and
  Information Exchange Between Systems Local and Metropolitan Area
  Networks--Specific Requirements Part 11: Wireless LAN Medium Access Control
  (MAC) and Physical Layer (PHY) Specifications - Redline},'' \emph{IEEE Std
  802.11-2012 (Revision of IEEE Std 802.11-2007) - Redline}, pp. 1--5229, March
  2012.

\bibitem{bertizzolo2019arena}
L.~Bertizzolo, L.~Bonati, E.~Demirors, and T.~Melodia, ``Arena: A 64-antenna
  sdr-based ceiling grid testbed for sub-6 ghz radio spectrum research,'' in
  \emph{Proc. of the 13th Intl Workshop on Wireless Network Testbeds}, 2019.

\bibitem{sundaram2019survey}
J.~P.~S. Sundaram, W.~Du, and Z.~Zhao, ``A survey on lora networking: Research
  problems, current solutions, and open issues,'' \emph{IEEE Communications
  Surveys \& Tutorials}, vol.~22, no.~1, pp. 371--388, 2019.

\end{thebibliography}
